%% file: main.tex
\newif\ifdouble
\newif\ifsingle
\newif\ifchange
\doubletrue

\ifdouble
  \documentclass[sigconf, usenames, dvipsnames]{acmart}
\else
  \documentclass[manuscript,anonymous]{acmart}
  \singletrue
\fi

\usepackage{dblfloatfix} % for figure placement at bottom
\usepackage{booktabs} % multi column
\usepackage{arydshln}
\usepackage{subfig}
\usepackage{soul}
\usepackage{booktabs}
\usepackage{pdfpages}
\copyrightyear{2024}
\acmYear{2024}
\setcopyright{acmlicensed}\acmConference[DIS '24]{Designing Interactive Systems Conference}{July 1--5, 2024}{IT University of Copenhagen, Denmark}
\acmBooktitle{Designing Interactive Systems Conference (DIS '24), July 1--5, 2024, IT University of Copenhagen, Denmark}
\acmDOI{10.1145/3643834.3661631}
\acmISBN{979-8-4007-0583-0/24/07}

\newcommand{\diff}[1]{\textcolor{purple}{#1}}

\renewcommand{\diff}[1]{#1}

\AtBeginDocument{%
  \providecommand\BibTeX{{%
    \normalfont B\kern-0.5em{\scshape i\kern-0.25em b}\kern-0.8em\TeX}}}

\newcommand{\system}{RealityEffects}

\begin{document}
\pagenumbering{arabic}
\pagestyle{plain}
\title{\system{}: Augmenting 3D Volumetric Videos with Object-Centric Annotation and Dynamic Visual Effects}

\author{Jian Liao}
\affiliation{%
  \institution{University of Calgary}
  \city{Calgary}
  \country{Canada}}
\email{jian.liao1@alumni.ucalgary.ca}

\author{Kevin Van}
\affiliation{%
  \institution{University of Calgary}
  \city{Calgary}
  \country{Canada}}
\email{kevin.van@ucalgary.ca}

\author{Zhijie Xia}
\affiliation{%
  \institution{University of Calgary}
  \city{Calgary}
  \country{Canada}}
\email{zhijie.xia@ucalgary.ca}

\author{Ryo Suzuki}
\affiliation{%
  \institution{University of Calgary}
  \city{Calgary}
  \country{Canada}}
\email{ryo.suzuki@ucalgary.ca}

\renewcommand{\shortauthors}{Liao, Van, Xia, and Suzuki}

\input{0-abstract}

\begin{teaserfigure}
\centering
\includegraphics[width=0.245\textwidth]{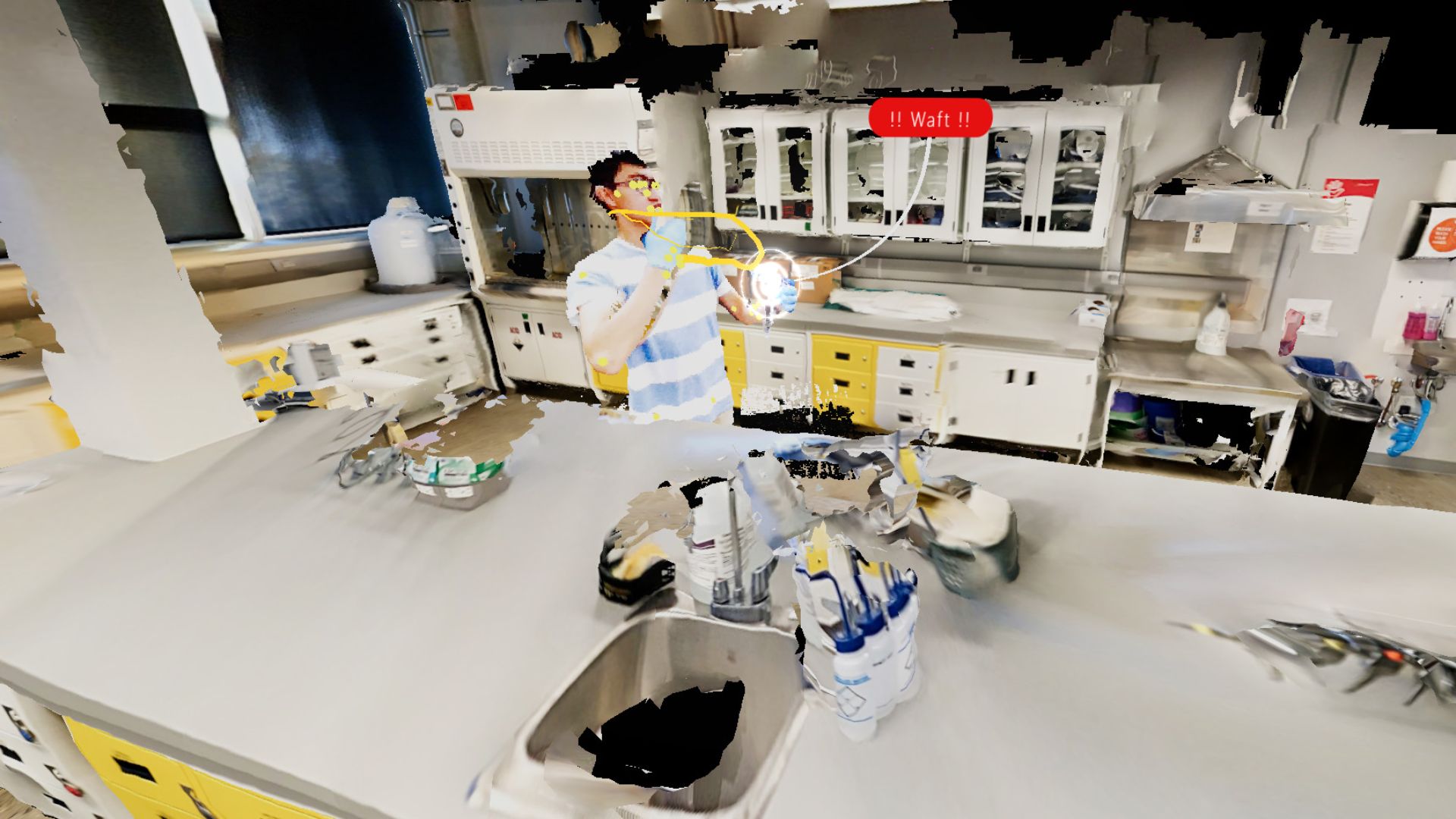}
\includegraphics[width=0.245\textwidth]{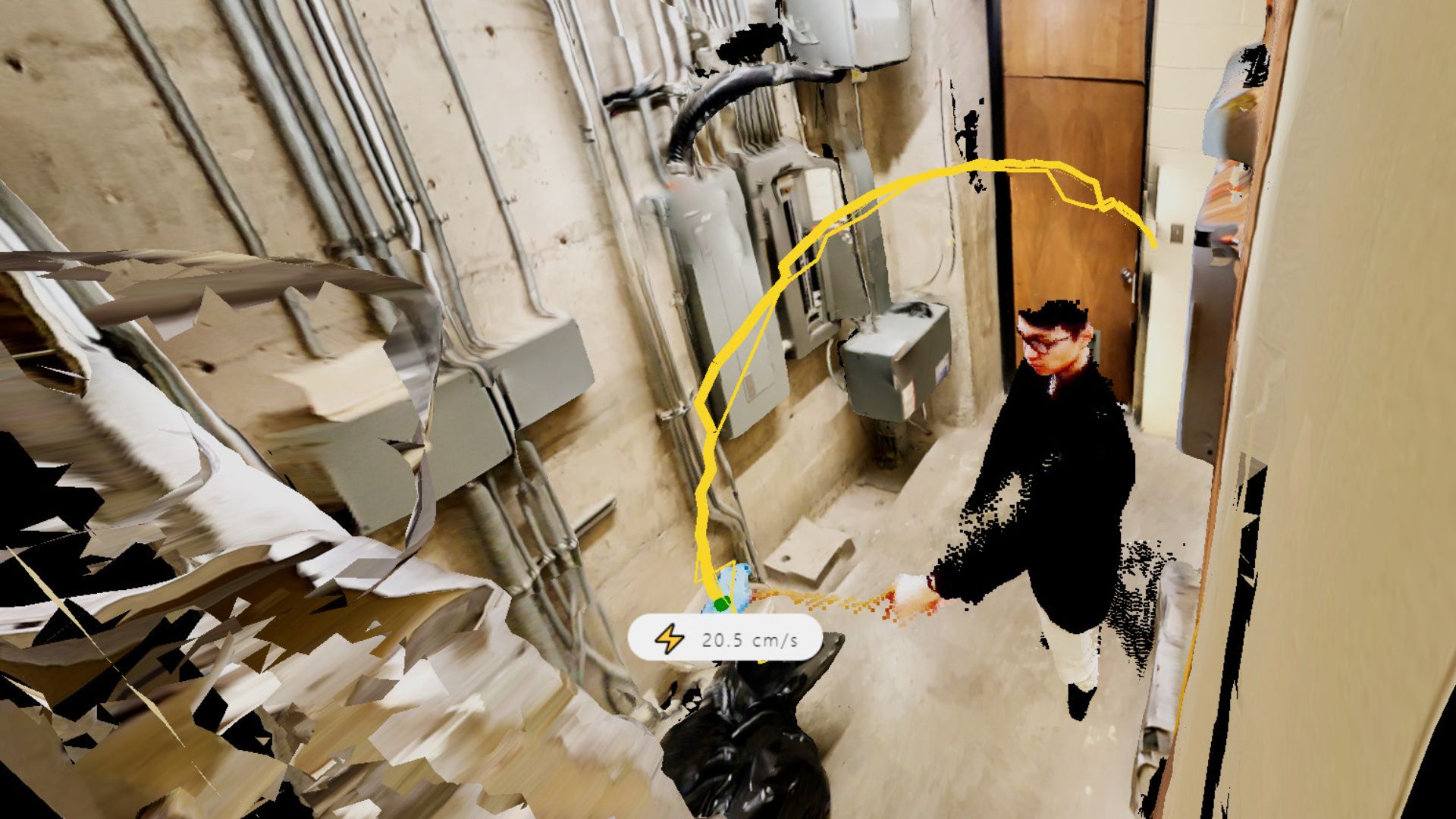}
\includegraphics[width=0.245\textwidth]{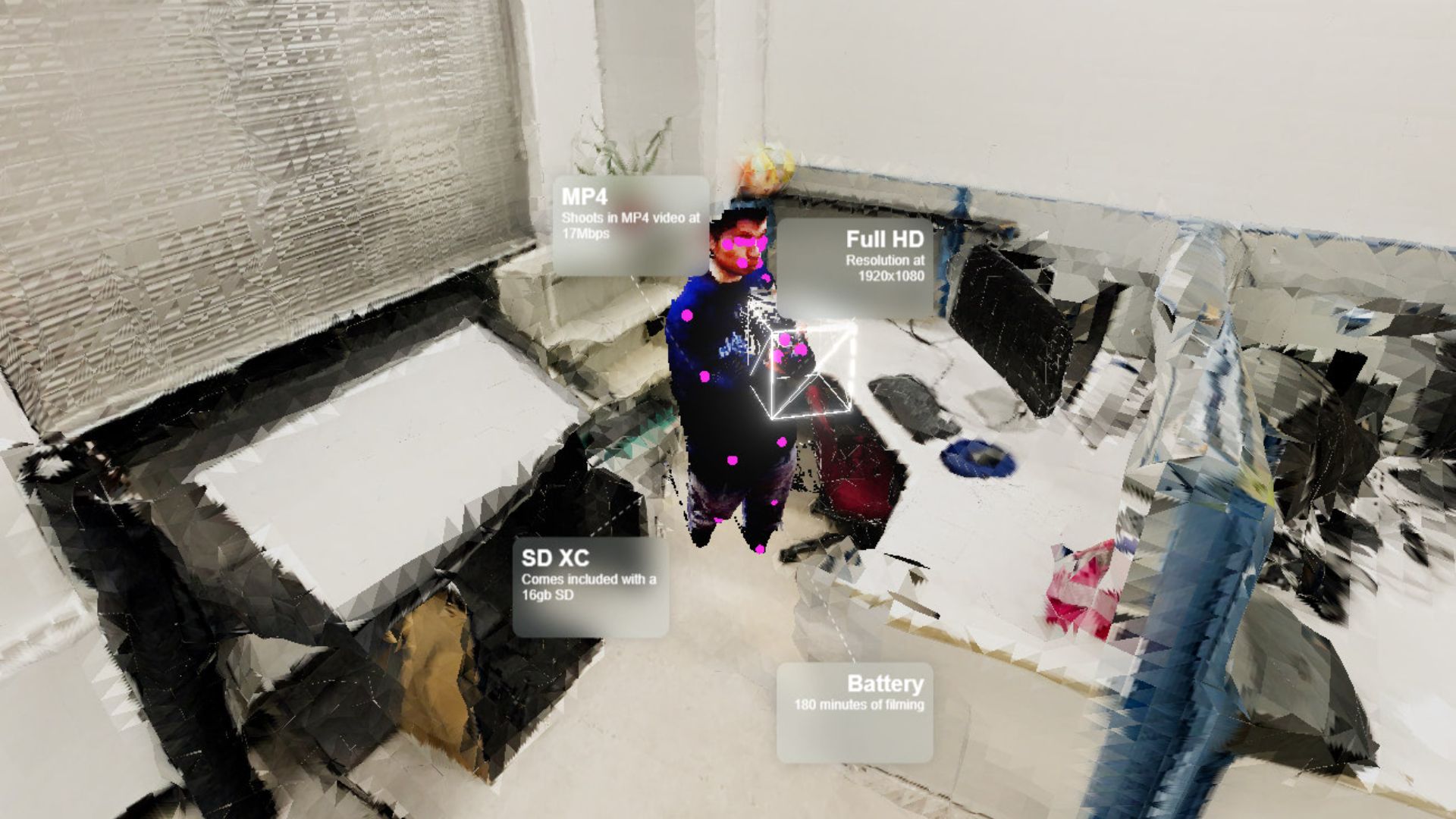}
\includegraphics[width=0.245\textwidth]{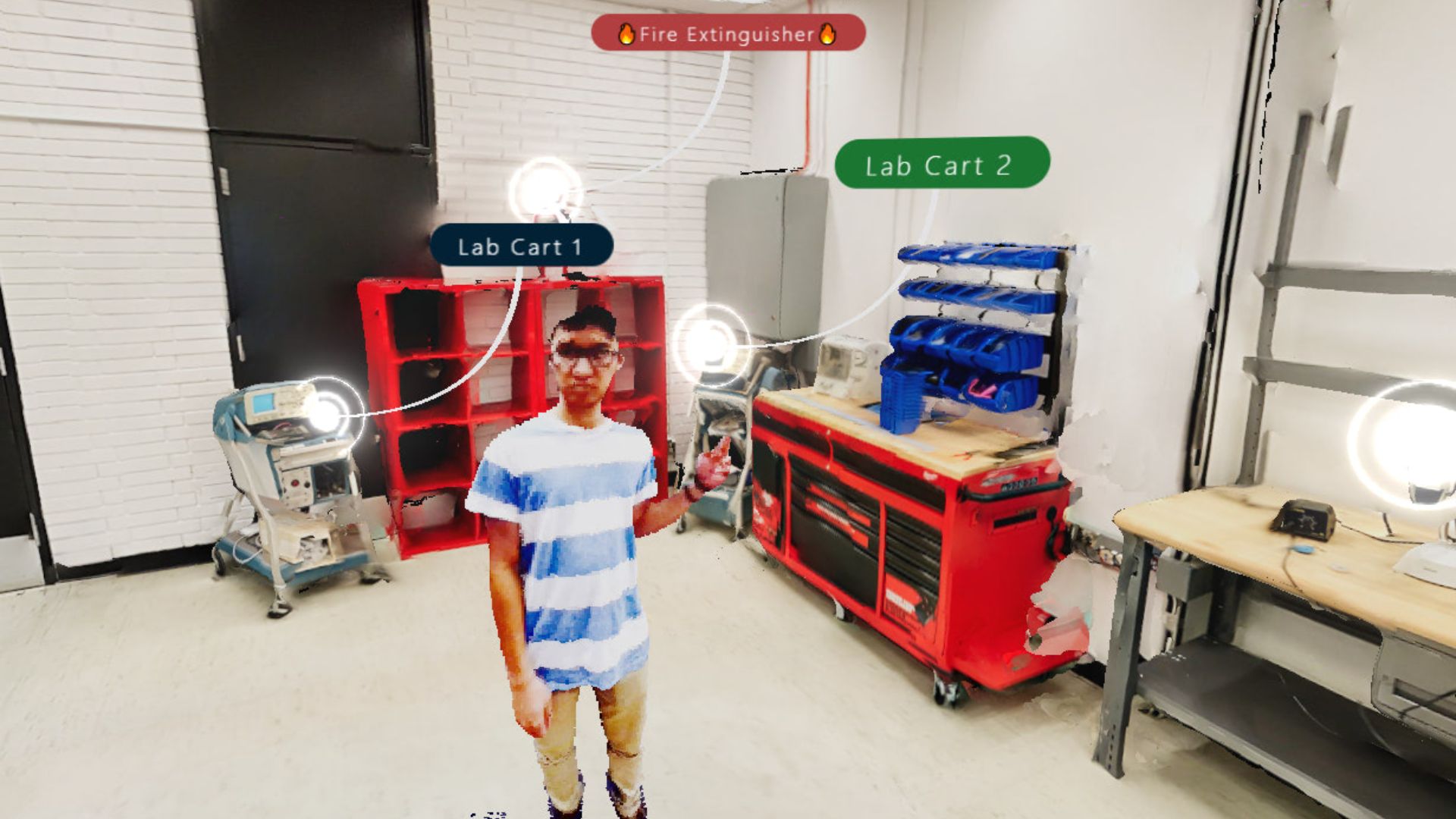}
\caption{\system{} is a desktop authoring interface to augment 3D volumetric videos with object-centric annotation and visual effects.}
\label{fig:teaser}
\end{teaserfigure}

\maketitle

\input{1-introduction}
\input{2-related-work}

\input{3-design-space}
\input{4-system}

\input{5-application}
\input{6-user-study}
\input{7-future-work}

\input{8-conclusion}

\input{acknowledgements}

\ifdouble
  \balance
\fi
\bibliographystyle{ACM-Reference-Format}
\bibliography{references}

\input{appendix}

\end{document}

%% file: 0-abstract.tex
\begin{abstract}
This paper introduces RealityEffects, a desktop authoring interface designed for editing and augmenting 3D volumetric videos with object-centric annotations and visual effects. RealityEffects enhances volumetric capture by introducing a novel method for augmenting captured physical motion with embedded, responsive visual effects, referred to as \textit{object-centric augmentation}. In RealityEffects, users can interactively attach various visual effects to physical objects within the captured 3D scene, enabling these effects to dynamically move and animate in sync with the corresponding physical motion and body movements. The primary contribution of this paper is the development of a taxonomy for such object-centric augmentations, which includes annotated labels, highlighted objects, ghost effects, and trajectory visualization. This taxonomy is informed by an analysis of 120 edited videos featuring object-centric visual effects. The findings from our user study confirm that our direct manipulation techniques lower the barriers to editing and annotating volumetric captures, thereby enhancing interactive and engaging viewing experiences of 3D volumetric videos.\end{abstract}


\ccsdesc[500]{Human-centered computing~Mixed / augmented reality}
% http://dl.acm.org/ccs.cfm

\keywords{Volumetric Video; Authoring Interface; Mixed Reality; Augmented Visual Effects; Object-Centric Annotation}

%% file: 1-introduction.tex
\section{Introduction}
In recent years, \textit{augmented videos}~\cite{kolbe2004augmented, saquib2019interactive, chen2021augmenting}---live or recorded 2D videos enhanced with embedded visual effects---have gained an increasing popularity in human-computer interaction (HCI).  
By seamlessly integrating visual effects with physical motion, augmented videos provide more interactive and engaging viewing experiences, similar to augmented and mixed reality, but on a screen. 
Traditionally, creating such augmented videos requires significant time and expertise using professional video-editing software like Adobe Premiere Pro. However, recent HCI research has enabled interactive and improvisational authoring experiences, simplifying the creation of these augmented live or recorded videos in various applications, such as sports analysis (e.g., \textit{VisCommentator}~\cite{chen2021augmenting}), classroom education (e.g., \textit{RealitySketch}~\cite{suzuki2020realitysketch}), storytelling (e.g., \textit{Interactive Body-Driven Graphics}~\cite{saquib2019interactive}), interactive data visualization (e.g., \textit{Augmented Chironomia}~\cite{hall2022augmented}), live presentation (e.g., \textit{RealityTalk}~\cite{liao2022realitytalk}), and entertainment (e.g., \textit{PoseTween}~\cite{liu2020posetween}).

However, these works primarily focus on augmented \textit{2D videos}, and to the best of our knowledge, no prior work has explored augmented \textit{3D volumetric videos}. \diff{Especially with the recent releases of sophisticated mixed reality headsets like the Apple Vision Pro, spatial and 3D volumetric videos become an emerging entertainment medium in the mainstream consumer market}.
Despite the recent proliferation of 3D volumetric capture technologies, such as point-cloud rendering or reconstructed 3D capture with depth cameras or LiDAR sensors, editing and augmenting these volumetric videos remains challenging. 
Existing tools like \textit{4Dfx}~\cite{4dfx}, \textit{DepthKit Studio}~\cite{depthkit}, and \textit{HoloEdit}~\cite{holoedit} offer only basic video touch-ups and timeline manipulation. Consequently, users must either edit these effects frame-by-frame, similar to the traditional 2D video-editing techniques, or program the behavior of visual effects within a 3D game environment, such as Unity or Unreal Engine.

In this paper, we present \system{}, a desktop authoring interface that supports the real-time and interactive creation of augmented 3D volumetric videos.
To augment the volumetric 3D scene, users can simply select and bind captured physical objects with annotated visual effects. The system then automatically tracks physical objects such that the embedded visual effects can move and respond dynamically with the corresponding physical motion and body movement. We call this approach \textit{object-centric augmentation}, which can significantly reduce the time and cost of creating augmented volumetric videos.
Unlike 2D videos, the augmented 3D scene allows free-viewpoint movement, enabling immersive viewing experiences.

To design our system, we collected 120 video examples utilizing the video-edited object-centric augmentation. Based on the observed common augmentation techniques, we contribute a taxonomy of object-centric augmentations for 3D volumetric videos, which includes annotated labels, highlighted objects, ghost effects, and trajectory visualization. 
Along with the novel direct manipulation authoring, \system{} extends the idea of previously explored volumetric augmentation (e.g., \textit{Remixed Reality}~\cite{lindlbauer2018remixed}) to support more comprehensive visual effects that can be used in a wide range of applications, such as sports analysis, physics education, classroom tutorials, and live presentations.
We evaluated our system with a lab-based usability study (N=19). Our study results suggest that object-centric augmentation is a promising way to lower the barrier to editing and annotating volumetric captures while allowing flexible and expressive video augmentation.

Finally, our paper contributes to: 
\begin{enumerate}
\item A taxonomy and design space of object-centric augmentation for 3D volumetric captures, based on the analysis of existing object-centric 2D video augmentation techniques.
\item \system{}, a tool for creating augmented 3D volumetric videos that leverage a novel direct manipulation technique to bind dynamic visual effects with corresponding physical motion.
\item Application demonstration and user evaluation of \system{}, which suggests the untapped potential of augmented volumetric captures for more interactive and engaging viewing experiences.
\end{enumerate}

%% file: 2-related-work.tex
\section{Related Work}

\subsection{Volumetric Capturing and Editing}
\subsubsection{Volumetric Capture and Its Applications}
Volumetric captures or videos refer to the technique of capturing 3D space and subsequently viewing it on a screen with free-viewpoint movement. These techniques have been explored since the 1990s (e.g., \textit{Virtualized Reality}~\cite{kanade1997virtualized}), but recent research has greatly advanced this domain in both high-quality 3D reconstruction (e.g., \textit{Fusion4D}~\cite{dou2016fusion4d}, \textit{Montage4D}~\cite{du2018montage4d}, \textit{Relightables}~\cite{guo2019relightables}, \textit{VolumeDeform}~\cite{innmann2016volumedeform}) and more accessible volumetric capturing with mobile phones (e.g., \textit{Kinect Fusion}~\cite{izadi2011kinectfusion}, \textit{DepthLab}~\cite{du2020depthlab}, \textit{Polycam}~\cite{polycam}). 
With recent advances in commercially-available depth cameras like Kinect, volumetric captures have been used in various applications such as telepresence (e.g., \textit{Holoportation}~\cite{orts2016holoportation}, \textit{JackInSpace}~\cite{komiyama2017jackin}, \textit{Project Starline}~\cite{lawrence2021project}, \textit{PhotoPortals}~\cite{kunert2014photoportals}), remote collaboration (e.g., \textit{RemoteFusion}~\cite{adcock2013remotefusion}, \textit{Mini-Me}~\cite{piumsomboon2018mini}, \textit{On the Shoulder of Giant}~\cite{piumsomboon2019shoulder}, \textit{Virtual Makerspaces}~\cite{radu2021virtual}), remote hands-on instruction (e.g., \textit{Loki}~\cite{thoravi2019loki}, \textit{BeThere}~\cite{sodhi2013bethere}, \textit{3D Helping Hands}~\cite{tecchia20123d}), and immersive tutorials for physical tasks (e.g., \textit{MobileTutAR}~\cite{cao2022mobiletutar}, \textit{ProcessAR}~\cite{chidambaram2021processar}, \textit{My Tai Chi Coaches}~\cite{han2017my}).
Past research has utilized static or live 3D reconstructed scenes for remote MR collaboration, facilitating more immersive interactions with remote users~\cite{tait2015effect, gao2016oriented, teo2019mixed}. 
Alternatively, live 3D reconstruction has been used to facilitate co-located communications for VR users (e.g., \textit{Slice of Light}~\cite{wang2020slice}, \textit{Asynchronous Reality}~\cite{fender2022causality}).
These captured 3D geometries are also used for anchoring virtual elements (e.g., \textit{SnapToReality}~\cite{nuernberger2016snaptoreality}, \textit{SemanticAdapt}~\cite{cheng2021semanticadapt}), creating virtual contents (e.g., \textit{SweepCanvas}~\cite{li2017sweepcanvas}, \textit{Window-Shaping}~\cite{huo2017window}), or generating virtual environments (e.g., \textit{VRoamer}~\cite{cheng2019vroamer}, \textit{Oasis}~\cite{sra2017oasis, sra2016procedurally}) by leveraging object detection and semantic segmentation of volumetric scenes (e.g., \textit{SemanticPaint}~\cite{valentin2015semanticpaint}, \textit{ScanNet}~\cite{dai2017scannet}).

\subsubsection{Augmenting and Editing Volumetric Capture}
More closely related to our work, past work has also explored further blending the virtual and physical worlds by augmenting captured volumetric scenes or the real world. 
By using VR/MR devices, systems can alternate the captured scene by erasing physical objects (e.g., \textit{SceneCtrl}~\cite{yue2017scenectrl}, \textit{Diminished Reality}~\cite{cheng2022towards, mori2017survey}) or replacing them with virtual ones (e.g., \textit{RealityCheck}~\cite{hartmann2019realitycheck}, \textit{TransforMR}~\cite{kari2021transformr}). 
Alternatively, previous work has used the depth information to blend virtual augmentation into the real-world with projection mapping (e.g., \textit{IllumiRoom}~\cite{jones2013illumiroom}, \textit{RoomAlive}~\cite{jones2014roomalive}, \textit{Room2Room}~\cite{pejsa2016room2room}, \textit{Dyadic Projected SAR}~\cite{benko2014dyadic}, \textit{OptiSpace}~\cite{fender2018optispace}). 
Systems like \textit{Mixed Voxel Reality}~\cite{regenbrecht2017mixed}, \textit{Remixed Reality}~\cite{lindlbauer2018remixed}, and \diff{\textit{Virtual Reality Annotator}~\cite{ribeiro2018virtual}} further advance this approach by augmenting the volumetric scene by leveraging both spatial manipulation (copy, erase, move), temporal modification (record, playback, loop), and \diff{volumetric annotation (sketches)} with a VR headset and live 3D reconstruction.

While these works partially demonstrated the visual augmenting of captured scenes, supported augmentation techniques remain simple (e.g., appearance change for color or texture). 
Moreover, since their focus is on the immersive experience of these modified scenes, the authoring aspect of these volumetric scenes and videos is not well explored in the literature.
Our focus is rather on the authoring interface, which can support more comprehensive visual augmentation for the volumetric scenes. 
This is because the current work on authoring tools or video-editing tools for volumetric capture is either focused on static scenes (e.g., \textit{DistanciAR}~\cite{wang2021distanciar}), timeline manipulation (e.g., \textit{4Dfx} \cite{4dfx}), or simple video touch-ups (e.g., \textit{DepthKit Studio} \cite{depthkit}, \textit{HoloEdit} \cite{holoedit}). 
In contrast, \system{} enables more expressive visual augmentation for dynamic volumetric scenes by leveraging \textit{object-centric augmentation}, which we take inspiration from 2D video authoring, as described next.

\subsection{Authoring Augmented 2D Videos}
In the context of 2D videos or mobile AR interfaces, \textit{augmented videos} refer to a live or recorded video in which embedded visuals are seamlessly coupled with captured physical objects~\cite{kolbe2004augmented, saquib2019interactive, chen2021augmenting}. 
Systems like \textit{PoseTween}~\cite{liu2020posetween}, and \textit{Interactive Body-Driven Graphics}~\cite{saquib2019interactive} demonstrate the interactive authoring tools for generating responsive graphics that can move with the corresponding body movement in the live or recorded video.
Such visual augmentation can provide more engaging experiences for live presentations (e.g., \textit{RealityTalk}~\cite{liao2022realitytalk}, \textit{Augmented Chironomia}~\cite{hall2022augmented}) sports training (e.g., \textit{VisCommentator}~\cite{chen2021augmenting}, \textit{EventAnchor}~\cite{deng2021eventanchor}, \textit{YouMove}~\cite{anderson2013youmove}), storytelling (e.g., \textit{RealityCanvas}~\cite{xia2023realitycanvas}), and education (e.g., \textit{HoloBoard}~\cite{gong2021holoboard}, \textit{Sketched Reality}~\cite{kaimoto2022sketched}). 
Moreover, augmented videos are also useful media for prototyping AR experiences (e.g., \textit{Pronto}~\cite{leiva2020pronto}, \textit{Rapido}~\cite{leiva2021rapido}, \textit{Teachable Reality}~\cite{monteiro2023teachable}) or remote collaboration (e.g., \textit{In-Touch with the Remote World}~\cite{gauglitz2014touch, gauglitz2014world}). 

Traditionally, these videos require professional video-editing skills, but HCI researchers have investigated end-user authoring tools to lower the barrier of expertise.
In particular, taking inspiration from object-based video navigation techniques~\cite{karrer2009pocketdragon, walther2012dragimation, nguyen2013direct, nguyen2012video, nguyen2014direct, santosa2013direct}, 
Goldman et al.~\cite{goldman2008video} and Silvia et al.~\cite{silva2012real} explored object-centric video annotation, which allows users to add dynamic annotation based on the tracked object in the 2D video. 
More recently, systems like \textit{RealitySketch}~\cite{suzuki2020realitysketch}, \textit{RealityCanvas}~\cite{xia2023realitycanvas}, \textit{VideoDoodles}~\cite{yu2023videodoodles}, and \textit{Graphiti}~\cite{saquib2022graphiti} have further expanded the object-centric augmentation for dynamic AR sketching interfaces. 
However, to the best of our knowledge, no prior work has explored these techniques for 3D volumetric videos, which introduce the additional interaction challenge of selecting or aligning objects in 3D scenes~\cite{hudson2016understanding, montano2020slicing}. 
This paper contributes to the first object-centric augmentation for \textit{3D volumetric video}, along with a taxonomy of possible augmentation design.

\subsection{Object-Centric Immersive Visualization}
Our design for spatial annotations and visual effects is also inspired by various object-centric immersive visualization and visual analytics techniques~\cite{decamp2010immersive}.
Previous works have explored various spatio-temporal visualization techniques, such as spatial and semantic object annotation (e.g., \textit{ReLive}~\cite{hubenschmid2022relive}, \textit{Skeletonotator}~\cite{lee2019semantic}), trajectories of objects (e.g.,  \textit{MIRIA}~\cite{buschel2021miria}), trajectories of human motion (e.g., \textit{AvatAR}~\cite{reipschlager2022avatar}, \textit{Reactive Video}~\cite{clarke2020reactive}, \textit{DemoDraw}~\cite{chi2016authoring}), ghost effects (e.g., \textit{GhostAR}~\cite{cao2019ghostar}), object and location highlights (e.g., Kepplinger et al.~\cite{kepplinger2020see}), and heatmap visualizations (e.g., \textit{HeatSpace}~\cite{fender2017heatspace}, \textit{EagleView}~\cite{brudy2018eagleview}).
These free-viewpoint movements and multi-viewpoint analyses can greatly improve the way we watch and analyze object- and body-related movements with deeper insights~\cite{sugita2018browsing, yu2020perspective, brudy2018eagleview, kloiber2020immersive}. 
While our tool is inspired by these works, our focus lies on the \textit{authoring aspect} of these dynamic effects and visualizations, rather than developing novel visualization systems. For example, we designed our system in a way that end-users can easily select, bind, and visualize motion data without any pre-defined programs or configurations.
We believe our tool along with the direct manipulation authoring approach, allows flexible and customizable volumetric video editing that can be used for broader applications beyond these visual analytics tools.

% room capture~\cite{sankar2017interactive}
% Soccer on Your Tabletop~\cite{rematas2018soccer}
% DollhouseVR~\cite{ibayashi2015dollhouse}
% Changing the Appearance~\cite{lindlbauer2017changing}
% AfterMath~\cite{leigh2015aftermath}

%% file: 3-design-space.tex
\section{A Taxonomy of Object-Centric Augmentation}
\subsection{A Taxonomy Analysis}
To better understand common practices and techniques for object-centric augmentations, we first collected and analyzed a set of 120 existing videos available on the Internet, most of which were created using professional video-editing software. These examples showcase a variety of techniques and collectively contribute to a \diff{preliminary taxonomy} of object-centric visual augmentation, helping the design of end-user systems for authoring these effects. 

\subsubsection{Definition of Object-Centric Augmentation}
To design our system feature, we first need to understand and investigate common practices for \textit{object-centric augmentation}. In this paper, object-centric augmentation refers to \textit{``a class of virtual elements 1) that are embedded and spatially integrated with objects in a scene, and 2) whose properties change, respond, and animate based on the behaviors of physical objects in the scene.''}.
Here, \textit{virtual elements} can include text, images, visual effects, and visualization; \textit{objects} can be physical objects, parts of the human body, or environments; and \textit{properties} can encompass location, orientation, scale, and other visual properties. 

\subsubsection{Motivation and Goal}
While object-centric augmentations are frequently used in many professional videos and several works explore this domain~\cite{goldman2008video}, these works lack a taxonomy analysis~\cite{suzuki2020realitysketch, liu2020posetween, goldman2008video, saquib2022graphiti} or focus on more specific domains such as presentations~\cite{liao2022realitytalk}, storytelling~\cite{saquib2019interactive} or robotics~\cite{suzuki2022augmented}, leaving a gap in the holistic understanding of possible designs, even for 2D videos and, certainly, for 3D volumetric videos.
\diff{The goal of this taxonomy analysis is to provide initial insights into object-centric augmentations. We have adapted methods from similar prior research papers~\cite{liao2022realitytalk, xia2023realitycanvas, monteiro2023teachable} to provide an initial and preliminary taxonomy of a representative subset of common practices, recognizing that conducting a systematic visual search of videos is more challenging than conducting a systematic search of research papers.}

\begin{figure*}[h!]
\includegraphics[width=\textwidth]{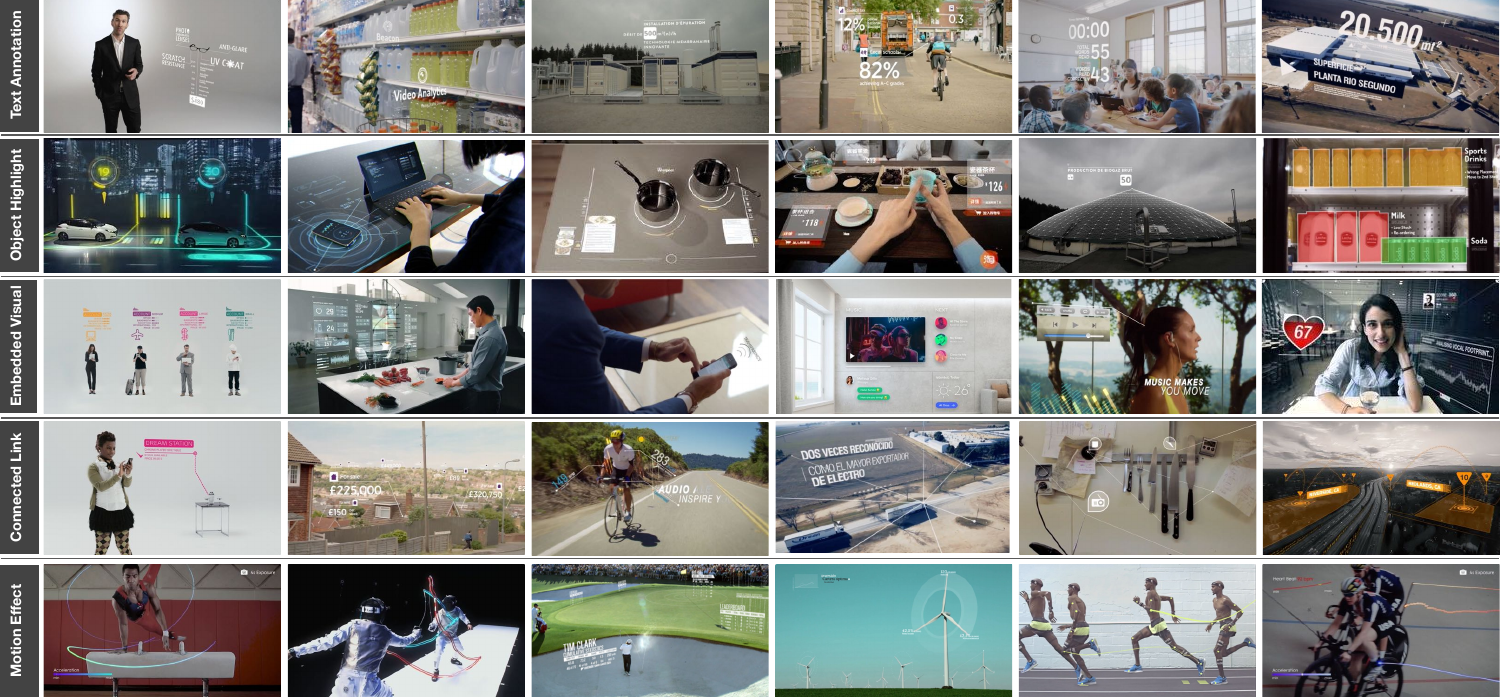}
\caption{Design Space Analysis: We collected a set of 120 existing edited videos and images and observed the five most common design techniques, namely, Text Annotation, Object Highlight, Embedded Visual, Connected Link, and Motion Effect. \diff{The screenshots are copyrighted by each video creator. We listed the link for each video in the Appendix.}}
\label{fig:design-space}
\end{figure*}

\subsubsection{Corpus and Dataset}
To collect the video examples, the authors \diff{(A1, A2, and A4)} manually searched popular video and image search platforms (e.g., YouTube, Pinterest, Vimeo, Behance, and Google Images), primarily relying on visual searches, as these videos are not associated with a specific keyword like \textit{``3D visual effects''}.
\diff{After some initial filtering, we started to identify some patterns in the visuals we collected, and with the help of the similar image suggestion feature on Pinterest, we expanded more visual search criteria like annotations, highlights, augmented effects, labels, floating text, floating screen, analysis, visualization, and motion.
We also did a reverse search to find the videos, through this process, we first collected 200 videos.}
Note that there is a much smaller proportion of examples for 3D volumetric videos, none of them are volumetric videos, while most of them feature 3D visual effects.
Then, the authors (\diff{A1, A2, and A4}) filter out by focusing only on the object-centric augmentation (e.g., removing videos that use entirely virtual effects without physical objects or visual effects that are not associated with the physical objects). After the filtering process, we obtained 120 videos \diff{that contain object-centric augmentation based on our definition}, 

\subsubsection{Coding Methodology}
\diff{We analyzed all 120 collected videos to identify snippets displaying annotations and visual effects, capturing screenshots of object-centric visual effects from each. Through this process, one of the authors (A1) led the collection of screenshots, with assistance from another author (A2), resulting in a total of 336 screenshots, averaging 2.8 screenshots per video. 
We chose screenshots over full videos as a coding corpus because each video may contain different techniques, and these screenshots served as representative keyframes for our taxonomy analysis.}
\diff{Subsequently, we conducted open coding to identify a preliminary taxonomy of object-centric visual effects. With the 336 collected representative images, author A1 led the initial open coding process to identify a first approximation of the dimensions and categories, then iterated with other authors (A2 and A4) on digital whiteboards (Miro and Google Slides). 
In this process, the three authors independently reviewed the collected screenshots and refined the taxonomy initially identified by A1.
Subsequently, all authors reflected on the initial design space to discuss the consistency and comprehensiveness of the categorization.
Finally, after systematic coding by authors A1 and A2, which involved individual tagging for the complete dataset, we reviewed the tagging to resolve discrepancies and obtain final coding results.} All authors then reflected on the design space and finalized the categorization by merging, expanding, and removing categories.

\subsubsection{Limitations}
\diff{We acknowledge several limitations in our current methodologies, including corpus selection and taxonomy analysis. First, our selected videos may not represent a comprehensive and exhaustive corpus. While we aimed to collect as diverse a dataset as possible, the nature of our visual search, rather than a systematic keyword search, limits our ability to claim comprehensive representation. Second, the taxonomy analysis might have benefited from the involvement of the video creators to better capture the design space from their perspectives. Despite these limitations, we believe this taxonomy can help identify common practices and techniques for object-centric augmentation, benefiting both our own and other HCI research.}

\subsection{Design Space of Object-Centric Augmentation}
Based on the analysis, we identified the following five most common augmentation techniques: 1) text annotations, 2) object highlights, 3) embedded visuals, 4) connected links, and 5) motion effects (Figure~\ref{fig:design-space}).

\subsubsection{Text Annotation}
Text annotation is one of the most common techniques identified. It involves attaching textual labels or descriptions to physical objects. These can be \textit{static descriptions}, providing information about the object, or \textit{dynamic data and parameters}, such as speed, distance, or price, akin to embedded data visualization~\cite{willett2016embedded}. The attached objects can be \textit{graspable physical items}, parts of the \textit{human body}, or \textit{stationary locations} like buildings or furniture.

\subsubsection{Object Highlight}
Object highlight is a technique used to visually attract an audience's attention to a specific object. For example, object highlight techniques include changing the \textit{color} of the object, highlighting the \textit{contour} of the object, adding highlighting \textit{marks} to the object, or changing the \textit{opacity} of other objects. These object highlights can be applied to either \textit{2D surfaces}, such as showing a colored circle on the ground around cars, phones, or pans, or \textit{3D objects}, such as displaying a bounding box and sphere or 3D mesh of the target object.

\subsubsection{Embedded Visual}
Embedded visuals are 2D images or visual information attached to describe objects, similar to text annotation but through static visuals or animations. Embedded visuals include \textit{simple icons} to describe the object, \textit{2D images and photos} to show the associated information, \textit{animation} to visually describe the behavior, \textit{screens} to display the associated website or user interfaces, and \textit{charts or graphs} to visualize the associated data.

\subsubsection{Connected Link}
Connected links are lines that indicate the relationship between two elements. These connected lines can be \textit{object to virtual elements}, linking text annotations or embedded visuals to a specific object to indicate which object is being described. Alternatively, the connected lines can be \textit{object to object}, explaining the relationship and association between multiple physical objects, such as indicating network communication between multiple IoT devices or visualizing the connection between different body parts like arms or legs. These connected links can dynamically move and animate whenever the physical objects move.

\subsubsection{Motion Effect}
Motion effects are techniques used to visualize the motion of physical objects. Most commonly, \textit{motion trajectories} are used to show the path a specific object moves, such as illustrating the trajectory of a golf swing, baseball batting swing, and body movement in gymnastics. Alternatively, some videos leverage slow-motion morphing effects or \textit{ghost effects} to depict the trajectory of the entire body or objects, similar to the famous bullet-time effects in the movie \textit{Matrix}.

\subsubsection{Others}
While much less common, we have also observed several other effects, such as particle effects and virtual 3D animation. However, since object-centric augmentation already leverages the dynamic motion of physical objects, simple visual augmentation can significantly make the video more expressive and enrich the viewing experience.

%% file: 4-system.tex
\section{\system{} System}

\subsection{System Overview}
This section introduces \system{}, a desktop authoring interface designed to support the real-time and interactive creation of augmented 3D volumetric videos, whether live or recorded. The goal of our system is to enable users to create augmented volumetric videos through direct manipulation, without the need for programming, by leveraging an object-centric augmentation approach. Given the design space exploration outlined above, \system{} allows users to easily embed text, visuals, highlight effects, and 3D objects, which can be bound to physical objects and bodies captured in the volumetric video. The following workflows are supported by \system{}:
\begin{enumerate}
\item [\textbf{Step 1.}] Track a captured object or body part by clicking the tracking points from the desktop 3D scene.
\item [\textbf{Step 2.}] Add visual effects that are automatically bound to the selected physical object.
\item [\textbf{Step 3.}] Obtain the dynamic data and parameters of the real-world motion.
\item [\textbf{Step 4.}] Bind and visualize the obtained dynamic parameter to create responsive graph plots or associated animation.
\end{enumerate}

\subsection{System Implementation}
\begin{figure}[h]
\centering
\includegraphics[width=0.9\linewidth]{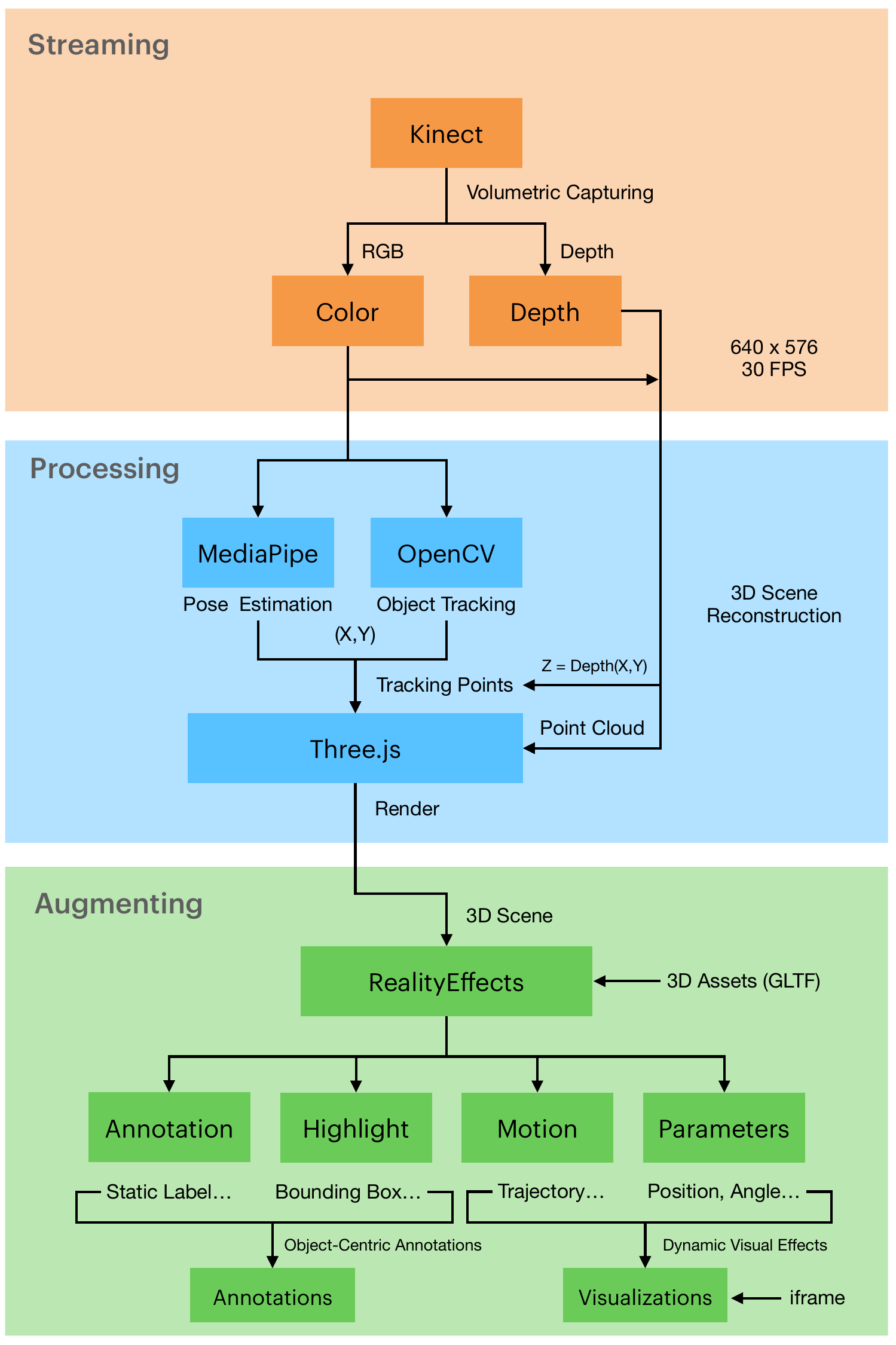}
\caption{ \system{} consists of three modules -- (1) Streaming: Azure Kinect SDK provides the depth camera data feed, (2) Processing: Mediapipe and OpenCV for body and color tracking (3) Augmenting: Three.js for visual rendering.}
\label{fig:system-implementation}
\end{figure}

\diff{As shown in Figure~\ref{fig:system-implementation}, \system{} is implemented across three main modules: streaming, processing, and augmenting. The entire application is written in JavaScript using React.js, React Three Fiber, and Electron.js. It runs on a desktop Windows machine, and we recommend using a desktop machine equipped with graphics cards to speed up rendering. The source code for our system implementation is available on GitHub~\footnote{\url{https://github.com/jlia0/RealityEffects}}}.

\subsubsection{Streaming Module}
\diff{The streaming module utilizes the off-the-shelf Azure Kinect depth camera SDK to capture volumetric point-cloud data. The data feed includes both RGB and Depth data in separate channels, each with a resolution of \textit{640 × 576} and a refresh rate of \textit{30} FPS. Both channels share the same \textit{(x,y)} coordinate data structure, enabling us to retrieve the depth information for any \textit{(x,y)} tracking point. The obtained RGB-D data is then passed to the processing module.}
% (CPU: [xxx], GPU: [xxx], RAM: [xxx]).

\subsubsection{Processing Module}
\diff{With the RGB-D data feed, the application performs 3D scene reconstruction by rendering the 3D point cloud data directly using Three.js, where \textit{z = Depth(x,y)} and \textit{RGB = Color(x,y)}. We utilize MediaPipe Pose Estimation for body tracking and OpenCV for object tracking. The application calculates the centroid by averaging the \textit{(x,y)} values, retrieves the depth information with the \textit{(x,y)} coordinates, and registers the centroid as the attachable object in the authoring interface for further augmentation.}

\subsubsection{Augmenting Module}
\diff{With the attachable object from the processing module, \system{} allows users to select objects from pose estimation and color tracking and augment them with object-centric annotations and dynamic visual effects. The object-centric annotations are essentially Three.js coded objects such as static text labels and bounding boxes, with a bloom pass to create glowing highlight effects. The dynamic visual effects are parameterized object motions that allow us to create visualizations from the motion and parameters, such as trajectory, position, distance, and angle. Users can augment the moving object with motion effects like a trajectory (a series of points) and a trailing effect~\footnote{\url{https://drei.pmnd.rs/?path=/docs/misc-trail--docs}}, and augment the motion with embedded visualizations using an iframe to create charts and interactive widgets. The augmentation can be applied to a real-time camera data feed, allowing users to review their own performance as it's being annotated, which unlocks several application scenarios like sport analysis and e-commerce live streaming. Users can also freely move or zoom the camera in 3D space through mouse movements. When using a recorded volumetric video, the system supports simple pause and play functionalities. Since we only use a single Kinect camera, capturing the entire room is challenging. Therefore, we also scan the room with a static 3D scanner (iPad Pro 12-inch with LiDAR camera and 3D Scanner App) and place it as a 3D volumetric background asset (glTF file) only for visual aesthetic purposes in most of the figure and video demonstrations.}

\begin{figure*}[h!]
\centering
\includegraphics[width=0.245\textwidth]{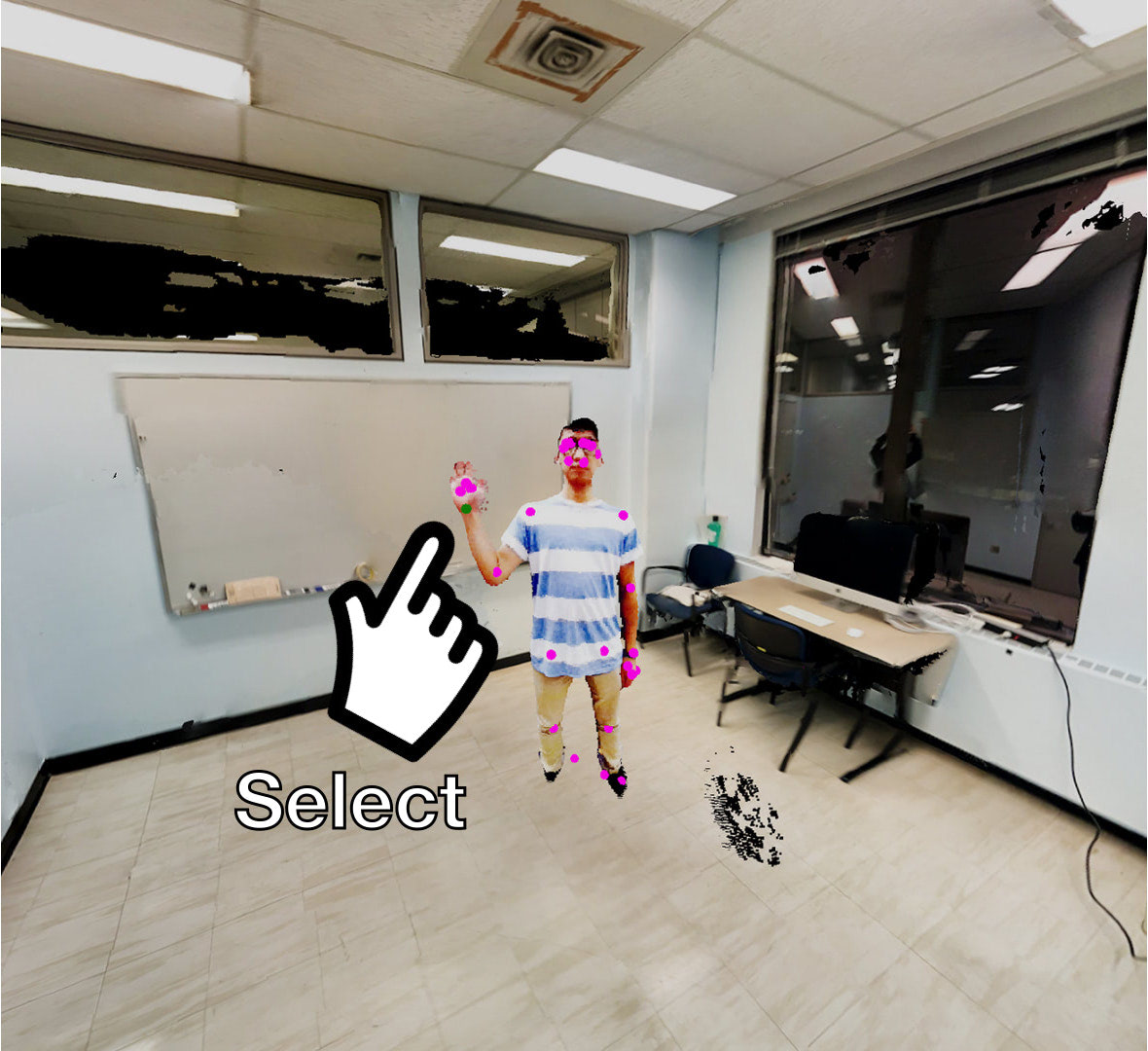}
\includegraphics[width=0.245\textwidth]{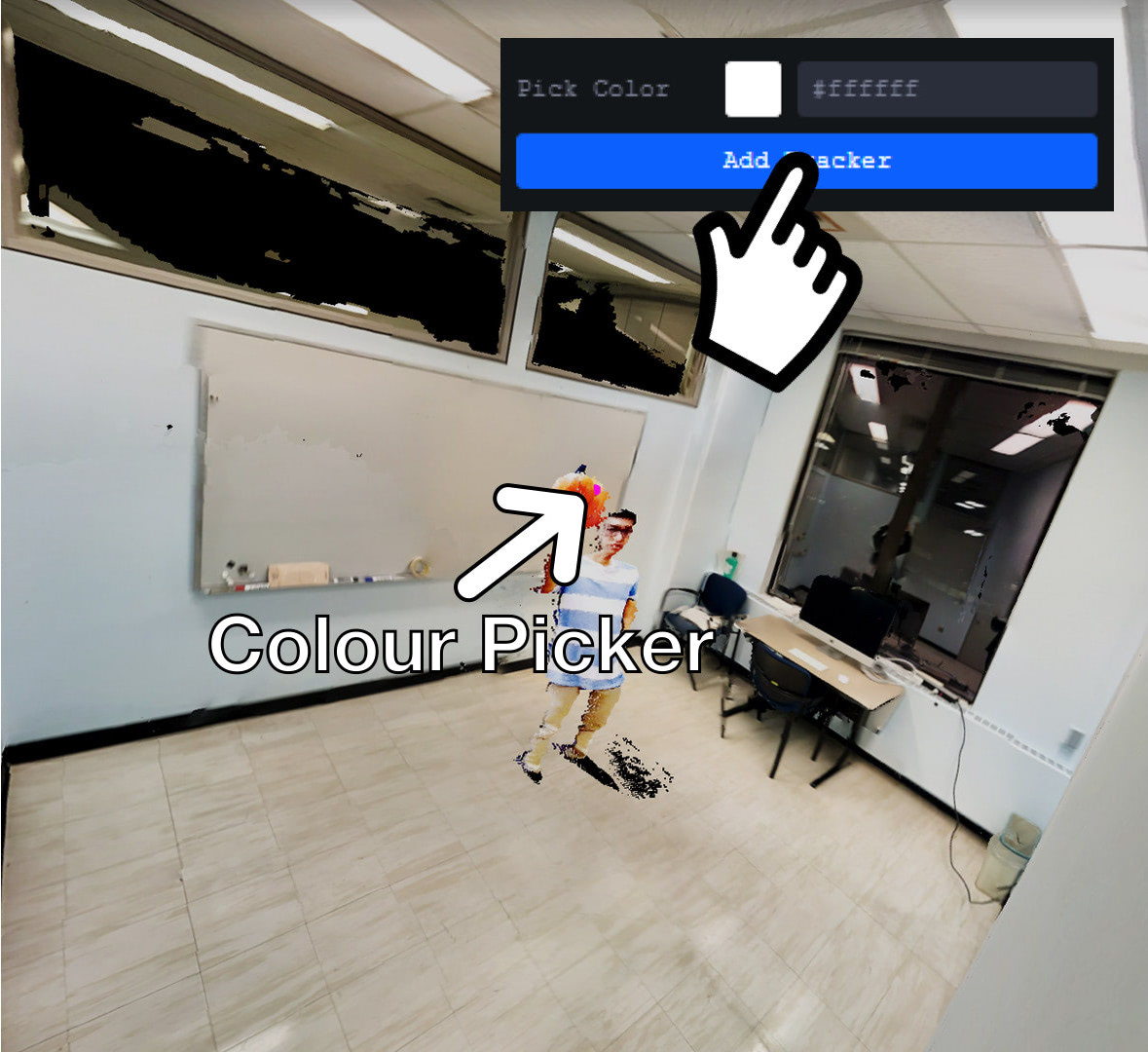}
\includegraphics[width=0.245\textwidth]{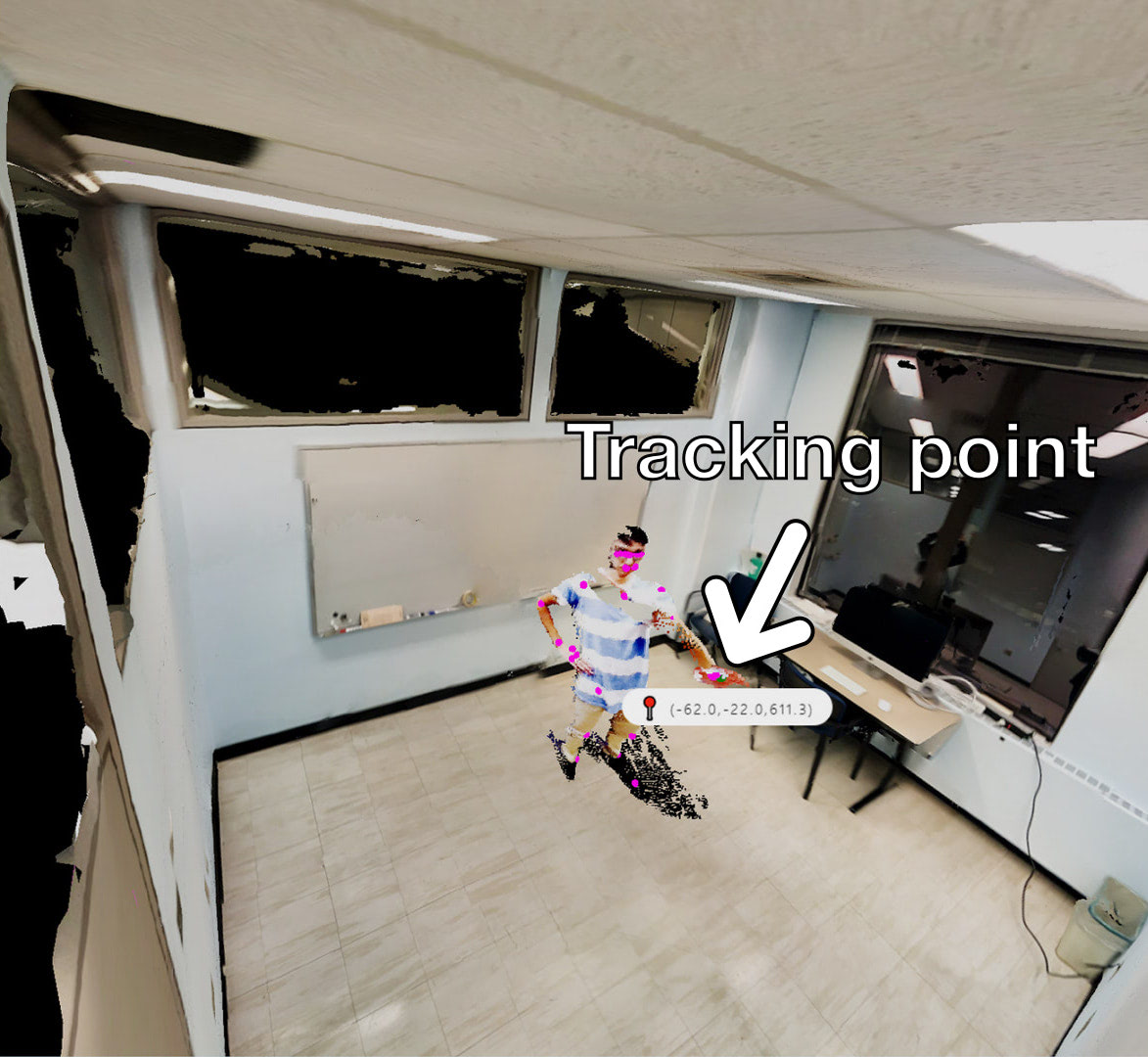}
\includegraphics[width=0.245\textwidth]{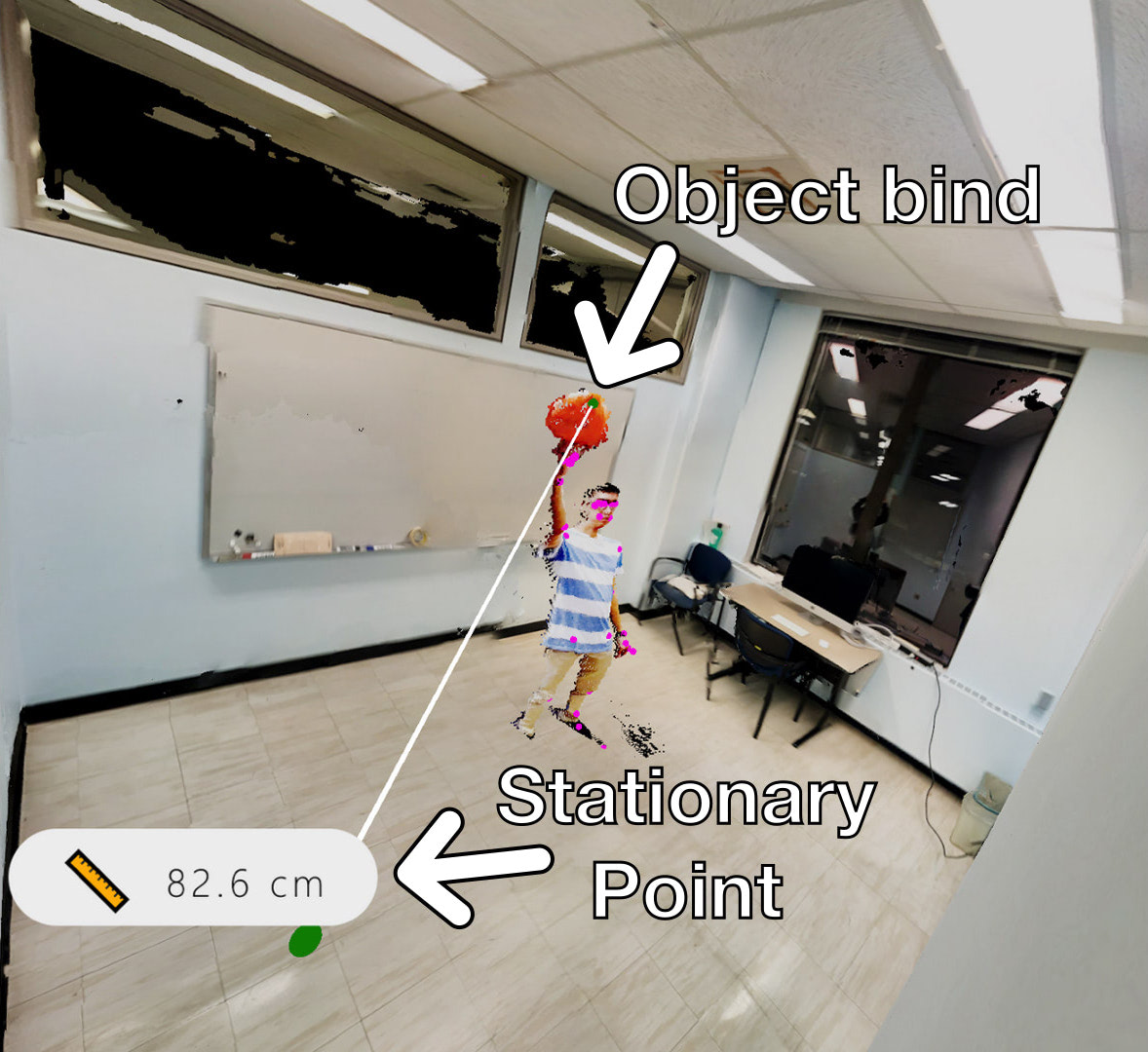}
\includegraphics[width=0.245\textwidth]{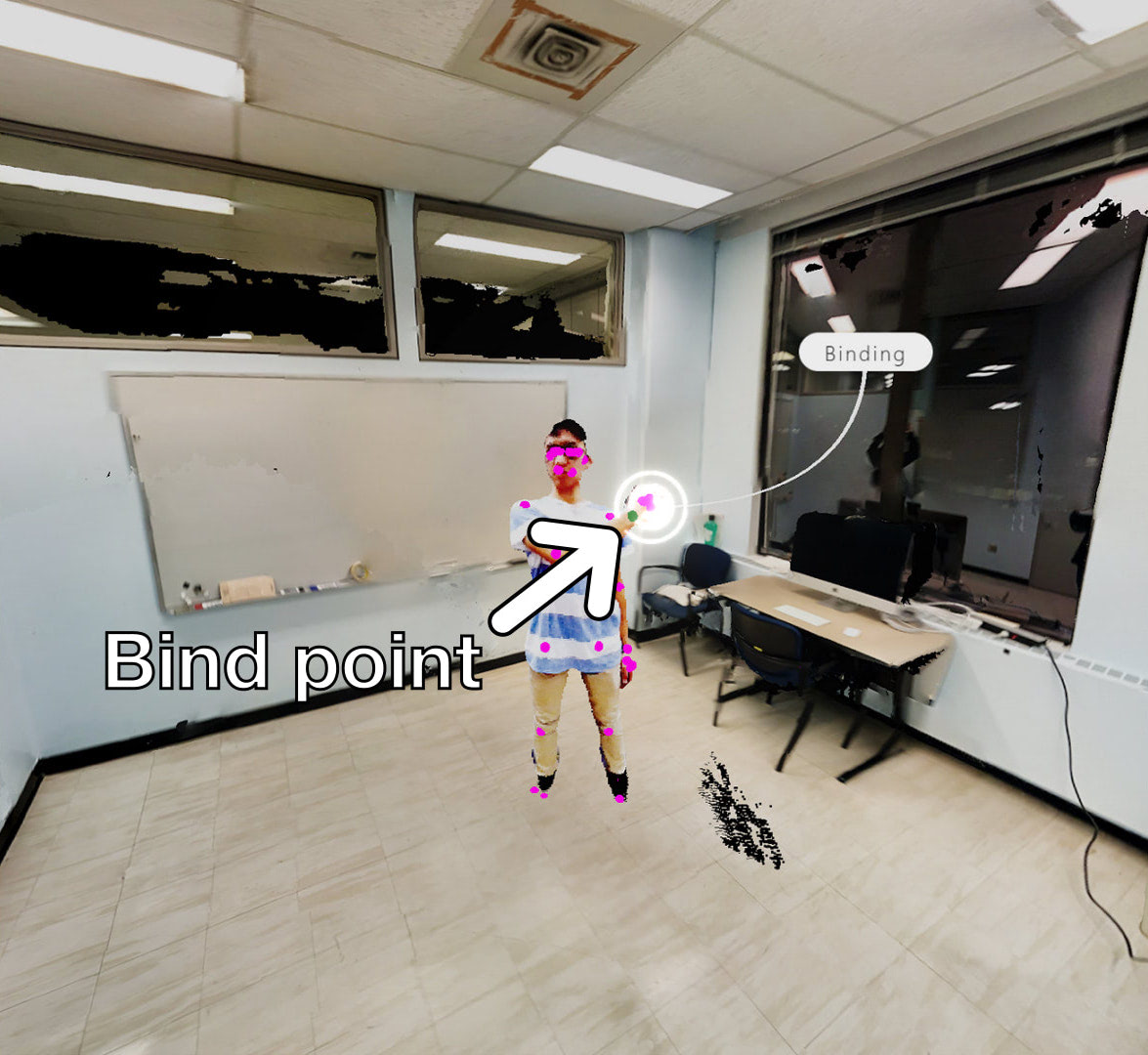}
\includegraphics[width=0.245\textwidth]{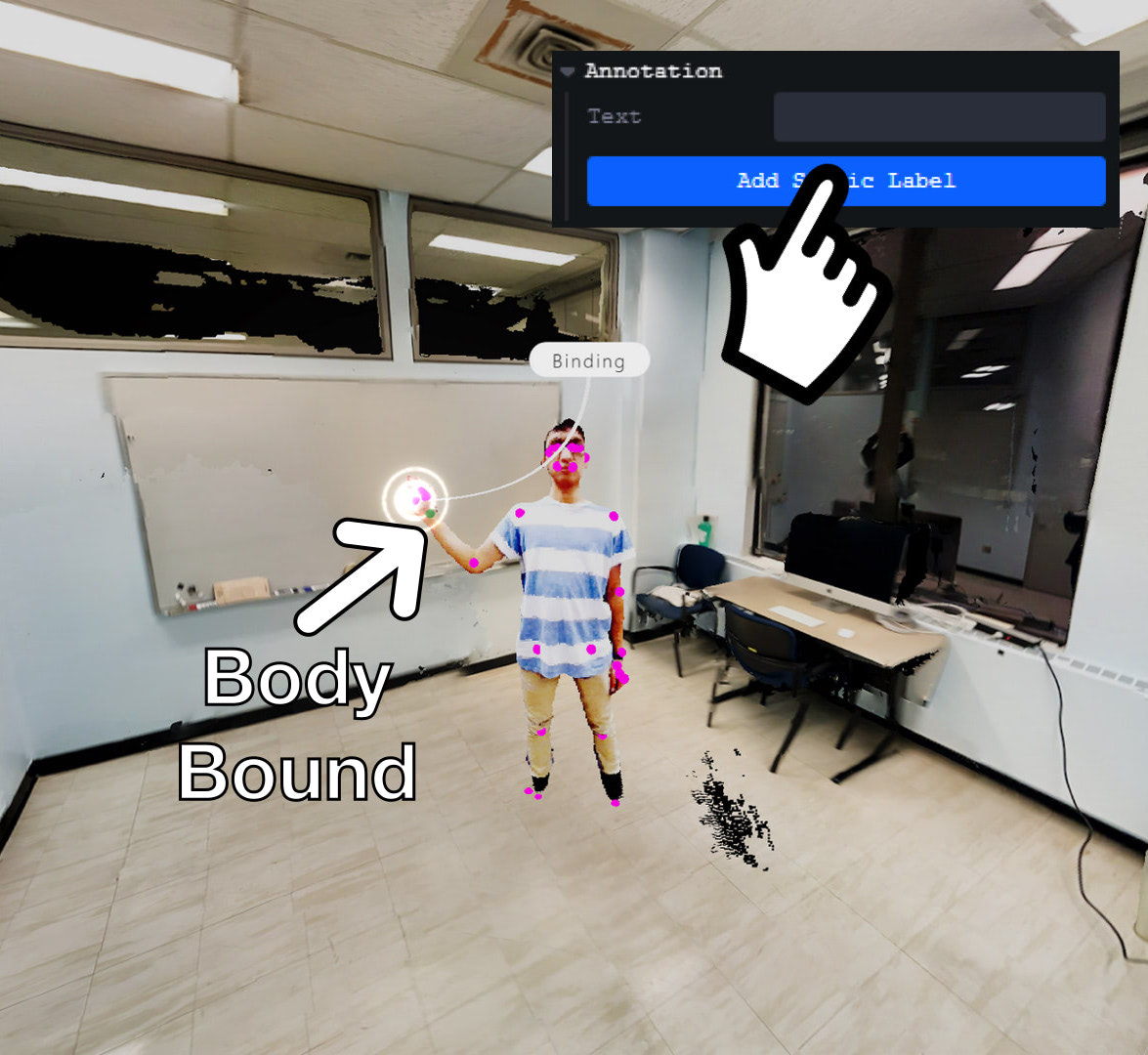}
\includegraphics[width=0.245\textwidth]{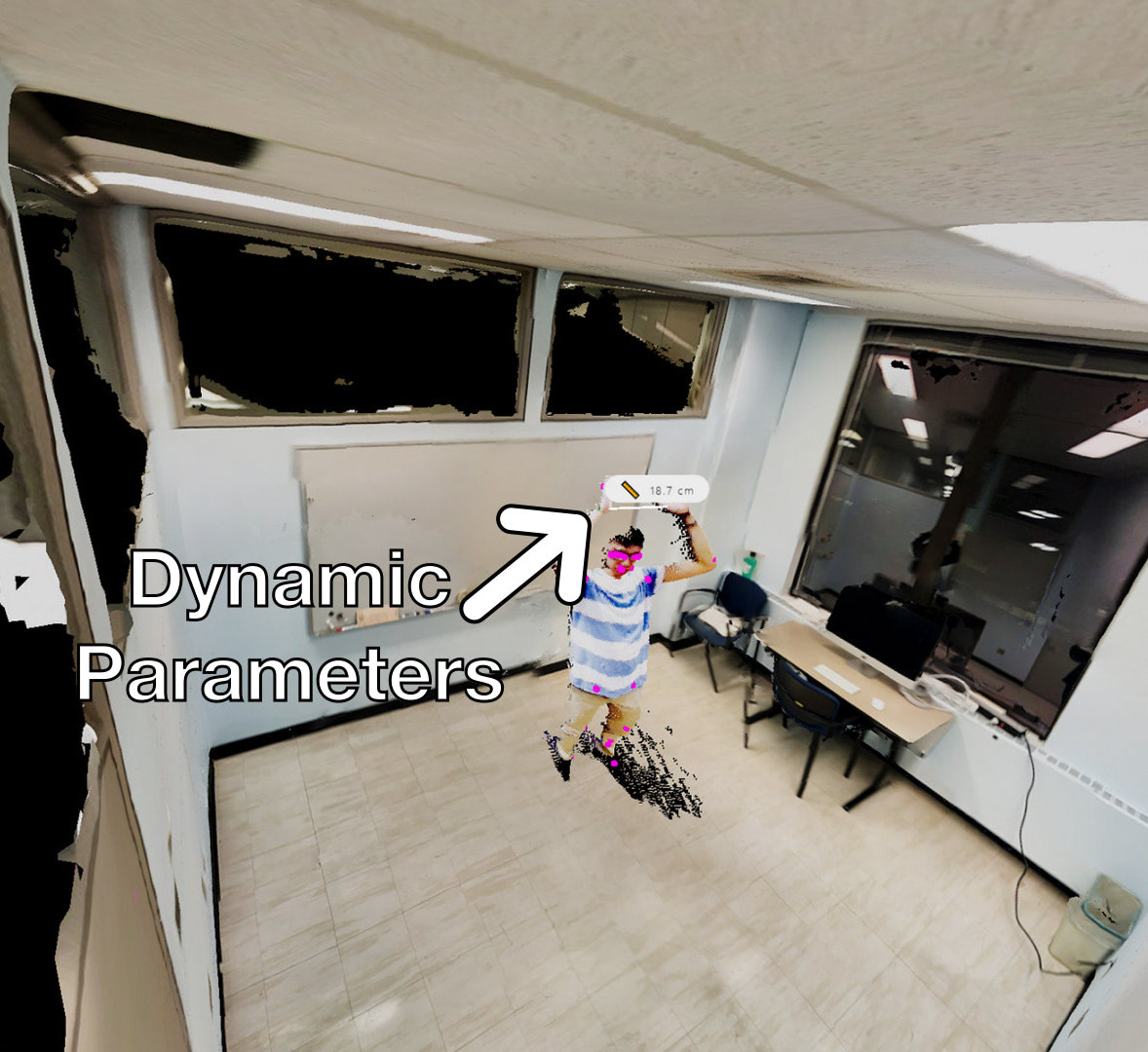}
\includegraphics[width=0.245\textwidth]{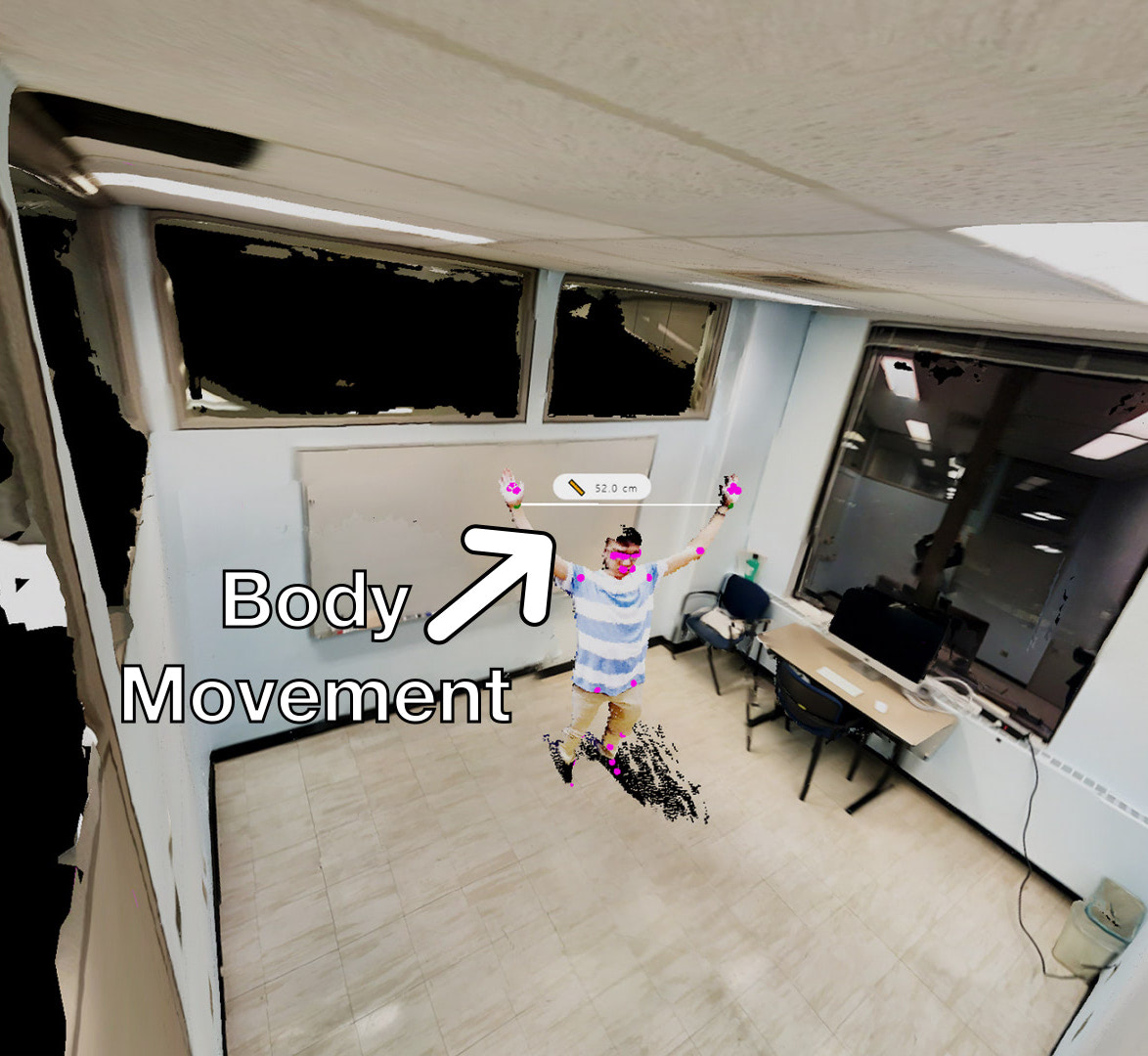}
\caption{Authoring Workflow: Collection of examples from \system{}'s workflow to demonstrate features such as dynamic parameters, highlighting and annotations.}
\label{fig:workflow-1}
\end{figure*}

\subsection*{Step 1. Object Selection and Tracking}
The first step is to select a captured physical object. 
For object-centric augmentation, all embedded annotations and visual effects should be tightly coupled with physical objects. 
Therefore, our system first allows the user to specify which objects to track and bind. 
To specify the object, the user can enter the selection mode and simply click the object in the scene. Then, the system automatically adds a tracking point in the 3D scene and tracks its location. 
For the tracking point, the system supports three categories: 1) \textit{physical object}, 2) \textit{body}, 3) \textit{stationary physical environment}. \diff{The performance of object tracking can be found in Table~\ref{tab:tracking-accuracy}. We evaluate the accuracy by counting the time duration of tracking target losses over a fixed period of a captured video while we freely move the objects around in space and at different angles. While this is a fairly simple evaluation, we found that pose estimation is fairly robust and close to its acclaimed benchmarks, whereas color tracking is unstable, especially for objects with reflective materials. Future improvements are needed using more robust methods such as \textit{SAM-Track}~\cite{cheng2023segment} and \textit{Track Anything}~\cite{yang2023track}.}

\begin{table}[htbp]
\centering
\begin{tabular}{@{}lll@{}}
\toprule
         & Pose Estimation & Color Tracking \\ \midrule
Accuracy & 91.3\%          & 65.7\%         \\ \bottomrule
\end{tabular}
\vspace{10pt}
\captionsetup{skip=1pt}
\caption{Pose Estimation and Color Tracking Accuracy}
\label{tab:tracking-accuracy}
\end{table}

\subsubsection{Physical Object}
First, for the colored physical object, the system tracks the object's 3D position based on the combination of color tracking and point-cloud information.
When the user clicks an object, the system gets the current RGB value of the clicked points in 2D screen. Then, the system captures a similar colors based on an upper and lower threshold range of RGB ($\pm$ 10) to obtain the largest contour in the scene, based on Node OpenCV library. Given the detected object in the 2D scene, the system raycasts to the volumetric scene to obtain the associated point-cloud depth information, which allows us to get the coordinated 3D position in the scene. 

\subsubsection{Body}
For post estimation and body tracking, we simply use MediaPipe to get the estimation of 33 body tracking points. We also tried Kinect-builtin body tracking feature, but the performance was not satisfying because of high latency and low accuracy. Similar to color tracking, the system allows the user to directly select one of the tracking points of the body skeleton, and the system automatically calibrate the 2D coordinates with the depth information. 
When the user enters the body selection mode, then the system shows the twenty body skeleton points which the user can select.
When the user selects a certain skeleton parts, then it becomes highlights and starts tracking in the 3D scene.

\subsubsection{Stationary Location}
For stationary location, the user can simply select a location in the scene and use a ray cast to obtain the stationary 3D position in the physical environment, such as floor or wall. 
The user can also place it in mid-air by moving the point with a mouse. 
In this selection, the tracked point is stationary, thus there is no dynamic movement. However, this tracked location can be used as a reference point, such as a distance from a certain location.

\subsection*{Step 2. Virtual Object Binding}
Once the system starts tracking the selected object, then the user can add virtual objects that can be bound to the tracked object.
Informed by the taxonomy analysis, the system supports the following four virtual 3D objects: 1) text annotation, 2) object highlight, and 3) embedded visual. 

\subsubsection{Text Annotation}
First, the user can bind the text label to the associated physical object in the volumetric 3D scene.
To place a text annotation, the user specifies the associated object and then clicks the text label button. 
Then, the system starts showing the 2D text label floating around the tracked object. 
Since the attached text label is bound to the object, the text label position moves when the object moves. 
The user can change the text value by typing the name in the menu window.
The user can also add a dynamic value by using a variable, based on the JavaScript variables such as \texttt{Date.now()}, or a user-defined variable based on the dynamic parameters, such as position, speed, angle, distance, etc., as we describe in Step 3. 
While the text is a 2D object, it always moves its orientation to face the camera. The user can also disable this to 

\subsubsection{Object Highlight}
The user can also add an object highlight bound to the tracked object. 
The system supports two basic object highlight options: 1) 3D primitive shapes, such as bounding box, bounding sphere, and bounding cylinder, and 2) 2D highlight shapes such as colored circles or rectangles.
To add the object highlights, the user first selects the tracked object and then chooses the object highlight button in the menu.
Then, the system lets the user choose the shape of the highlight (default: 3D sphere), then the object highlight is added to the scene. 
Unlike text annotation, the object highlight is placed in the center of the tracked object.
The user can also change the scale, offset, orientation, and color accordingly through a direct manipulation interface. 

\subsubsection{Embedded Visual}
The user can also add embedded 2D visuals.
Informed by the taxonomy analysis, the system supports images, icons, videos, and embedded websites as the associated visual aids. 
From the technical point of view, all the embedded visual is implemented as embedded iframe in Three.js.
Therefore, the image, YouTube video, or website can be embedded as an iframe by specifying the URL or local file. 
To add the embedded visual, the user can also select the object and enter the embedded visual menu. Then the user can enter the URL or file directory. Once loaded, the added visual elements start following based on the object's movement. 
Again, the user can also change the size, orientation, and opacity of these elements.
Since the embedded visual is an interactive HTML, the user can also interact with the screen such as buttons or links. 
By leveraging this feature, we can also embed dynamic graphs and charts by associating the dynamic parameter, as discussed in Step 4. 
By default, the 2D visual always changes its orientation to face the camera, but the user can also change it by disabling it.

\subsection*{Step 3. Parameterize the Real World}
The user can also parameterize the real world to obtain the dynamic data value associated with the captured motion. 
The system obtains these real-time values based on 3D reconstructed information. 
The system supports the following parameterized values: 1) X, Y, and Z position of the tracked object, 2) speed of the tracked object, 3) distance between two tracked objects, 4) angle between three tracked objects, and 5) 2D area of three or more tracked objects. 

\subsubsection{Position}
The system can obtain the 3D position of the tracked object by simply getting the current position value.
The user can use this dynamic value in text labels or dynamic graphs by using the specific variable.
In the system, the user can use this value by using the variable like \texttt{obj\_1.x}. 

\subsubsection{Speed}
The system also obtains the speed for all the tracked objects, by calculating 
\begin{equation}
\text{Speed} = \frac{\sqrt{(x_1 - x_0)^2 + (y_1 - y_0)^2 + (z_1 - z_0)^2}}{t_1 - t_0}
\end{equation}
where $t_1 - t_0$ is every 0.5 second (every 15 frames with 30 FPS).
The user can access this information by using the variable like \texttt{obj\_1.speed.x}

\subsubsection{Distance}
When the user selects multiple objects, the system also calculates the distance between the two objects. 
If the user also wants to show the line between the two points, the user can enter the geometry menu, and then add the connected line between two tracked points.
Then, the line geometry automatically changes its length and orientation based on the tracked objects. 
The endpoint of the line is not the dynamic object but can be also a stationary location such as a specific point in the scene. 
The user can use the distance information by using the variable like \texttt{distance\_1}

\subsubsection{Angle}
When selecting the three tracked objects, then the system also calculates the angle between the two lines. 
In this case, the angle is calculated given the two 3D vectors $a, b$ through $\theta = cos^{-1}[ (a \cdot b) / \|a\| \cdot \|b\|)]$, where $cos^{-1}$ is arc cosine and $a \cdot b$ is a dot product.
The user can use the angle data by using the variable like \texttt{angle\_1}

\subsubsection{Area} 
In the same way, the user can also obtain the dynamic parameter for the area of three points (the area of a connected triangle) or four points (the area of a connected rectangle).
The user can use the area by using the variable like \texttt{area\_1}

\subsection*{Step 4. Visualize the Dynamic Motion}
By default, the user can create object-centric augmentation by simply binding virtual elements to the tracked object in Step 2.
However, the user can even create more expressive dynamic effects by using or visualizing the dynamic parameters based on the variables defined in Step 3.
To that end, the system supports the following three parameter-based dynamic visual effects: 1) dynamic text annotation, 2) dynamic visual appearance, and 3) dynamic graph.
The system also supports two motion-related visual effects: 4) motion trajectory, and 5) ghost effects. 

\subsubsection{Dynamic Text Annotation}
Dynamic text annotation is the text annotation described in Step 1 but uses the parameterized value in the 3D scene. 
For example, if the user types the text value as \texttt{PositionX: \$\{obj\_1.x\}\$}, then the system shows the parameterized value in the text label, which is shown as like \texttt{PositionX: 34.23}.

\subsubsection{Dynamic Visual Appearance}
Similarly, the user can also bind the dynamic parameter to the visual property of the embedded virtual objects, such as scale, rotation, position, opacity, and color. 
For example, if the user associates the scale of the virtual object with the position of the tracked object, then the embedded virtual object's size changes in response to the position of the tracked object.

\subsubsection{Dynamic Graph}
The user can also show the dynamic graph by associating the dynamic value with the charts. 
As we mentioned, we can embed the interactive 2D data visualization with iframe. We prepare several basic graphs such as line charts, bar charts, or pie charts, based on the Chart.js library. 
For example, if the user associates the y value of the line graph as an angle of between the arm and body, then the system shows the real-time line chart to show the tracked parameter. 

\subsubsection{Motion Trajectories}
Alternatively, the system can also show the motion effects with several prepared visual effects. 
For example, the user can show the motion trajectory of the tracked object. 
To do so, the user selects the motion trajectory option in the menu, then the user selects the object. 
Then, the system starts the trajectory path of the motion, based on the object location.
To implement this, we simply place a small sphere in the position of the tracked object for each frame, then disappear for a certain duration (5 seconds).

\subsubsection{Ghost Effects}
Finally, the system also supports the ghost effects by duplicating the tracked object's geometry. 
To do this, we simply clone the entire tracked object for every second, so that the user can see the ghost effect.

%% file: 5-application.tex
\section{Applications}

\subsection{Product Showcase and Advertisement}
\begin{figure}[h!]
\centering
\includegraphics[width=0.49\linewidth]{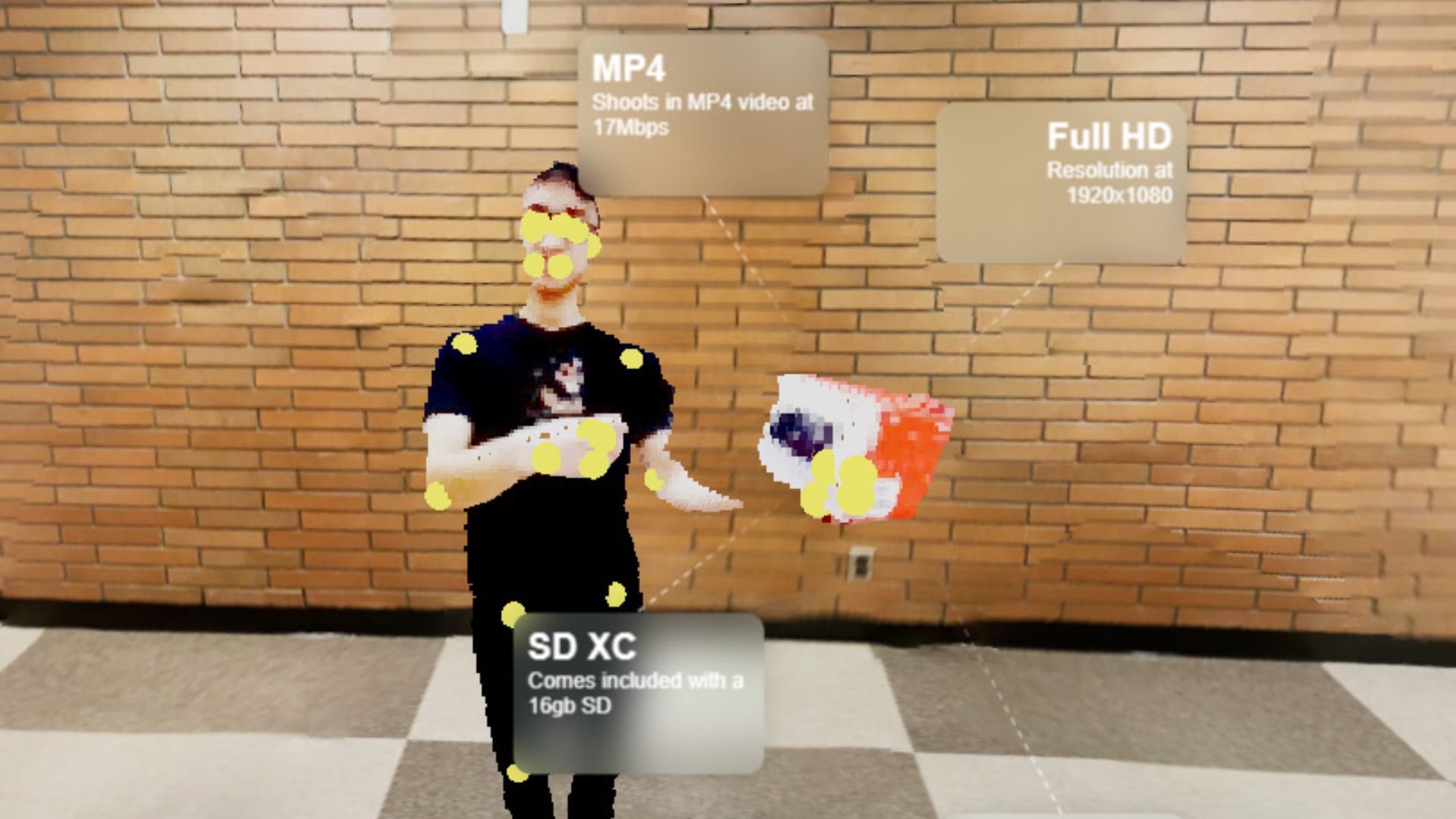}
\includegraphics[width=0.49\linewidth]{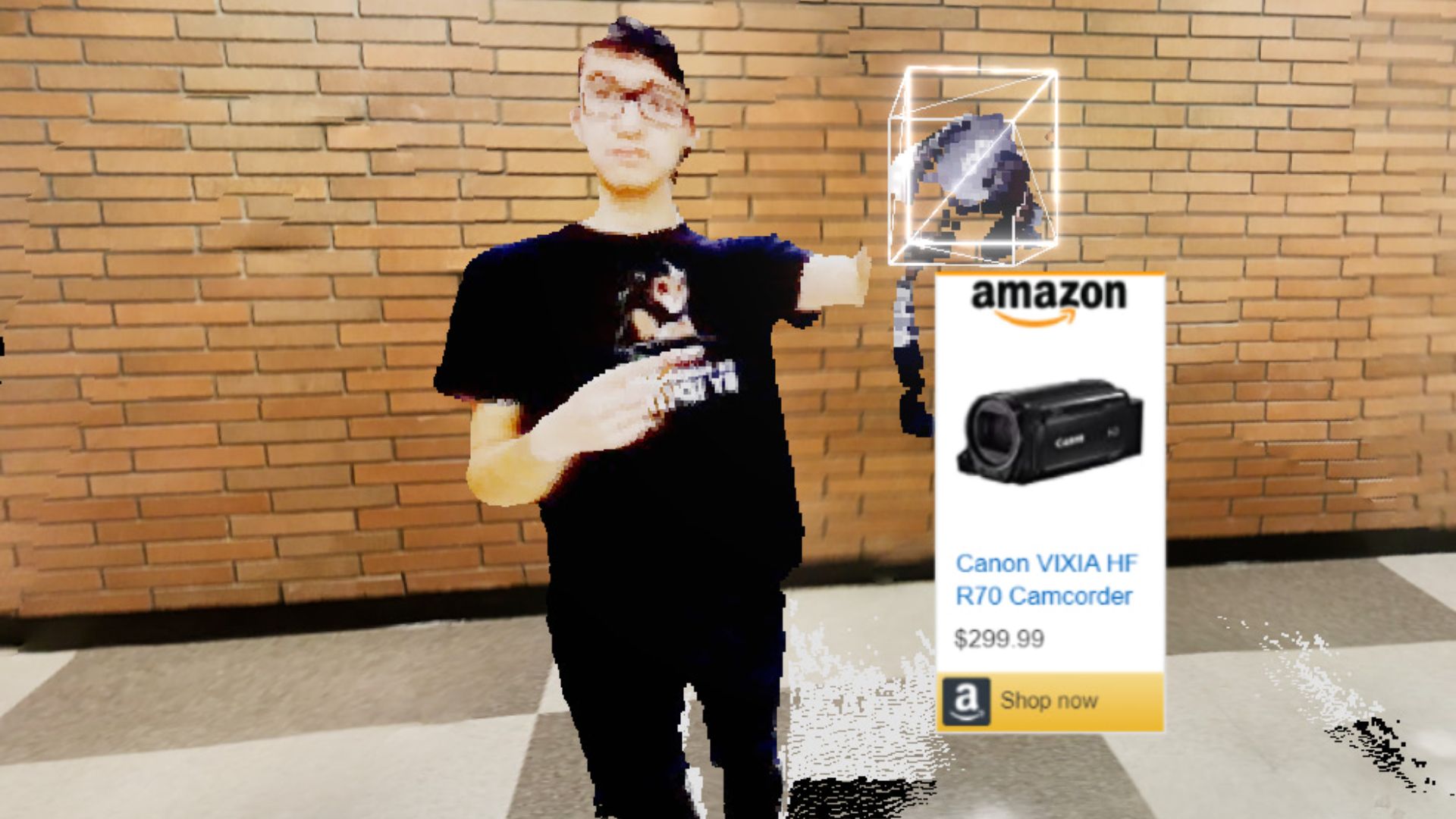}
\caption{Product Showcase: Use case scenario demonstrating a sales pitch for a handheld camera, where labels, visualizations, and highlights are used to enhance the product's appeal and provide information.}
\label{fig:product-showcase}
\end{figure}

Social e-commerce, which gained prominence during COVID, has popularized remote selling and virtual sales. Our system can be utilized for e-commerce live streaming or recorded product showcases. For example, Figure~\ref{fig:product-showcase} illustrates a virtual sale presentation using our system. Initially, the presenter showcases the camera, and then the user can annotate the product with labeled annotations. By using the embedded website feature, the user can also add an Amazon link directly in the 3D scene. These embedded websites are interactive and clickable, enabling the audience to directly access the shopping website.

\subsection{Tutorial and Instruction}
When conducting experiments in a lab, safety is crucial. \system{} can assist in maintaining safety standards, for instance, in a chemical lab where preventing cross-contamination of chemicals is essential. A user can define a specific space or surface for \system{} to monitor. Based on the data, such as the duration of interaction or movement, the system can augment visualizations to display a heatmap showing levels of contamination seconds after the area or surface is touched.

\begin{figure*}[h!]
\centering
\includegraphics[width=0.245\linewidth]{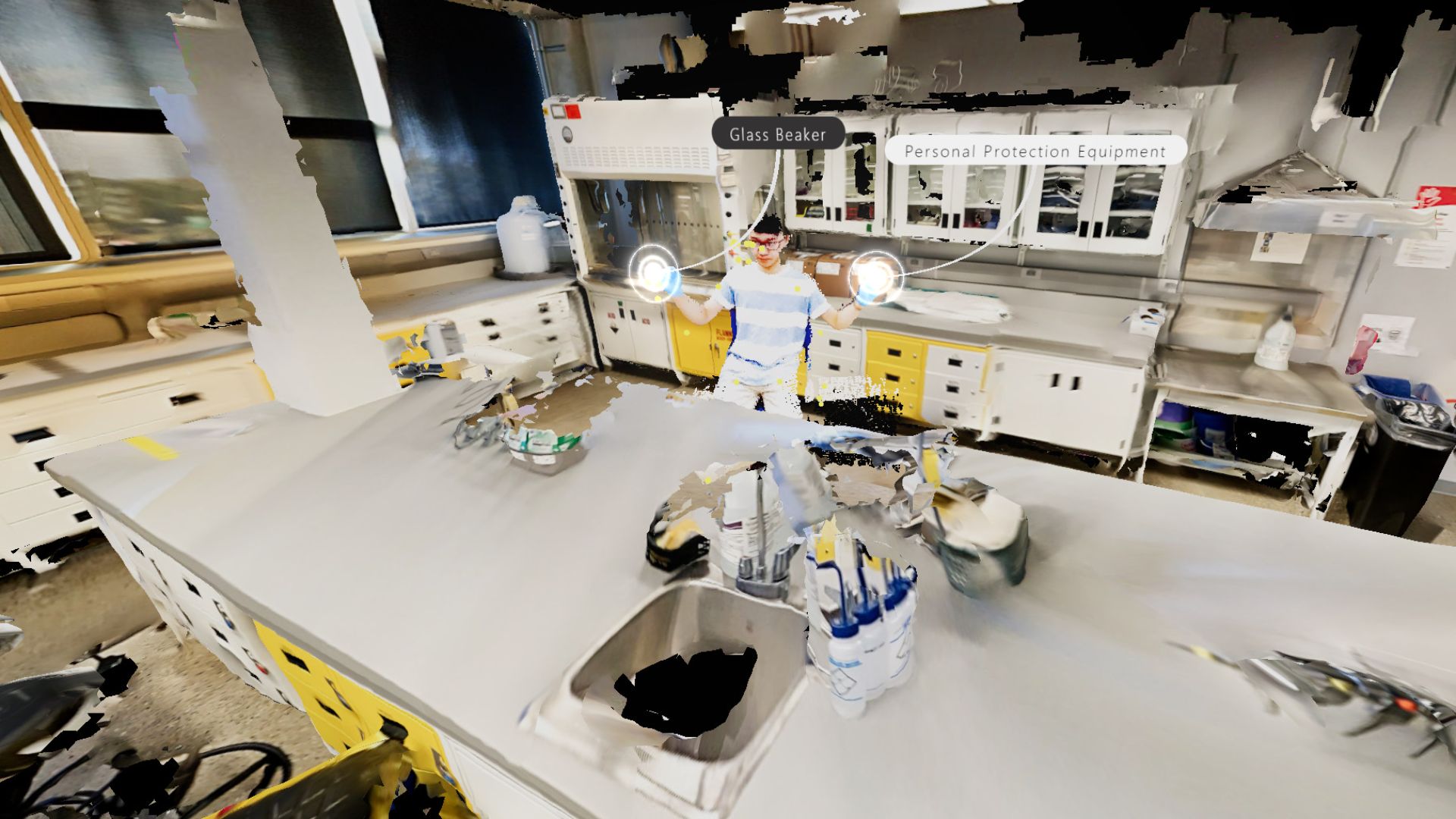}
\includegraphics[width=0.245\linewidth]{figures/chem-2.jpg}
\includegraphics[width=0.245\linewidth]{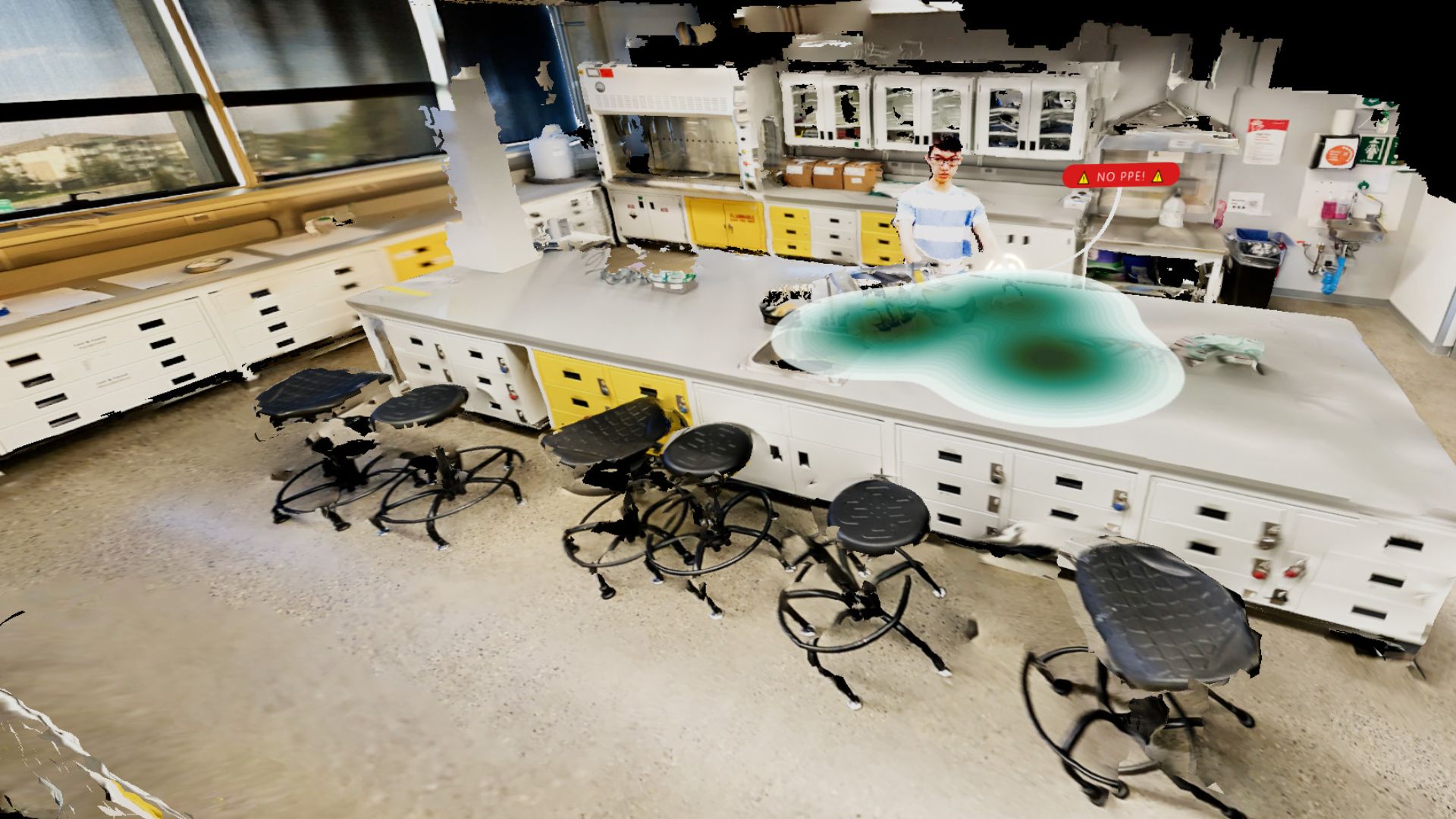}
\includegraphics[width=0.245\linewidth]{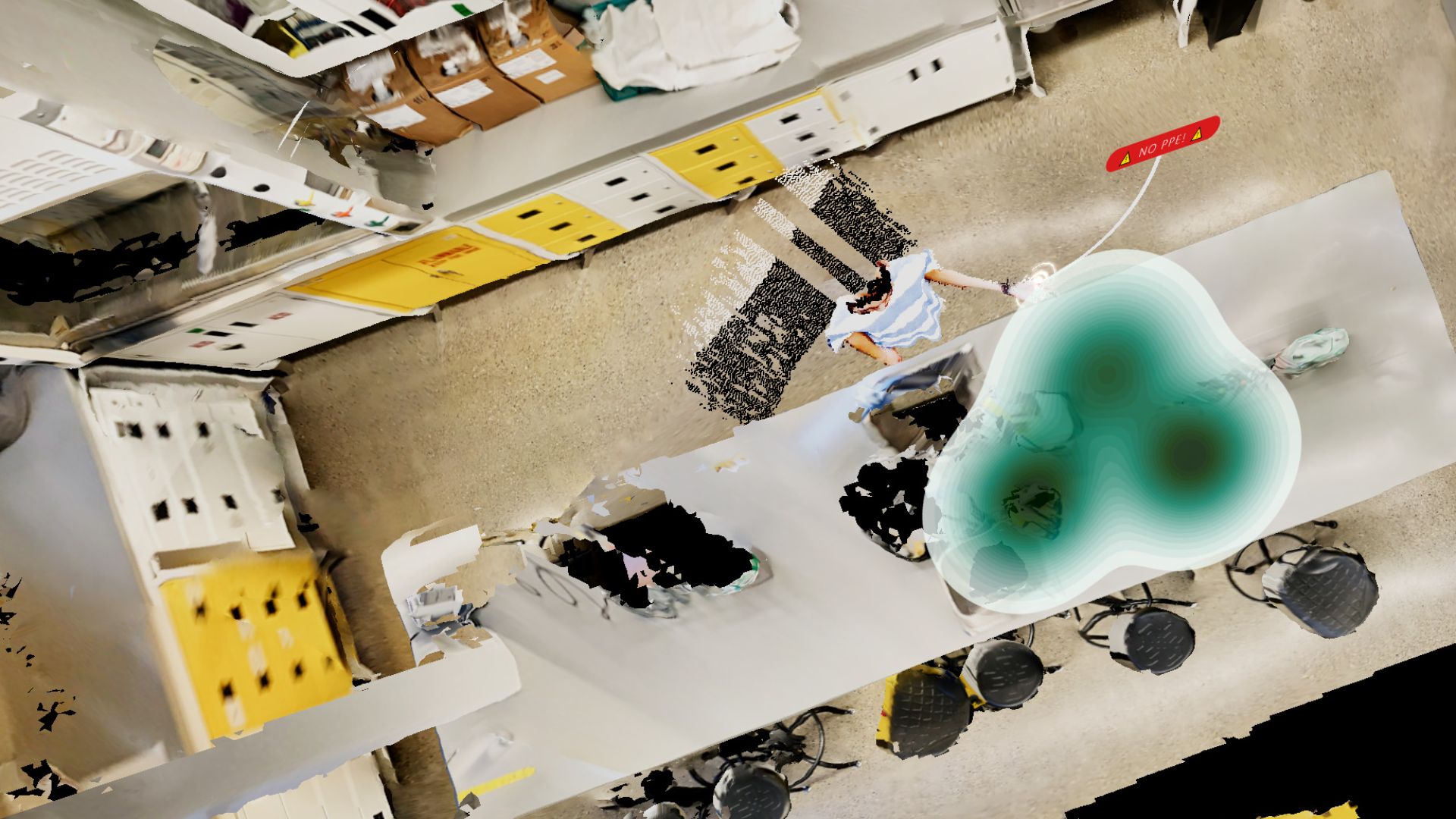}
\caption{Chemistry Lab Training: Use case scenario highlighting safe practices, necessary precautions, and potential lab dangers related to cross-contamination.}
\label{fig:chemistry-lab}
\end{figure*}

\begin{figure*}[h!]
\centering
\includegraphics[width=0.245\linewidth]{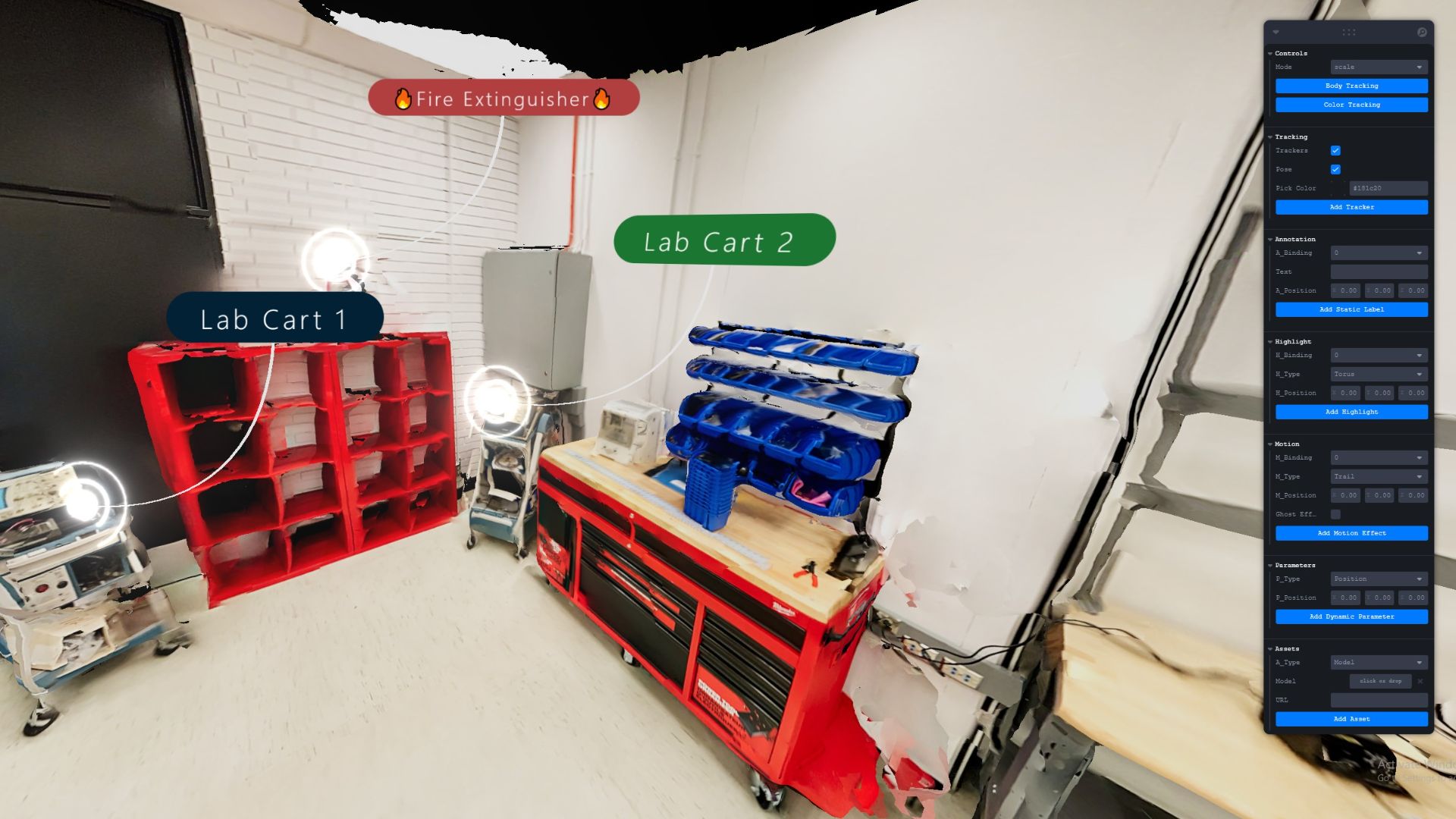}
\includegraphics[width=0.245\linewidth]{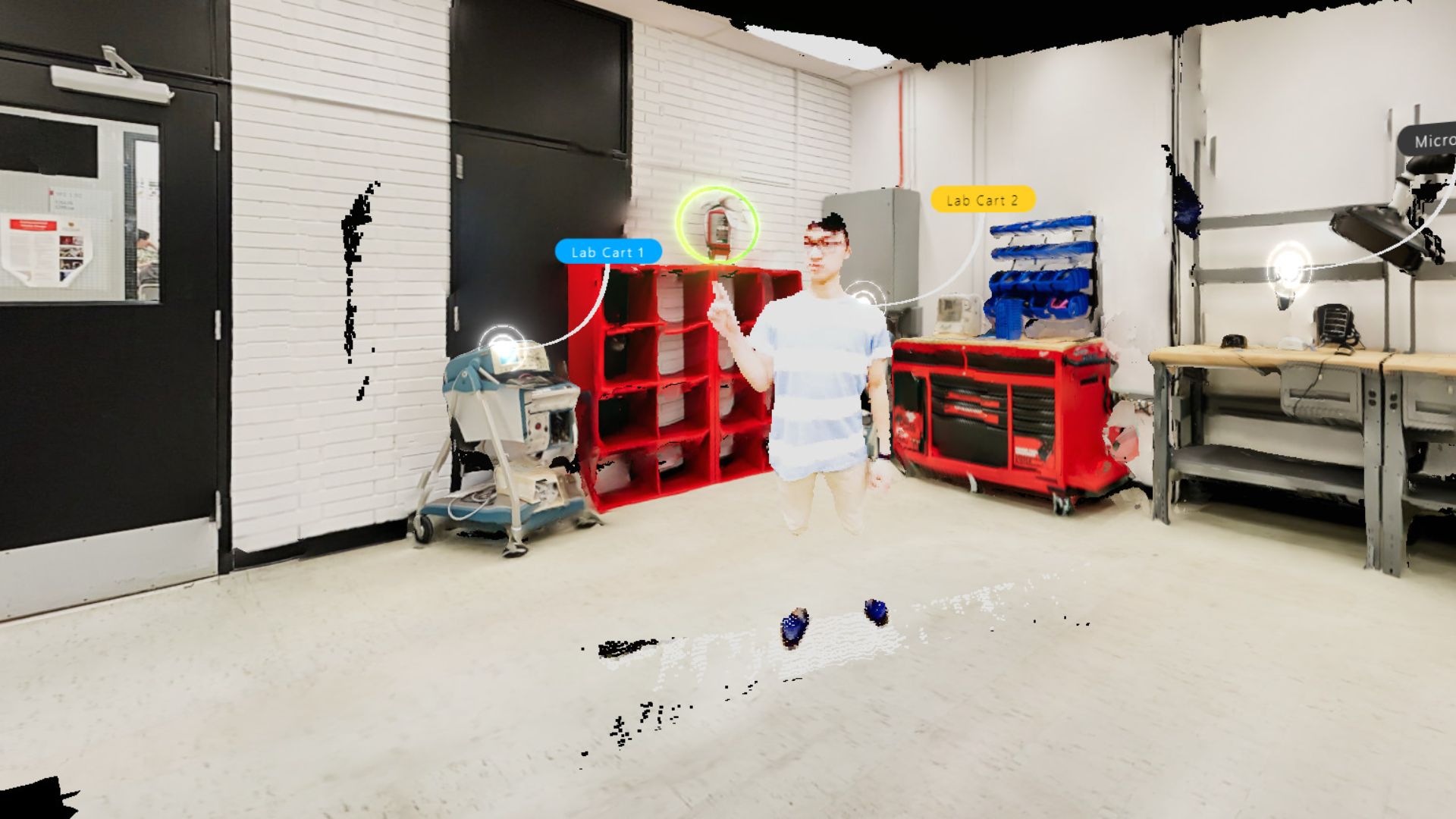}
\includegraphics[width=0.245\linewidth]{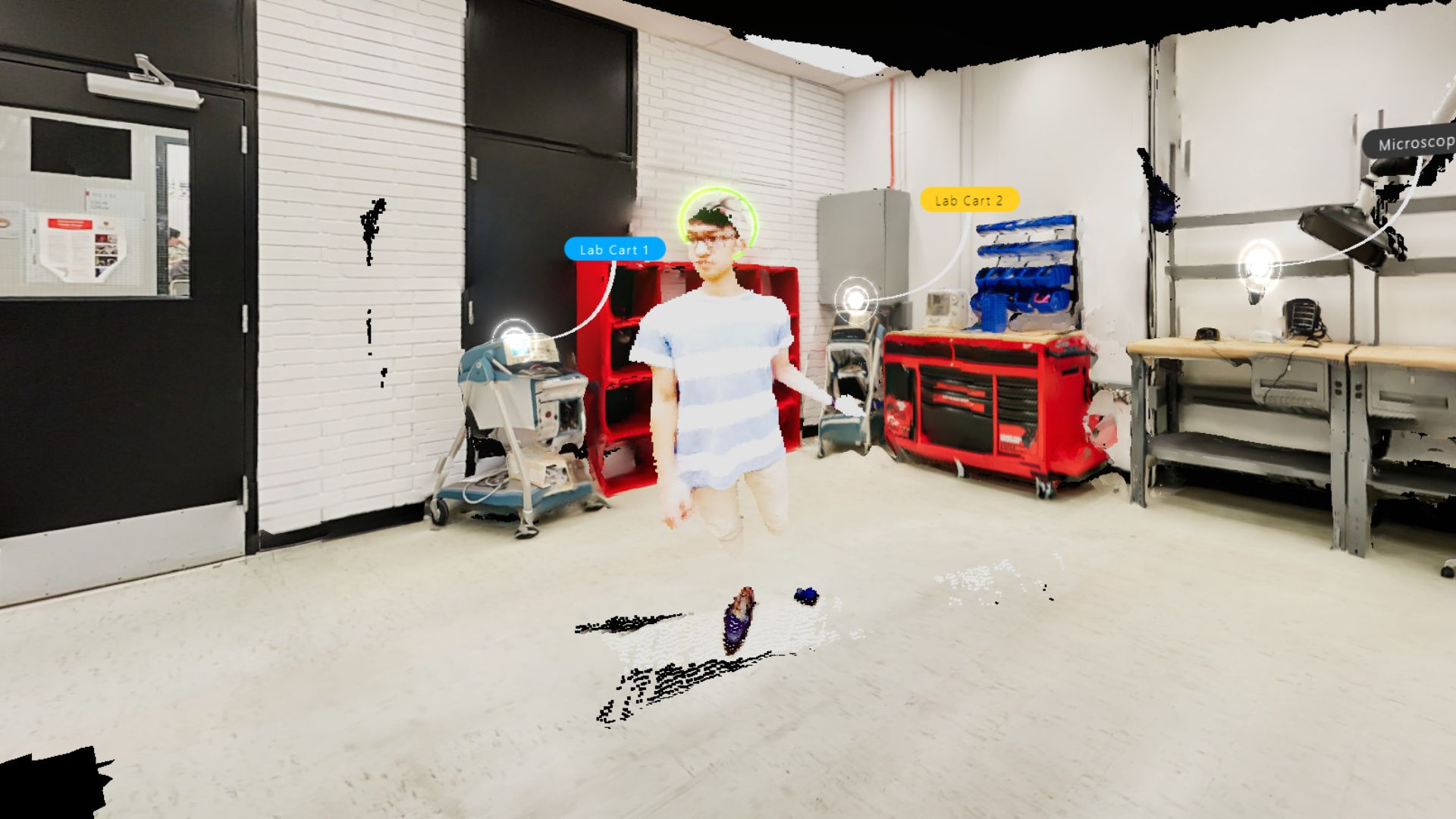}
\includegraphics[width=0.245\linewidth]{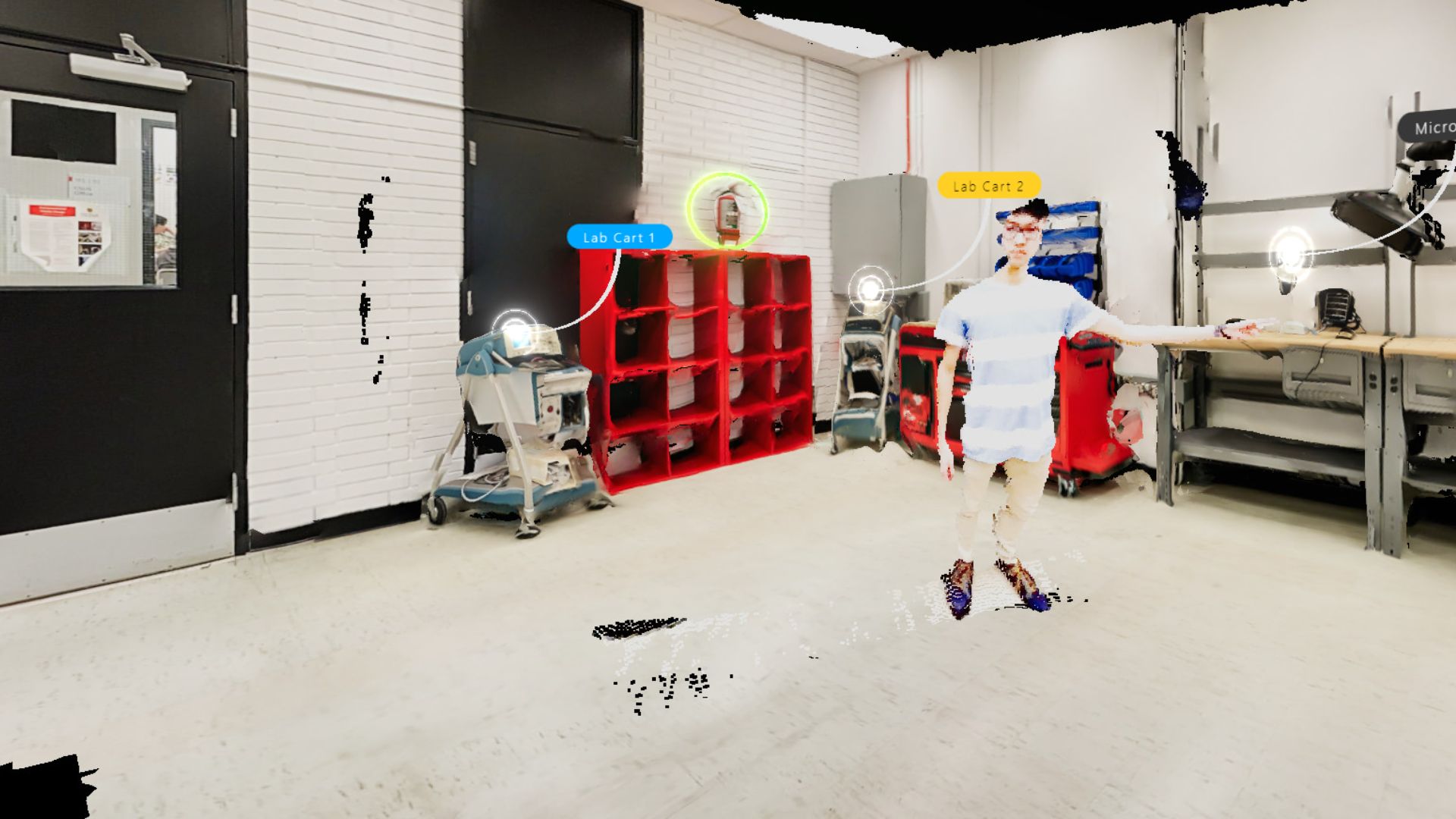}
\caption{Maker Space Introduction: Use case scenario providing an orientation to a creative space. Labels and highlights are utilized to identify equipment and safety measures.}
\label{fig:makerspace-intro}
\end{figure*}

\subsection{Physical Training and Sport Analysis}
Our system is also suitable for sports analysis. By augmenting volumetric videos of sports activities, \system{} can enhance the understanding of athletic actions. For example, in a soccer game, the system can annotate or highlight players to increase their visibility, focus on individual players by binding objects to them, or use highlighting features. Features such as object-object binding or trajectory augmentation can display visual lines between players to indicate player positioning or the trajectory of the ball during the game. Additionally, the system can generate user-defined data visualizations to display statistics such as player speed or heat maps of areas with frequent movement or activity.

\begin{figure*}[h!]
\centering
\includegraphics[width=0.245\linewidth]{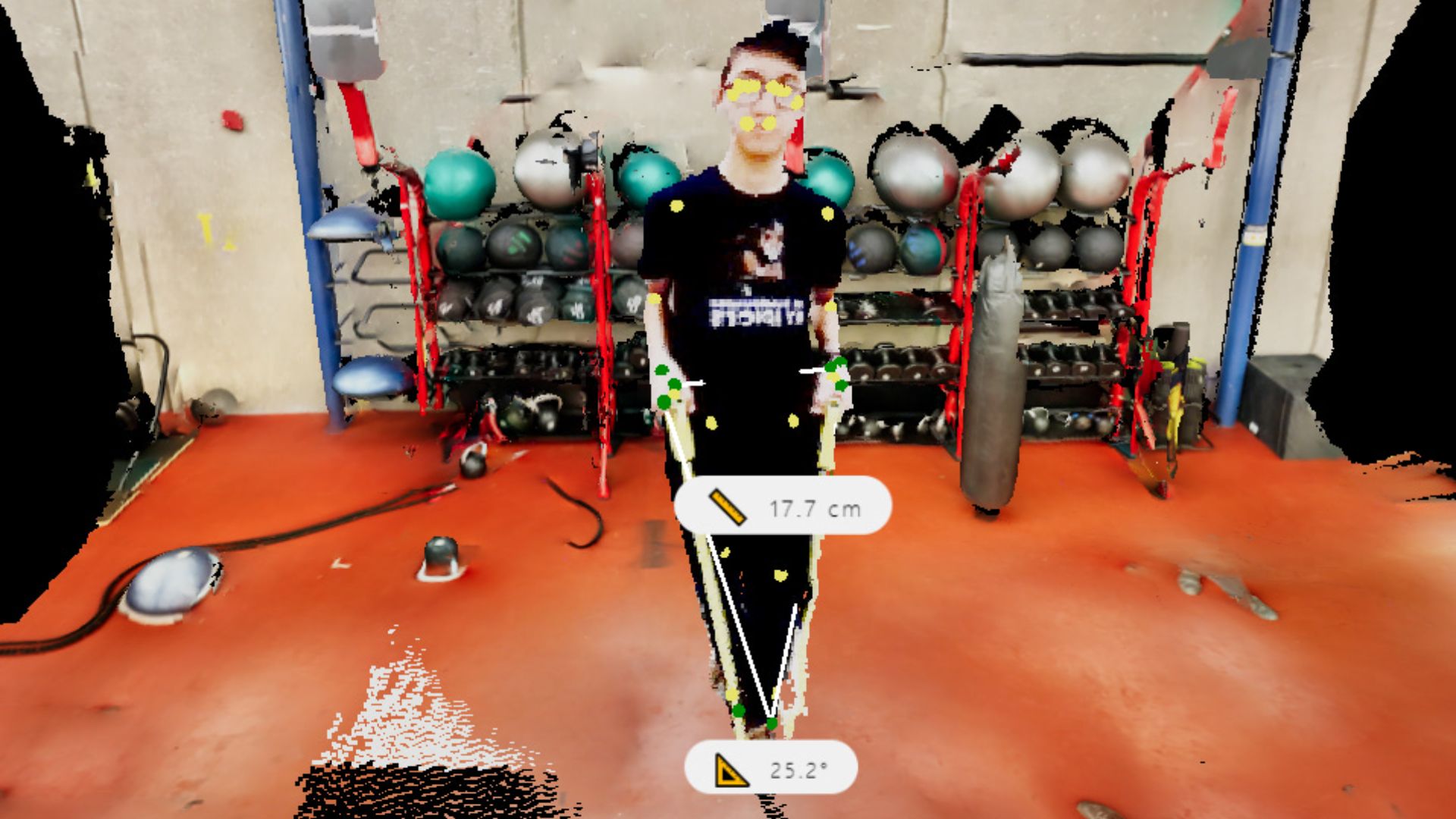}
\includegraphics[width=0.245\linewidth]{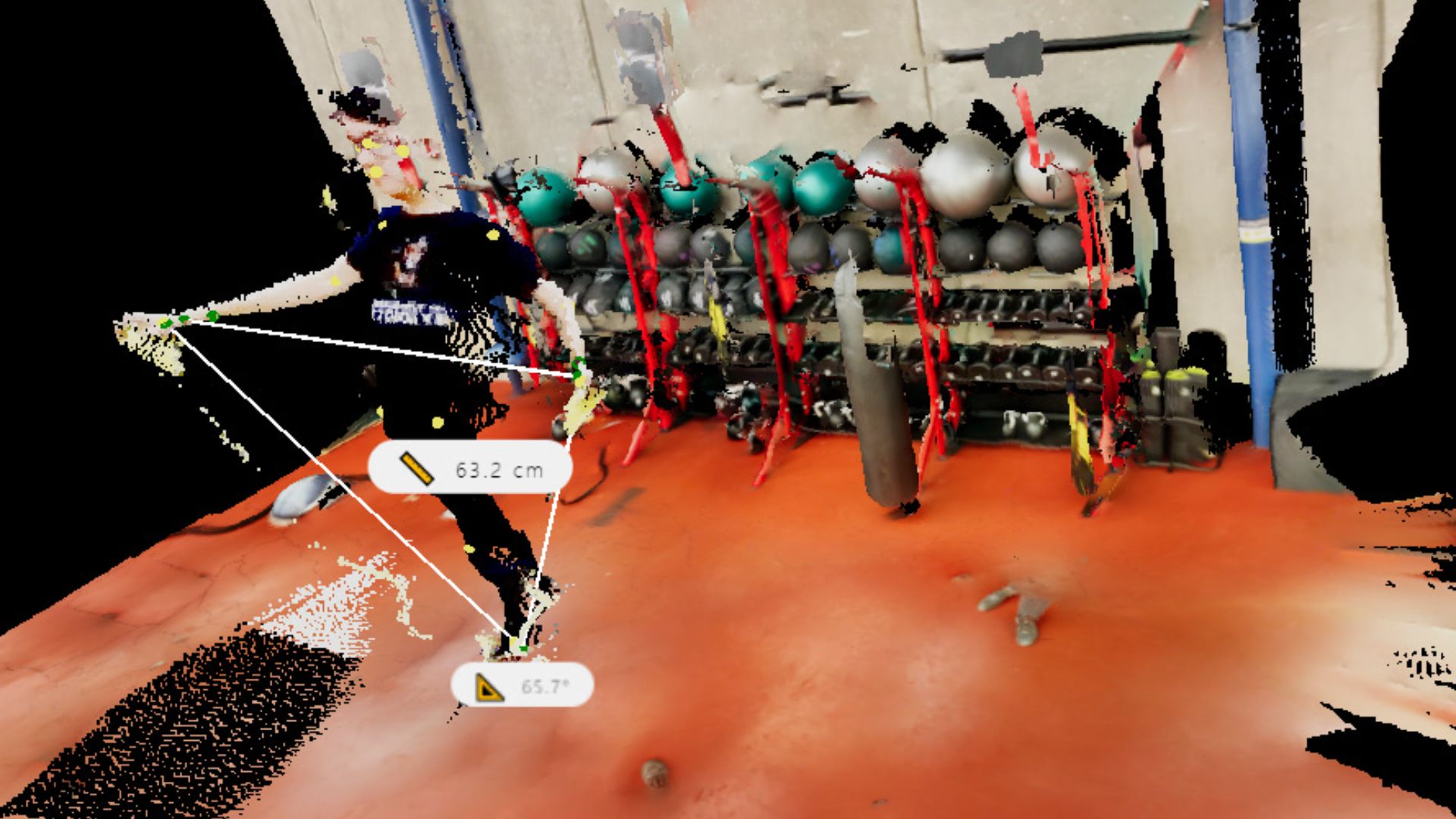}
\includegraphics[width=0.245\linewidth]{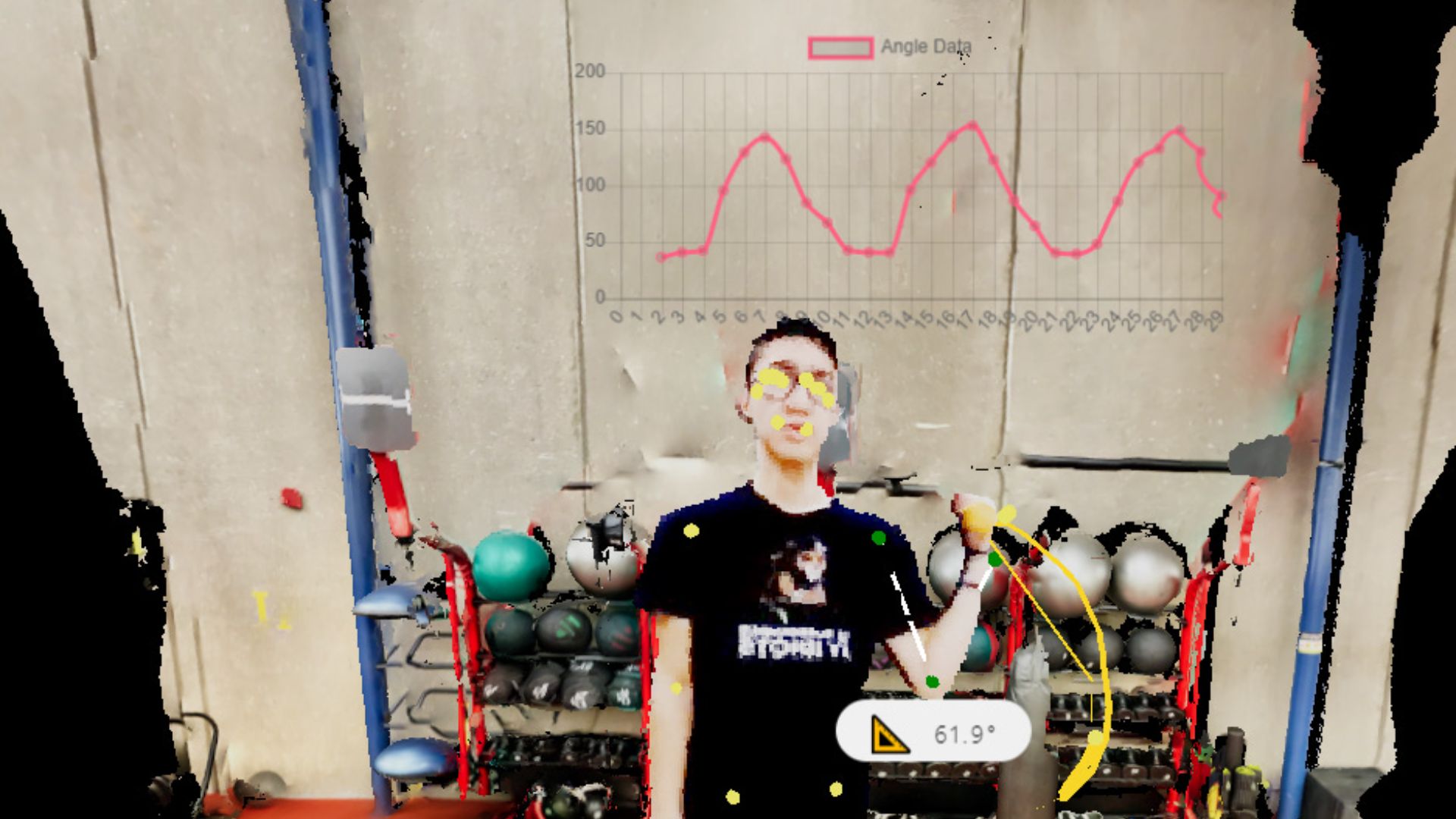}
\includegraphics[width=0.245\linewidth]{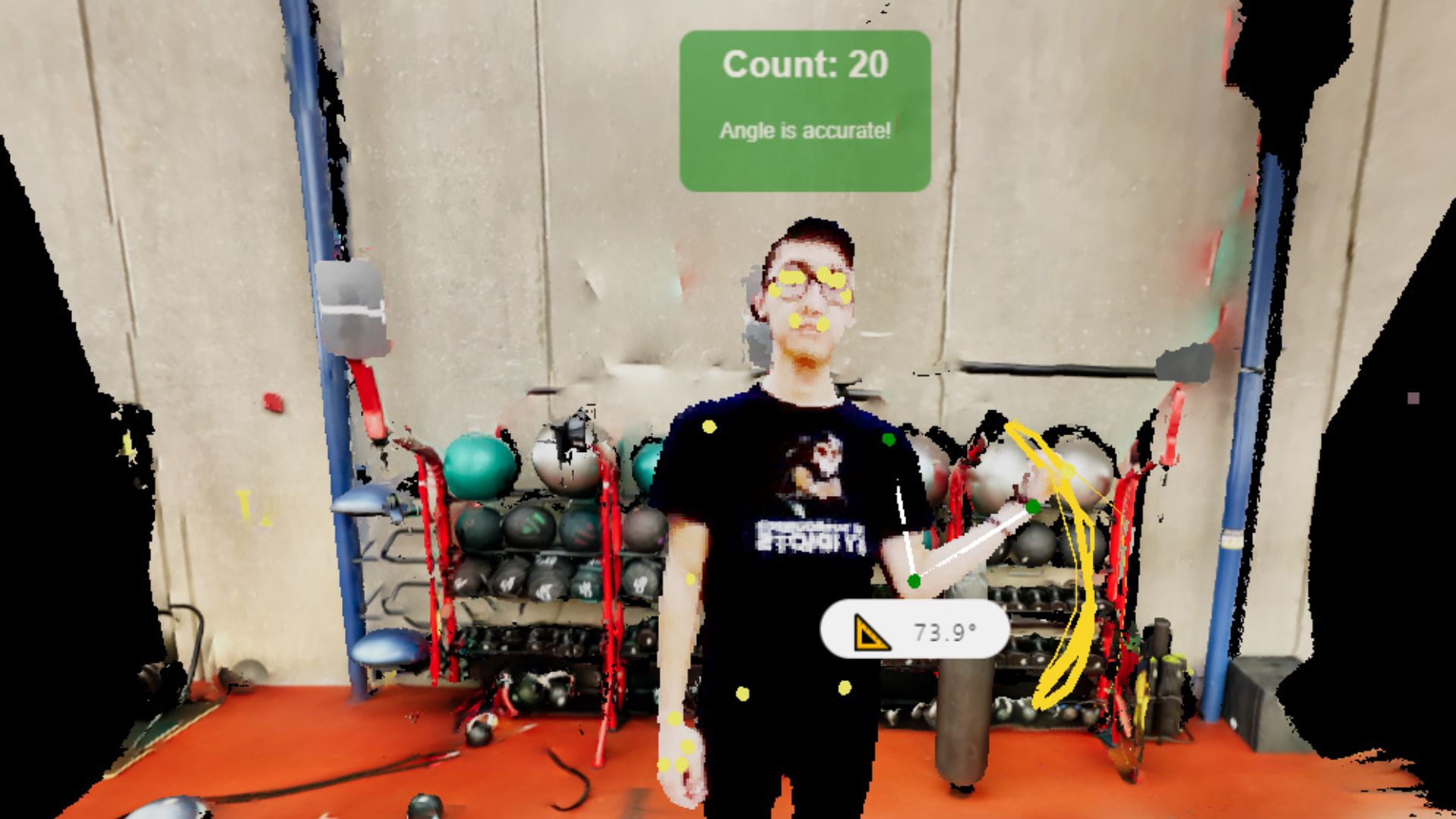}
\caption{Physical Training and Analysis: Use case scenario demonstrating a physical workout routine by measuring repetitions and bodily motion to display them as visualizations.}
\label{fig:physical-training}
\end{figure*}

These applications demonstrate the versatility of \system{} in enhancing interactive and dynamic visual experiences across various domains, from commercial showcases to educational settings and athletic training.

%% file: 6-user-study.tex
\newcommand{\subsub}[1]{\noindent\textbf{\textit{- #1:}}}

\section{User Study}
\diff{To evaluate the effectiveness and user satisfaction of our system, we conducted a lab-based usability study} with 19 participants (13 males, 6 females; aged between 19 and 29) from our local community, \diff{consisting of university students and working professionals on campus}. Each participant was compensated with a \$10 Amazon gift card for their involvement in the user study. \diff{Our study was structured around the ``usage evaluation'' framework proposed by Ledo et al.~\cite{ledo2018evaluation}. The primary purpose of conducting usability studies with end-users is to assess the creative freedom, ease of use, learnability, and overall usability of the system. We also measured which design features were useful in helping participants achieve their goals. Given that our system introduces a novel authoring tool for 3D volumetric videos, we lack a direct comparison against established baselines. To overcome this challenge, we employed a combination of lab-based usability studies and in-depth interviews with users. This approach allowed us to uncover the strengths and weaknesses of our system and provided valuable insights that will inform future research.}

\subsection{Method and Study Protocol}

\subsubsection{Method}
\diff{The study was structured into two sessions: the first aimed at evaluating the prototype's usability to ascertain its effectiveness, ease of use, and learnability through structured tasks and a survey}. Before the first session, we inquired about participants' experience with 3D graphics software development tools like Unity 3D, Blender, and Unreal Engine. Identifying experienced participants helped in gaining deeper insights during the follow-up conversational interviews. \diff{The second session involved an in-depth interview to discuss the system's benefits, limitations, and potential improvements.}

\subsubsection{Study Protocol}
\diff{The user study was designed to measure specific usability factors such as learnability, satisfaction, and ease of use. The total duration for the study was between 45 to 60 minutes per participant, structured as follows:}

\subsub{Introduction (3-5 minutes)}
Participants were introduced to the project's goals and the underlying technology. An online whiteboard presentation outlined the system's design and features, and participants were briefed on the concept of volumetric video, setting the stage for the tasks they would perform.

\subsub{Demonstration and Application (24-30 minutes)}
The demonstration phase was split into two parts to cover different aspects of the system. Initially, participants followed a guided tutorial with slides on how to attach static annotations for a product advertisement scenario. \diff{This task aimed to assess the system’s learnability and ease of use. Subsequently, participants engaged in a more complex task involving motion tracking, simulating a physical training scenario to evaluate the system’s performance under dynamic conditions. This helped in assessing the robustness and responsiveness of the system.}

\subsub{Survey (15-20 minutes)}
Finally, participants completed a Google Form questionnaire to provide feedback on their experience. \diff{The survey included questions designed to measure user satisfaction and identify usability issues, thereby providing qualitative and quantitative data to support the usability assessment.}

\subsection{Results}

\subsubsection{Demographics}
\begin{figure}[h]
\centering
\includegraphics[width=1\linewidth]{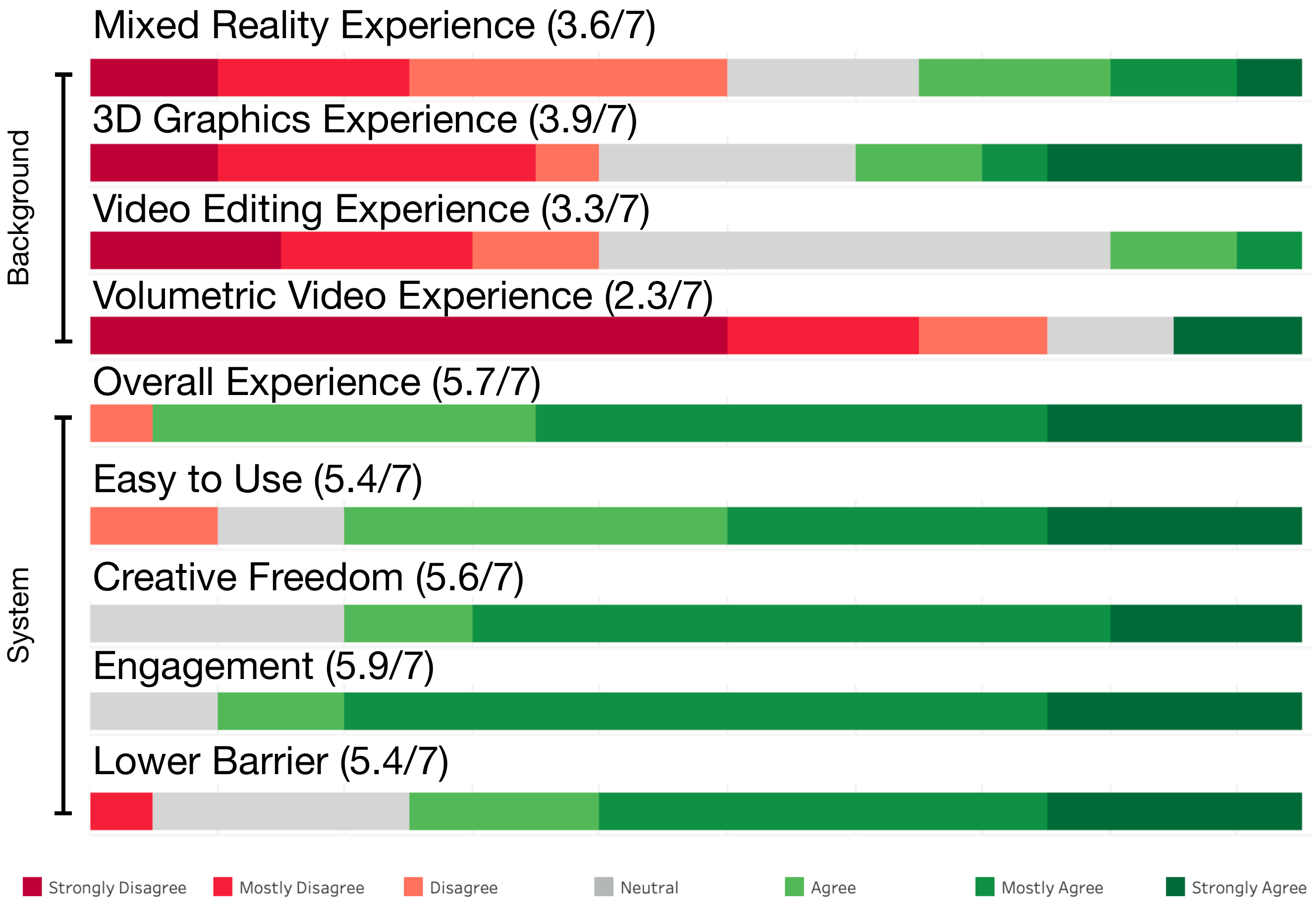}
\caption{User Study Results -- A graph summarizing the 7-point Likert scale responses of demographics and overall experiences from 19 participants.}
\label{fig:user-study-1}
\end{figure}

We asked the participants at the beginning of our survey to better learn about their background and demographics towards mixed reality experience, 3D graphics software, video editing experience, and volumetric video experience. The collected demographic information is shown in Figure \ref{fig:user-study-1}. Non-surprisingly, many of our participants do not have 3D graphics development experience or volumetric video experience. Many of them also mentioned that it was their first time hearing about 3D video or 3D volumetric video.

\subsubsection{Overall Experiences} As shown from the Figure \ref{fig:user-study-1}, which summarized 

The Figure ~\ref{fig:user-study-1} outlines the study results. Overall, the vast majority of participants had positive responses. Participant scores of the overall experience averaged 5.7/7. \textit{"The user experience was fun and interesting. It was pretty intuitive as well"}(P9) and  \textit{"It is such a fresh and interesting experience for me to play a 3D reality product."} (P13). Some participants also had optimistic views towards the system \textit{Seems interesting and could have potential uses for video editing software}(P16).
%Many comments about poor video quality

%Kevin - Unsure if the ease of use subsection should fall under the overall experience of authoring
\subsubsection{Ease of Use}
The system was determined to be fairly easy to use, averaging 5.4/7. P8 found \textit{"It was intuitive"} and said \textit{"the user panel was accessible"}. However, a (P5, P10, P13) found it hard to make selections. However, others(P9, 19) with experience with 3D software found it easy. P13 who was unfamiliar with using software with 3D spaces found it difficult to navigate. P3 put up concerns with certain demographics unfamiliar with 3D software and that \textit{it might seem overwhelming}. However, they did find the streamlined interface made \textit{animation and editing so much faster} P3.
%There were many suggestions for improving the ease of use

\subsubsection{Flexibility and Creative Freedom}
The creative freedom of the system was reported to be flexible, with an average score of 5.6/7. Regarding the variety of the features, P1 declared that they could \textit{imagine multiple uses for them}. Another said \textit{The trail feature and ghost effect inspired my creativity} P10. Body motion features resonated more and had positive feedback in terms of creative freedom, P19 said \textit{the ghosting and trailing effects were features they would personally use for creative reasons}.

\begin{figure}[h]
\centering
\includegraphics[width=1\linewidth]{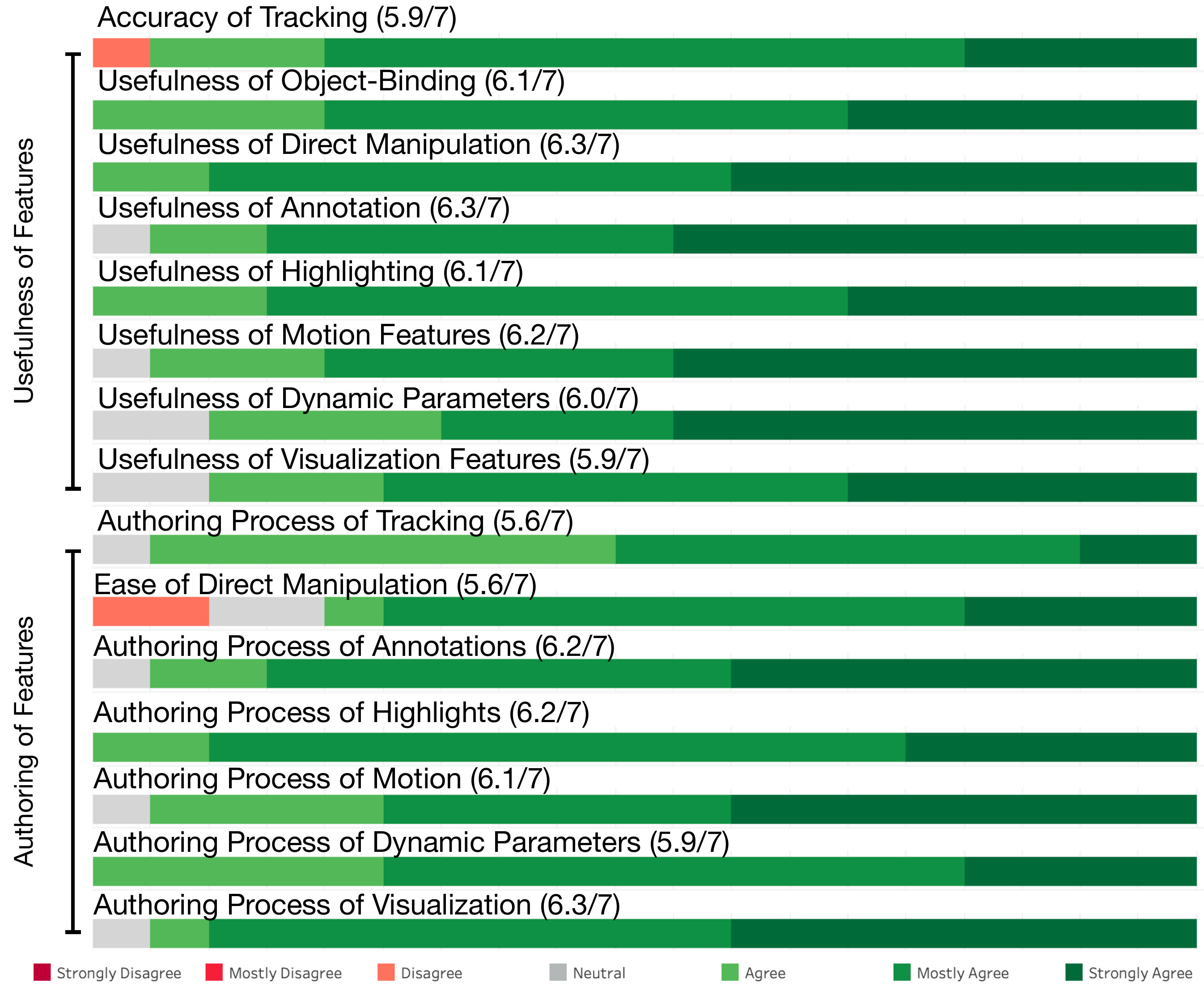}
\caption{User Study Results -- A graph summarizing the 7-point Likert scale responses of the usefulness of features and authoring experiences from 19 participants.}
\label{fig:user-study-2}
\end{figure}

\subsubsection{Usefulness of features}

One declared the tracking and binding features were necessary, as \textit{labels and highlights would likely not work without them.} (P2)
From the questionnaire, highlighting had positive feedback and that the torus \textit{seems very useful for highlighting objects in the scene} (P8, P11)

%These were categories in the questionnaire, should they be condensed somehow or individually review features(Probably not a good idea to do individually as it will be lengthy)?
% \subsub{Tracking}
% \subsub{Object binding}
% \subsub{Annotation}
% \subsub{Highlighting}
% \subsub{Motion}
% One participant stated, "They are pretty good effects for entertaining purposes".
% \subsub{Dynamic parameters}
% \subsub{Visualization}

%Many comments about the novelty of the 3D environment and its usefulness

\subsubsection{Potential Applications and Use Scenarios}
When asked about the potential applications and use cases, many participants see the potential for sports analysis (P3, P10) and data visualization (P7, P14). 
For example, P3 says \textit{"On the note of sports, like for example, in ballet, your position is very important. So if you're doing plays like, for example, videos where you highlight certain points like knees they have to be at a certain angle."}.
As the participants say, our tool allows the user not only to analyze the sports from different angles, but also to measure the trajectory or posture in improvisational and interactive ways. 
Also, the participants see the future potential of video streaming. 
For example, one of P4 says that \textit{"I would use this in my stream in some way to incorporate effects. And maybe even have audience members triggering different things in my space"}
The feedback points to a future when 3D volumetric videos become mainstream. The participants see that tools like \system{} allow such a video streaming medium more interactive and engaging.

\subsubsection{Limitations and Future Work}
While the feedback was generally positive, several limitations were noted. P7 mentioned that the tool's functionality is currently limited, particularly the types of 3D shapes available for highlighting objects, which are restricted to simple geometric forms. Many participants pointed out the need for improvement in tracking accuracy, especially with the color-based tracking system's susceptibility to slight environmental changes, such as lighting conditions or shadow occlusion. Future work should explore alternative tracking methods for 3D objects, given that volumetric object tracking remains an active research area and advances in this field could significantly enhance object-centric 3D video augmentation.

System limitations also include the requirement for physical interaction or assistance from another person, as stated by P4. P3 elaborated on this by mentioning the extensive setup required for video capture, including the need for an open area, camera rig, and trackable objects. Additionally, the tool currently lacks complex time manipulation features found in traditional video-editing tools. Integrating our features into volumetric video-editing platforms could create richer experiences.

Participants also criticized the quality of 3D capturing. Using only a single Azure Kinect depth camera limits the capture area and fails to record occluded regions. Although we integrated a static 3D scene as a background to mitigate this, the limited capture area restricts applications requiring dynamic movement across larger areas. A potential solution could involve using multiple Kinect cameras, similar to approaches like Remixed Reality~\cite{lindlbauer2018remixed}. While this would allow more immersive visualization, it would also increase computational demands and the complexity of the calibration process. Future work should consider incorporating multiple depth cameras to support activities requiring broader interaction spaces.

\diff{Direct manipulation in \system{} enables users to feel an immediate connection with the digital content, fostering a sense of control and ownership over the creative process. However, we recognize the importance of considering alternative approaches that could complement or enhance the user experience. Automatic annotation, for example, could offer efficiency benefits by reducing the manual effort required to label and annotate volumetric data. This method could automatically identify and label objects within a scene using advanced machine learning algorithms, which would be particularly useful in complex scenes or for users who require quicker workflows. A suggestion-based interface is another compelling alternative that could blend the strengths of direct manipulation with the efficiencies of automation. By providing users with intelligent suggestions based on context, previous actions, or common patterns, this approach could accelerate the editing process while still allowing users the freedom to make final decisions. Future work could explore these alternatives to support a wider range of user preferences.}

Currently, we focus on desktop authoring interfaces due to the complexity of interactions and manipulations involved. However, future investigations could explore opportunities within immersive environments using mixed reality or virtual reality headsets. Such environments would present unique design and technical challenges, such as selecting objects and streaming large amounts of data between the host computer and the headset. Addressing these issues could lead to innovative solutions for immersive augmentation, and we are keen on developing these capabilities for devices like the Hololens.

%% file: 7-future-work.tex
% \section{Limitations and Future Work}

% Limited number of features 
% cannot customize the shape 

%% file: 8-conclusion.tex
\section{Conclusion}
This paper presents \system{}, a desktop authoring interface designed to edit and augment 3D volumetric videos with object-centric annotations and visual effects. We introduce a novel approach to augment captured physical motion with embedded and responsive visual effects. The primary contribution of this paper is the development of a taxonomy of augmentation techniques. We demonstrate various augmentation techniques, including annotated labels, highlighted objects, ghost effects, and trajectory visualization. The results of our user study indicate that our direct manipulation techniques significantly lower the barrier to annotating volumetric videos. Based on the feedback received, we also discuss potential future work.

%% file: acknowledgements.tex
\begin{acks}
This work was partially supported by NSERC Discovery Grant and JST PRESTO Grant Number JPMJPR23I5, Japan.
\end{acks}

%% file: appendix.tex
\section{Appendix}

\diff{The following videos, which are a subset of the 120 videos analyzed for this taxonomy, are used as examples of object-centric augmentation techniques in Figure~\ref{fig:design-space}. Each letter indicates the category of the augmentations: T for text annotation, O for object highlight, E for embedded visual, C for connected link, and M for motion effect. Each number represents the order from left to right. The screenshots in Figure~\ref{fig:design-space} are copyrighted by each video creator.}

\begin{enumerate}
\item[{[T1]}] Clearly Contacts "Saving Money" \textcopyright \ Copyright by Giant Ant \\ \url{https://vimeo.com/10904876}
\item[{[T2]}] NORTH: Analytics for the real world — Symphoni \textcopyright \ Copyright by PwC Digital Experience Center \\ \url{https://vimeo.com/121175225}
\item[{[T3]}] GRTgaz biomethane \textcopyright \ Copyright by la famille \\ \url{https://vimeo.com/40092864}
\item[{[T4]}] Zoopla TV advert "Smart Knows" \textcopyright \ Copyright by Zoopla \\ \url{https://www.youtube.com/watch?v=jkADFJdYakY}
\item[{[T5]}] inBloom vision video \textcopyright \ Copyright by Intentional Futures \\ \url{https://vimeo.com/60661666}
\item[{[T6]}] DREAN // Motion Tracking + layouts \textcopyright \ Copyright by Estudio Ánimo \\ \url{https://vimeo.com/68242831}
\item[{[O1]}] Live from Tokyo: 2018 Nissan LEAF Launch \textcopyright \ Copyright by 
George P. Johnson \\ \url{https://www.youtube.com/watch?v=EoMU3SuZ-uw}
\item[{[O2]}] Device UI in Realtime \textcopyright \ Copyright by Dennis Schaefer \\ \url{https://vimeo.com/165467760}
\item[{[O3]}] Whirlpool Interactive Cooktop \textcopyright \ Copyright by The Hobbs Report \\ \url{https://www.youtube.com/watch?v=Efj6gKw3wKc}
\item[{[O4]}] Alibaba brings AR, VR, and virtual influencers to online shopping \textcopyright \ Copyright by TechNode \\ \url{https://www.youtube.com/watch?v=xLQAxYMYxlU}
\item[{[O5]}] GRTgaz biomethane \textcopyright \ Copyright by la famille \\ \url{https://vimeo.com/40092864}
\item[{[O6]}] NORTH: Analytics for the real world — Symphoni \textcopyright \ Copyright by PwC Digital Experience Center \\ \url{https://vimeo.com/121175225}
\item[{[E1]}] Ericsson - Business Users Survey - Commercial \textcopyright \ Copyright by Erik Nordlund, FSF \\ \url{https://vimeo.com/20168424}
\item[{[E2]}] NTT Data - Future Experiences \textcopyright \ Copyright by Designit \\ \url{https://vimeo.com/142118168}
\item[{[E3]}] Crafting Brands for Future Life \textcopyright \ Copyright by Ben Collier-Marsh \\ \url{https://vimeo.com/196708386}
\item[{[E4]}] Mixed Reality - Home Kit \textcopyright \ Copyright by Sertan Helvacı \\ \url{https://dribbble.com/shots/6172560-Mixed-Reality-Home-Kit}
\item[{[E5]}] Scosche myTrek :: 2011 [Evlab] \textcopyright \ Copyright by Greg Del Savio \\ \url{https://vimeo.com/27620294}
\item[{[E6]}] Sight \textcopyright \ Copyright by Eran May-Raz and Daniel Lazo\\ \url{https://www.youtube.com/watch?v=OstCyV0nOGs}
\item[{[C1]}] Ericsson - Business Users Survey - Commercial \textcopyright \ Copyright by Erik Nordlund, FSF \\ \url{https://vimeo.com/20168424}
\item[{[C2]}] Zoopla TV advert "Smart Knows" \textcopyright \ Copyright by Zoopla \\ \url{https://www.youtube.com/watch?v=jkADFJdYakY}
\item[{[C3]}] Scosche myTrek :: 2011 [Evlab] \textcopyright \ Copyright by Greg Del Savio \\ \url{https://vimeo.com/27620294}
\item[{[C4]}] DREAN // Motion Tracking + layouts \textcopyright \ Copyright by Estudio Ánimo \\ \url{https://vimeo.com/68242831}
\item[{[C5]}] La Boulangerie Delannay \textcopyright \ Copyright by Julien Loth \\ \url{https://vimeo.com/45055294}
\item[{[C6]}] Thomson // Reuters \textcopyright \ Copyright by Rushes Creative, Domhnall Ó Maoleoin, BT CORCORAN, Tania Nunes, and Guy Hancock \\ \url{https://www.behance.net/gallery/54032303/Thomson-Reuters}
\item[{[M1]}] Writing Performance in the Language of Light \textcopyright \ Copyright by GE Lighting, a Savant company \\ \url{https://www.youtube.com/watch?v=G9cBpSRT500}
\item[{[M2]}] Yuki Ota Fencing Visualized Project - MORE ENJOY FENCING (English Ver.) \textcopyright \ Copyright by fencing visualized project \\ \url{https://www.youtube.com/watch?v=h2DXCAWI8gU}
\item[{[M3]}] IBM PGA \textcopyright \ Copyright by LOS YORK \\ \url{https://vimeo.com/40882289}
\item[{[M4]}] FeelCapital corporate video \textcopyright \ Copyright by democràcia \\ \url{https://vimeo.com/98023574}
\item[{[M5]}] Nike: Pegasus 31 \textcopyright \ Copyright by Tad Greenough \\ \url{https://vimeo.com/132446809}
\item[{[M6]}] Writing Performance in the Language of Light \textcopyright \ Copyright by GE Lighting, a Savant company \\ \url{https://www.youtube.com/watch?v=G9cBpSRT500}
\end{enumerate}

%% file: main.bbl
%%% -*-BibTeX-*-
%%% Do NOT edit. File created by BibTeX with style
%%% ACM-Reference-Format-Journals [18-Jan-2012].

\begin{thebibliography}{101}

%%% ====================================================================
%%% NOTE TO THE USER: you can override these defaults by providing
%%% customized versions of any of these macros before the \bibliography
%%% command.  Each of them MUST provide its own final punctuation,
%%% except for \shownote{}, \showDOI{}, and \showURL{}.  The latter two
%%% do not use final punctuation, in order to avoid confusing it with
%%% the Web address.
%%%
%%% To suppress output of a particular field, define its macro to expand
%%% to an empty string, or better, \unskip, like this:
%%%
%%% \newcommand{\showDOI}[1]{\unskip}   % LaTeX syntax
%%%
%%% \def \showDOI #1{\unskip}           % plain TeX syntax
%%%
%%% ====================================================================

\ifx \showCODEN    \undefined \def \showCODEN     #1{\unskip}     \fi
\ifx \showDOI      \undefined \def \showDOI       #1{#1}\fi
\ifx \showISBNx    \undefined \def \showISBNx     #1{\unskip}     \fi
\ifx \showISBNxiii \undefined \def \showISBNxiii  #1{\unskip}     \fi
\ifx \showISSN     \undefined \def \showISSN      #1{\unskip}     \fi
\ifx \showLCCN     \undefined \def \showLCCN      #1{\unskip}     \fi
\ifx \shownote     \undefined \def \shownote      #1{#1}          \fi
\ifx \showarticletitle \undefined \def \showarticletitle #1{#1}   \fi
\ifx \showURL      \undefined \def \showURL       {\relax}        \fi
% The following commands are used for tagged output and should be
% invisible to TeX
\providecommand\bibfield[2]{#2}
\providecommand\bibinfo[2]{#2}
\providecommand\natexlab[1]{#1}
\providecommand\showeprint[2][]{arXiv:#2}

\bibitem[\protect\citeauthoryear{4DViews}{4DViews}{[n.d.]}]%
        {4dfx}
\bibfield{author}{\bibinfo{person}{4DViews}.} \bibinfo{year}{[n.d.]}\natexlab{}.
\newblock \bibinfo{title}{4Dfx}.
\newblock \bibinfo{howpublished}{\url{https://www.4dviews.com/volumetric-software}}.
\newblock


\bibitem[\protect\citeauthoryear{Adcock, Anderson, and Thomas}{Adcock et~al\mbox{.}}{2013}]%
        {adcock2013remotefusion}
\bibfield{author}{\bibinfo{person}{Matt Adcock}, \bibinfo{person}{Stuart Anderson}, {and} \bibinfo{person}{Bruce Thomas}.} \bibinfo{year}{2013}\natexlab{}.
\newblock \showarticletitle{RemoteFusion: real time depth camera fusion for remote collaboration on physical tasks}. In \bibinfo{booktitle}{\emph{Proceedings of the 12th ACM SIGGRAPH international conference on virtual-reality continuum and its applications in industry}}. \bibinfo{pages}{235--242}.
\newblock


\bibitem[\protect\citeauthoryear{Anderson, Grossman, Matejka, and Fitzmaurice}{Anderson et~al\mbox{.}}{2013}]%
        {anderson2013youmove}
\bibfield{author}{\bibinfo{person}{Fraser Anderson}, \bibinfo{person}{Tovi Grossman}, \bibinfo{person}{Justin Matejka}, {and} \bibinfo{person}{George Fitzmaurice}.} \bibinfo{year}{2013}\natexlab{}.
\newblock \showarticletitle{YouMove: enhancing movement training with an augmented reality mirror}. In \bibinfo{booktitle}{\emph{Proceedings of the 26th annual ACM symposium on User interface software and technology}}. \bibinfo{pages}{311--320}.
\newblock


\bibitem[\protect\citeauthoryear{Benko, Wilson, and Zannier}{Benko et~al\mbox{.}}{2014}]%
        {benko2014dyadic}
\bibfield{author}{\bibinfo{person}{Hrvoje Benko}, \bibinfo{person}{Andrew~D Wilson}, {and} \bibinfo{person}{Federico Zannier}.} \bibinfo{year}{2014}\natexlab{}.
\newblock \showarticletitle{Dyadic projected spatial augmented reality}. In \bibinfo{booktitle}{\emph{Proceedings of the 27th annual ACM symposium on User interface software and technology}}. \bibinfo{pages}{645--655}.
\newblock


\bibitem[\protect\citeauthoryear{Brudy, Suwanwatcharachat, Zhang, Houben, and Marquardt}{Brudy et~al\mbox{.}}{2018}]%
        {brudy2018eagleview}
\bibfield{author}{\bibinfo{person}{Frederik Brudy}, \bibinfo{person}{Suppachai Suwanwatcharachat}, \bibinfo{person}{Wenyu Zhang}, \bibinfo{person}{Steven Houben}, {and} \bibinfo{person}{Nicolai Marquardt}.} \bibinfo{year}{2018}\natexlab{}.
\newblock \showarticletitle{Eagleview: A video analysis tool for visualising and querying spatial interactions of people and devices}. In \bibinfo{booktitle}{\emph{Proceedings of the 2018 ACM International Conference on Interactive Surfaces and Spaces}}. \bibinfo{pages}{61--72}.
\newblock


\bibitem[\protect\citeauthoryear{B{\"u}schel, Lehmann, and Dachselt}{B{\"u}schel et~al\mbox{.}}{2021}]%
        {buschel2021miria}
\bibfield{author}{\bibinfo{person}{Wolfgang B{\"u}schel}, \bibinfo{person}{Anke Lehmann}, {and} \bibinfo{person}{Raimund Dachselt}.} \bibinfo{year}{2021}\natexlab{}.
\newblock \showarticletitle{Miria: A mixed reality toolkit for the in-situ visualization and analysis of spatio-temporal interaction data}. In \bibinfo{booktitle}{\emph{Proceedings of the 2021 CHI Conference on Human Factors in Computing Systems}}. \bibinfo{pages}{1--15}.
\newblock


\bibitem[\protect\citeauthoryear{Cao, Fuste, and Heun}{Cao et~al\mbox{.}}{2022}]%
        {cao2022mobiletutar}
\bibfield{author}{\bibinfo{person}{Yuanzhi Cao}, \bibinfo{person}{Anna Fuste}, {and} \bibinfo{person}{Valentin Heun}.} \bibinfo{year}{2022}\natexlab{}.
\newblock \showarticletitle{MobileTutAR: a Lightweight Augmented Reality Tutorial System using Spatially Situated Human Segmentation Videos}. In \bibinfo{booktitle}{\emph{CHI Conference on Human Factors in Computing Systems Extended Abstracts}}. \bibinfo{pages}{1--8}.
\newblock


\bibitem[\protect\citeauthoryear{Cao, Wang, Qian, Rao, Wadhawan, Huo, and Ramani}{Cao et~al\mbox{.}}{2019}]%
        {cao2019ghostar}
\bibfield{author}{\bibinfo{person}{Yuanzhi Cao}, \bibinfo{person}{Tianyi Wang}, \bibinfo{person}{Xun Qian}, \bibinfo{person}{Pawan~S Rao}, \bibinfo{person}{Manav Wadhawan}, \bibinfo{person}{Ke Huo}, {and} \bibinfo{person}{Karthik Ramani}.} \bibinfo{year}{2019}\natexlab{}.
\newblock \showarticletitle{GhostAR: A time-space editor for embodied authoring of human-robot collaborative task with augmented reality}. In \bibinfo{booktitle}{\emph{Proceedings of the 32nd Annual ACM Symposium on User Interface Software and Technology}}. \bibinfo{pages}{521--534}.
\newblock


\bibitem[\protect\citeauthoryear{Chen, Ye, Chu, Xia, Zhang, Qu, and Wu}{Chen et~al\mbox{.}}{2021}]%
        {chen2021augmenting}
\bibfield{author}{\bibinfo{person}{Zhutian Chen}, \bibinfo{person}{Shuainan Ye}, \bibinfo{person}{Xiangtong Chu}, \bibinfo{person}{Haijun Xia}, \bibinfo{person}{Hui Zhang}, \bibinfo{person}{Huamin Qu}, {and} \bibinfo{person}{Yingcai Wu}.} \bibinfo{year}{2021}\natexlab{}.
\newblock \showarticletitle{Augmenting sports videos with viscommentator}.
\newblock \bibinfo{journal}{\emph{IEEE Transactions on Visualization and Computer Graphics}} \bibinfo{volume}{28}, \bibinfo{number}{1} (\bibinfo{year}{2021}), \bibinfo{pages}{824--834}.
\newblock


\bibitem[\protect\citeauthoryear{Cheng, Ofek, Holz, and Wilson}{Cheng et~al\mbox{.}}{2019}]%
        {cheng2019vroamer}
\bibfield{author}{\bibinfo{person}{Lung-Pan Cheng}, \bibinfo{person}{Eyal Ofek}, \bibinfo{person}{Christian Holz}, {and} \bibinfo{person}{Andrew~D Wilson}.} \bibinfo{year}{2019}\natexlab{}.
\newblock \showarticletitle{Vroamer: generating on-the-fly VR experiences while walking inside large, unknown real-world building environments}. In \bibinfo{booktitle}{\emph{2019 IEEE Conference on Virtual Reality and 3D User Interfaces (VR)}}. IEEE, \bibinfo{pages}{359--366}.
\newblock


\bibitem[\protect\citeauthoryear{Cheng, Li, Xu, Li, Yang, Wang, and Yang}{Cheng et~al\mbox{.}}{2023}]%
        {cheng2023segment}
\bibfield{author}{\bibinfo{person}{Yangming Cheng}, \bibinfo{person}{Liulei Li}, \bibinfo{person}{Yuanyou Xu}, \bibinfo{person}{Xiaodi Li}, \bibinfo{person}{Zongxin Yang}, \bibinfo{person}{Wenguan Wang}, {and} \bibinfo{person}{Yi Yang}.} \bibinfo{year}{2023}\natexlab{}.
\newblock \showarticletitle{Segment and track anything}.
\newblock \bibinfo{journal}{\emph{arXiv preprint arXiv:2305.06558}} (\bibinfo{year}{2023}).
\newblock


\bibitem[\protect\citeauthoryear{Cheng, Yan, Yi, Shi, and Lindlbauer}{Cheng et~al\mbox{.}}{2021}]%
        {cheng2021semanticadapt}
\bibfield{author}{\bibinfo{person}{Yifei Cheng}, \bibinfo{person}{Yukang Yan}, \bibinfo{person}{Xin Yi}, \bibinfo{person}{Yuanchun Shi}, {and} \bibinfo{person}{David Lindlbauer}.} \bibinfo{year}{2021}\natexlab{}.
\newblock \showarticletitle{Semanticadapt: Optimization-based adaptation of mixed reality layouts leveraging virtual-physical semantic connections}. In \bibinfo{booktitle}{\emph{The 34th Annual ACM Symposium on User Interface Software and Technology}}. \bibinfo{pages}{282--297}.
\newblock


\bibitem[\protect\citeauthoryear{Cheng, Yin, Yan, Gugenheimer, and Lindlbauer}{Cheng et~al\mbox{.}}{2022}]%
        {cheng2022towards}
\bibfield{author}{\bibinfo{person}{Yi~Fei Cheng}, \bibinfo{person}{Hang Yin}, \bibinfo{person}{Yukang Yan}, \bibinfo{person}{Jan Gugenheimer}, {and} \bibinfo{person}{David Lindlbauer}.} \bibinfo{year}{2022}\natexlab{}.
\newblock \showarticletitle{Towards Understanding Diminished Reality}. In \bibinfo{booktitle}{\emph{CHI Conference on Human Factors in Computing Systems}}. \bibinfo{pages}{1--16}.
\newblock


\bibitem[\protect\citeauthoryear{Chi, Vogel, Dontcheva, Li, and Hartmann}{Chi et~al\mbox{.}}{2016}]%
        {chi2016authoring}
\bibfield{author}{\bibinfo{person}{Pei-Yu Chi}, \bibinfo{person}{Daniel Vogel}, \bibinfo{person}{Mira Dontcheva}, \bibinfo{person}{Wilmot Li}, {and} \bibinfo{person}{Bj{\"o}rn Hartmann}.} \bibinfo{year}{2016}\natexlab{}.
\newblock \showarticletitle{Authoring illustrations of human movements by iterative physical demonstration}. In \bibinfo{booktitle}{\emph{Proceedings of the 29th Annual Symposium on User Interface Software and Technology}}. \bibinfo{pages}{809--820}.
\newblock


\bibitem[\protect\citeauthoryear{Chidambaram, Huang, He, Qian, Villanueva, Redick, Stuerzlinger, and Ramani}{Chidambaram et~al\mbox{.}}{2021}]%
        {chidambaram2021processar}
\bibfield{author}{\bibinfo{person}{Subramanian Chidambaram}, \bibinfo{person}{Hank Huang}, \bibinfo{person}{Fengming He}, \bibinfo{person}{Xun Qian}, \bibinfo{person}{Ana~M Villanueva}, \bibinfo{person}{Thomas~S Redick}, \bibinfo{person}{Wolfgang Stuerzlinger}, {and} \bibinfo{person}{Karthik Ramani}.} \bibinfo{year}{2021}\natexlab{}.
\newblock \showarticletitle{Processar: An augmented reality-based tool to create in-situ procedural 2d/3d ar instructions}. In \bibinfo{booktitle}{\emph{Designing Interactive Systems Conference 2021}}. \bibinfo{pages}{234--249}.
\newblock


\bibitem[\protect\citeauthoryear{Clarke, Cavdir, Chiu, Denoue, and Kimber}{Clarke et~al\mbox{.}}{2020}]%
        {clarke2020reactive}
\bibfield{author}{\bibinfo{person}{Christopher Clarke}, \bibinfo{person}{Doga Cavdir}, \bibinfo{person}{Patrick Chiu}, \bibinfo{person}{Laurent Denoue}, {and} \bibinfo{person}{Don Kimber}.} \bibinfo{year}{2020}\natexlab{}.
\newblock \showarticletitle{Reactive video: adaptive video playback based on user motion for supporting physical activity}. In \bibinfo{booktitle}{\emph{Proceedings of the 33rd Annual ACM Symposium on User Interface Software and Technology}}. \bibinfo{pages}{196--208}.
\newblock


\bibitem[\protect\citeauthoryear{Dai, Chang, Savva, Halber, Funkhouser, and Nie{\ss}ner}{Dai et~al\mbox{.}}{2017}]%
        {dai2017scannet}
\bibfield{author}{\bibinfo{person}{Angela Dai}, \bibinfo{person}{Angel~X Chang}, \bibinfo{person}{Manolis Savva}, \bibinfo{person}{Maciej Halber}, \bibinfo{person}{Thomas Funkhouser}, {and} \bibinfo{person}{Matthias Nie{\ss}ner}.} \bibinfo{year}{2017}\natexlab{}.
\newblock \showarticletitle{Scannet: Richly-annotated 3d reconstructions of indoor scenes}. In \bibinfo{booktitle}{\emph{Proceedings of the IEEE conference on computer vision and pattern recognition}}. \bibinfo{pages}{5828--5839}.
\newblock


\bibitem[\protect\citeauthoryear{DeCamp, Shaw, Kubat, and Roy}{DeCamp et~al\mbox{.}}{2010}]%
        {decamp2010immersive}
\bibfield{author}{\bibinfo{person}{Philip DeCamp}, \bibinfo{person}{George Shaw}, \bibinfo{person}{Rony Kubat}, {and} \bibinfo{person}{Deb Roy}.} \bibinfo{year}{2010}\natexlab{}.
\newblock \showarticletitle{An immersive system for browsing and visualizing surveillance video}. In \bibinfo{booktitle}{\emph{Proceedings of the 18th ACM international conference on Multimedia}}. \bibinfo{pages}{371--380}.
\newblock


\bibitem[\protect\citeauthoryear{Deng, Wu, Wang, Wu, Xie, Zhou, Zhang, Zhang, and Wu}{Deng et~al\mbox{.}}{2021}]%
        {deng2021eventanchor}
\bibfield{author}{\bibinfo{person}{Dazhen Deng}, \bibinfo{person}{Jiang Wu}, \bibinfo{person}{Jiachen Wang}, \bibinfo{person}{Yihong Wu}, \bibinfo{person}{Xiao Xie}, \bibinfo{person}{Zheng Zhou}, \bibinfo{person}{Hui Zhang}, \bibinfo{person}{Xiaolong Zhang}, {and} \bibinfo{person}{Yingcai Wu}.} \bibinfo{year}{2021}\natexlab{}.
\newblock \showarticletitle{EventAnchor: reducing human interactions in event annotation of racket sports videos}. In \bibinfo{booktitle}{\emph{Proceedings of the 2021 CHI Conference on Human Factors in Computing Systems}}. \bibinfo{pages}{1--13}.
\newblock


\bibitem[\protect\citeauthoryear{DepthKit}{DepthKit}{[n.d.]}]%
        {depthkit}
\bibfield{author}{\bibinfo{person}{DepthKit}.} \bibinfo{year}{[n.d.]}\natexlab{}.
\newblock \bibinfo{title}{DepthKit Studio}.
\newblock \bibinfo{howpublished}{\url{https://www.depthkit.tv/depthkit-studio}}.
\newblock


\bibitem[\protect\citeauthoryear{Dou, Khamis, Degtyarev, Davidson, Fanello, Kowdle, Escolano, Rhemann, Kim, Taylor, et~al\mbox{.}}{Dou et~al\mbox{.}}{2016}]%
        {dou2016fusion4d}
\bibfield{author}{\bibinfo{person}{Mingsong Dou}, \bibinfo{person}{Sameh Khamis}, \bibinfo{person}{Yury Degtyarev}, \bibinfo{person}{Philip Davidson}, \bibinfo{person}{Sean~Ryan Fanello}, \bibinfo{person}{Adarsh Kowdle}, \bibinfo{person}{Sergio~Orts Escolano}, \bibinfo{person}{Christoph Rhemann}, \bibinfo{person}{David Kim}, \bibinfo{person}{Jonathan Taylor}, {et~al\mbox{.}}} \bibinfo{year}{2016}\natexlab{}.
\newblock \showarticletitle{Fusion4d: Real-time performance capture of challenging scenes}.
\newblock \bibinfo{journal}{\emph{ACM Transactions on Graphics (ToG)}} \bibinfo{volume}{35}, \bibinfo{number}{4} (\bibinfo{year}{2016}), \bibinfo{pages}{1--13}.
\newblock


\bibitem[\protect\citeauthoryear{Du, Chuang, Chang, Hoppe, and Varshney}{Du et~al\mbox{.}}{2018}]%
        {du2018montage4d}
\bibfield{author}{\bibinfo{person}{Ruofei Du}, \bibinfo{person}{Ming Chuang}, \bibinfo{person}{Wayne Chang}, \bibinfo{person}{Hugues Hoppe}, {and} \bibinfo{person}{Amitabh Varshney}.} \bibinfo{year}{2018}\natexlab{}.
\newblock \showarticletitle{Montage4d: Interactive seamless fusion of multiview video textures}.
\newblock  (\bibinfo{year}{2018}).
\newblock


\bibitem[\protect\citeauthoryear{Du, Turner, Dzitsiuk, Prasso, Duarte, Dourgarian, Afonso, Pascoal, Gladstone, Cruces, et~al\mbox{.}}{Du et~al\mbox{.}}{2020}]%
        {du2020depthlab}
\bibfield{author}{\bibinfo{person}{Ruofei Du}, \bibinfo{person}{Eric Turner}, \bibinfo{person}{Maksym Dzitsiuk}, \bibinfo{person}{Luca Prasso}, \bibinfo{person}{Ivo Duarte}, \bibinfo{person}{Jason Dourgarian}, \bibinfo{person}{Joao Afonso}, \bibinfo{person}{Jose Pascoal}, \bibinfo{person}{Josh Gladstone}, \bibinfo{person}{Nuno Cruces}, {et~al\mbox{.}}} \bibinfo{year}{2020}\natexlab{}.
\newblock \showarticletitle{DepthLab: Real-time 3D interaction with depth maps for mobile augmented reality}. In \bibinfo{booktitle}{\emph{Proceedings of the 33rd Annual ACM Symposium on User Interface Software and Technology}}. \bibinfo{pages}{829--843}.
\newblock


\bibitem[\protect\citeauthoryear{Fender, Herholz, Alexa, and M{\"u}ller}{Fender et~al\mbox{.}}{2018}]%
        {fender2018optispace}
\bibfield{author}{\bibinfo{person}{Andreas Fender}, \bibinfo{person}{Philipp Herholz}, \bibinfo{person}{Marc Alexa}, {and} \bibinfo{person}{J{\"o}rg M{\"u}ller}.} \bibinfo{year}{2018}\natexlab{}.
\newblock \showarticletitle{Optispace: automated placement of interactive 3D projection mapping content}. In \bibinfo{booktitle}{\emph{Proceedings of the 2018 CHI Conference on Human Factors in Computing Systems}}. \bibinfo{pages}{1--11}.
\newblock


\bibitem[\protect\citeauthoryear{Fender, Lindlbauer, Herholz, Alexa, and M{\"u}ller}{Fender et~al\mbox{.}}{2017}]%
        {fender2017heatspace}
\bibfield{author}{\bibinfo{person}{Andreas Fender}, \bibinfo{person}{David Lindlbauer}, \bibinfo{person}{Philipp Herholz}, \bibinfo{person}{Marc Alexa}, {and} \bibinfo{person}{J{\"o}rg M{\"u}ller}.} \bibinfo{year}{2017}\natexlab{}.
\newblock \showarticletitle{Heatspace: Automatic placement of displays by empirical analysis of user behavior}. In \bibinfo{booktitle}{\emph{Proceedings of the 30th Annual ACM Symposium on User Interface Software and Technology}}. \bibinfo{pages}{611--621}.
\newblock


\bibitem[\protect\citeauthoryear{Fender and Holz}{Fender and Holz}{2022}]%
        {fender2022causality}
\bibfield{author}{\bibinfo{person}{Andreas~Rene Fender} {and} \bibinfo{person}{Christian Holz}.} \bibinfo{year}{2022}\natexlab{}.
\newblock \showarticletitle{Causality-preserving Asynchronous Reality}. In \bibinfo{booktitle}{\emph{CHI Conference on Human Factors in Computing Systems}}. \bibinfo{pages}{1--15}.
\newblock


\bibitem[\protect\citeauthoryear{Gao, Bai, Lee, and Billinghurst}{Gao et~al\mbox{.}}{2016}]%
        {gao2016oriented}
\bibfield{author}{\bibinfo{person}{Lei Gao}, \bibinfo{person}{Huidong Bai}, \bibinfo{person}{Gun Lee}, {and} \bibinfo{person}{Mark Billinghurst}.} \bibinfo{year}{2016}\natexlab{}.
\newblock \showarticletitle{An oriented point-cloud view for MR remote collaboration}.
\newblock In \bibinfo{booktitle}{\emph{SIGGRAPH ASIA 2016 Mobile Graphics and Interactive Applications}}. \bibinfo{pages}{1--4}.
\newblock


\bibitem[\protect\citeauthoryear{Gauglitz, Nuernberger, Turk, and H{\"o}llerer}{Gauglitz et~al\mbox{.}}{2014a}]%
        {gauglitz2014touch}
\bibfield{author}{\bibinfo{person}{Steffen Gauglitz}, \bibinfo{person}{Benjamin Nuernberger}, \bibinfo{person}{Matthew Turk}, {and} \bibinfo{person}{Tobias H{\"o}llerer}.} \bibinfo{year}{2014}\natexlab{a}.
\newblock \showarticletitle{In touch with the remote world: Remote collaboration with augmented reality drawings and virtual navigation}. In \bibinfo{booktitle}{\emph{Proceedings of the 20th ACM Symposium on Virtual Reality Software and Technology}}. \bibinfo{pages}{197--205}.
\newblock


\bibitem[\protect\citeauthoryear{Gauglitz, Nuernberger, Turk, and H{\"o}llerer}{Gauglitz et~al\mbox{.}}{2014b}]%
        {gauglitz2014world}
\bibfield{author}{\bibinfo{person}{Steffen Gauglitz}, \bibinfo{person}{Benjamin Nuernberger}, \bibinfo{person}{Matthew Turk}, {and} \bibinfo{person}{Tobias H{\"o}llerer}.} \bibinfo{year}{2014}\natexlab{b}.
\newblock \showarticletitle{World-stabilized annotations and virtual scene navigation for remote collaboration}. In \bibinfo{booktitle}{\emph{Proceedings of the 27th annual ACM symposium on User interface software and technology}}. \bibinfo{pages}{449--459}.
\newblock


\bibitem[\protect\citeauthoryear{Goldman, Gonterman, Curless, Salesin, and Seitz}{Goldman et~al\mbox{.}}{2008}]%
        {goldman2008video}
\bibfield{author}{\bibinfo{person}{Dan~B Goldman}, \bibinfo{person}{Chris Gonterman}, \bibinfo{person}{Brian Curless}, \bibinfo{person}{David Salesin}, {and} \bibinfo{person}{Steven~M Seitz}.} \bibinfo{year}{2008}\natexlab{}.
\newblock \showarticletitle{Video object annotation, navigation, and composition}. In \bibinfo{booktitle}{\emph{Proceedings of the 21st annual ACM symposium on User interface software and technology}}. \bibinfo{pages}{3--12}.
\newblock


\bibitem[\protect\citeauthoryear{Gong, Han, Guo, Li, Zha, Zhang, Tian, Wang, and Rui}{Gong et~al\mbox{.}}{2021}]%
        {gong2021holoboard}
\bibfield{author}{\bibinfo{person}{Jiangtao Gong}, \bibinfo{person}{Teng Han}, \bibinfo{person}{Siling Guo}, \bibinfo{person}{Jiannan Li}, \bibinfo{person}{Siyu Zha}, \bibinfo{person}{Liuxin Zhang}, \bibinfo{person}{Feng Tian}, \bibinfo{person}{Qianying Wang}, {and} \bibinfo{person}{Yong Rui}.} \bibinfo{year}{2021}\natexlab{}.
\newblock \showarticletitle{Holoboard: A large-format immersive teaching board based on pseudo holographics}. In \bibinfo{booktitle}{\emph{The 34th Annual ACM Symposium on User Interface Software and Technology}}. \bibinfo{pages}{441--456}.
\newblock


\bibitem[\protect\citeauthoryear{Guo, Lincoln, Davidson, Busch, Yu, Whalen, Harvey, Orts-Escolano, Pandey, Dourgarian, et~al\mbox{.}}{Guo et~al\mbox{.}}{2019}]%
        {guo2019relightables}
\bibfield{author}{\bibinfo{person}{Kaiwen Guo}, \bibinfo{person}{Peter Lincoln}, \bibinfo{person}{Philip Davidson}, \bibinfo{person}{Jay Busch}, \bibinfo{person}{Xueming Yu}, \bibinfo{person}{Matt Whalen}, \bibinfo{person}{Geoff Harvey}, \bibinfo{person}{Sergio Orts-Escolano}, \bibinfo{person}{Rohit Pandey}, \bibinfo{person}{Jason Dourgarian}, {et~al\mbox{.}}} \bibinfo{year}{2019}\natexlab{}.
\newblock \showarticletitle{The relightables: Volumetric performance capture of humans with realistic relighting}.
\newblock \bibinfo{journal}{\emph{ACM Transactions on Graphics (ToG)}} \bibinfo{volume}{38}, \bibinfo{number}{6} (\bibinfo{year}{2019}), \bibinfo{pages}{1--19}.
\newblock


\bibitem[\protect\citeauthoryear{Hall, Bartram, and Brehmer}{Hall et~al\mbox{.}}{2022}]%
        {hall2022augmented}
\bibfield{author}{\bibinfo{person}{Brian~D Hall}, \bibinfo{person}{Lyn Bartram}, {and} \bibinfo{person}{Matthew Brehmer}.} \bibinfo{year}{2022}\natexlab{}.
\newblock \showarticletitle{Augmented Chironomia for Presenting Data to Remote Audiences}.
\newblock \bibinfo{journal}{\emph{arXiv preprint arXiv:2208.04451}} (\bibinfo{year}{2022}).
\newblock


\bibitem[\protect\citeauthoryear{Han, Chen, Zhong, Wang, and Hung}{Han et~al\mbox{.}}{2017}]%
        {han2017my}
\bibfield{author}{\bibinfo{person}{Ping-Hsuan Han}, \bibinfo{person}{Yang-Sheng Chen}, \bibinfo{person}{Yilun Zhong}, \bibinfo{person}{Han-Lei Wang}, {and} \bibinfo{person}{Yi-Ping Hung}.} \bibinfo{year}{2017}\natexlab{}.
\newblock \showarticletitle{My Tai-Chi coaches: an augmented-learning tool for practicing Tai-Chi Chuan}. In \bibinfo{booktitle}{\emph{Proceedings of the 8th Augmented Human International Conference}}. \bibinfo{pages}{1--4}.
\newblock


\bibitem[\protect\citeauthoryear{Hartmann, Holz, Ofek, and Wilson}{Hartmann et~al\mbox{.}}{2019}]%
        {hartmann2019realitycheck}
\bibfield{author}{\bibinfo{person}{Jeremy Hartmann}, \bibinfo{person}{Christian Holz}, \bibinfo{person}{Eyal Ofek}, {and} \bibinfo{person}{Andrew~D Wilson}.} \bibinfo{year}{2019}\natexlab{}.
\newblock \showarticletitle{Realitycheck: Blending virtual environments with situated physical reality}. In \bibinfo{booktitle}{\emph{Proceedings of the 2019 CHI Conference on Human Factors in Computing Systems}}. \bibinfo{pages}{1--12}.
\newblock


\bibitem[\protect\citeauthoryear{Hubenschmid, Wieland, Fink, Batch, Zagermann, Elmqvist, and Reiterer}{Hubenschmid et~al\mbox{.}}{2022}]%
        {hubenschmid2022relive}
\bibfield{author}{\bibinfo{person}{Sebastian Hubenschmid}, \bibinfo{person}{Jonathan Wieland}, \bibinfo{person}{Daniel~Immanuel Fink}, \bibinfo{person}{Andrea Batch}, \bibinfo{person}{Johannes Zagermann}, \bibinfo{person}{Niklas Elmqvist}, {and} \bibinfo{person}{Harald Reiterer}.} \bibinfo{year}{2022}\natexlab{}.
\newblock \showarticletitle{ReLive: Bridging In-Situ and Ex-Situ Visual Analytics for Analyzing Mixed Reality User Studies}. In \bibinfo{booktitle}{\emph{CHI Conference on Human Factors in Computing Systems}}. \bibinfo{pages}{1--20}.
\newblock


\bibitem[\protect\citeauthoryear{Hudson, Alcock, and Chilana}{Hudson et~al\mbox{.}}{2016}]%
        {hudson2016understanding}
\bibfield{author}{\bibinfo{person}{Nathaniel Hudson}, \bibinfo{person}{Celena Alcock}, {and} \bibinfo{person}{Parmit~K Chilana}.} \bibinfo{year}{2016}\natexlab{}.
\newblock \showarticletitle{Understanding newcomers to 3D printing: Motivations, workflows, and barriers of casual makers}. In \bibinfo{booktitle}{\emph{Proceedings of the 2016 CHI conference on human factors in computing systems}}. \bibinfo{pages}{384--396}.
\newblock


\bibitem[\protect\citeauthoryear{Huo and Ramani}{Huo and Ramani}{2017}]%
        {huo2017window}
\bibfield{author}{\bibinfo{person}{Ke Huo} {and} \bibinfo{person}{Karthik Ramani}.} \bibinfo{year}{2017}\natexlab{}.
\newblock \showarticletitle{Window-shaping: 3d design ideation by creating on, borrowing from, and looking at the physical world}. In \bibinfo{booktitle}{\emph{Proceedings of the Eleventh International Conference on Tangible, Embedded, and Embodied Interaction}}. \bibinfo{pages}{37--45}.
\newblock


\bibitem[\protect\citeauthoryear{Innmann, Zollh{\"o}fer, Nie{\ss}ner, Theobalt, and Stamminger}{Innmann et~al\mbox{.}}{2016}]%
        {innmann2016volumedeform}
\bibfield{author}{\bibinfo{person}{Matthias Innmann}, \bibinfo{person}{Michael Zollh{\"o}fer}, \bibinfo{person}{Matthias Nie{\ss}ner}, \bibinfo{person}{Christian Theobalt}, {and} \bibinfo{person}{Marc Stamminger}.} \bibinfo{year}{2016}\natexlab{}.
\newblock \showarticletitle{Volumedeform: Real-time volumetric non-rigid reconstruction}. In \bibinfo{booktitle}{\emph{European conference on computer vision}}. Springer, \bibinfo{pages}{362--379}.
\newblock


\bibitem[\protect\citeauthoryear{Izadi, Kim, Hilliges, Molyneaux, Newcombe, Kohli, Shotton, Hodges, Freeman, Davison, et~al\mbox{.}}{Izadi et~al\mbox{.}}{2011}]%
        {izadi2011kinectfusion}
\bibfield{author}{\bibinfo{person}{Shahram Izadi}, \bibinfo{person}{David Kim}, \bibinfo{person}{Otmar Hilliges}, \bibinfo{person}{David Molyneaux}, \bibinfo{person}{Richard Newcombe}, \bibinfo{person}{Pushmeet Kohli}, \bibinfo{person}{Jamie Shotton}, \bibinfo{person}{Steve Hodges}, \bibinfo{person}{Dustin Freeman}, \bibinfo{person}{Andrew Davison}, {et~al\mbox{.}}} \bibinfo{year}{2011}\natexlab{}.
\newblock \showarticletitle{KinectFusion: real-time 3D reconstruction and interaction using a moving depth camera}. In \bibinfo{booktitle}{\emph{Proceedings of the 24th annual ACM symposium on User interface software and technology}}. \bibinfo{pages}{559--568}.
\newblock


\bibitem[\protect\citeauthoryear{Jones, Sodhi, Murdock, Mehra, Benko, Wilson, Ofek, MacIntyre, Raghuvanshi, and Shapira}{Jones et~al\mbox{.}}{2014}]%
        {jones2014roomalive}
\bibfield{author}{\bibinfo{person}{Brett Jones}, \bibinfo{person}{Rajinder Sodhi}, \bibinfo{person}{Michael Murdock}, \bibinfo{person}{Ravish Mehra}, \bibinfo{person}{Hrvoje Benko}, \bibinfo{person}{Andrew Wilson}, \bibinfo{person}{Eyal Ofek}, \bibinfo{person}{Blair MacIntyre}, \bibinfo{person}{Nikunj Raghuvanshi}, {and} \bibinfo{person}{Lior Shapira}.} \bibinfo{year}{2014}\natexlab{}.
\newblock \showarticletitle{Roomalive: Magical experiences enabled by scalable, adaptive projector-camera units}. In \bibinfo{booktitle}{\emph{Proceedings of the 27th annual ACM symposium on User interface software and technology}}. \bibinfo{pages}{637--644}.
\newblock


\bibitem[\protect\citeauthoryear{Jones, Benko, Ofek, and Wilson}{Jones et~al\mbox{.}}{2013}]%
        {jones2013illumiroom}
\bibfield{author}{\bibinfo{person}{Brett~R Jones}, \bibinfo{person}{Hrvoje Benko}, \bibinfo{person}{Eyal Ofek}, {and} \bibinfo{person}{Andrew~D Wilson}.} \bibinfo{year}{2013}\natexlab{}.
\newblock \showarticletitle{IllumiRoom: peripheral projected illusions for interactive experiences}. In \bibinfo{booktitle}{\emph{Proceedings of the SIGCHI Conference on Human Factors in Computing Systems}}. \bibinfo{pages}{869--878}.
\newblock


\bibitem[\protect\citeauthoryear{Kaimoto, Monteiro, Faridan, Li, Farajian, Kakehi, Nakagaki, and Suzuki}{Kaimoto et~al\mbox{.}}{2022}]%
        {kaimoto2022sketched}
\bibfield{author}{\bibinfo{person}{Hiroki Kaimoto}, \bibinfo{person}{Kyzyl Monteiro}, \bibinfo{person}{Mehrad Faridan}, \bibinfo{person}{Jiatong Li}, \bibinfo{person}{Samin Farajian}, \bibinfo{person}{Yasuaki Kakehi}, \bibinfo{person}{Ken Nakagaki}, {and} \bibinfo{person}{Ryo Suzuki}.} \bibinfo{year}{2022}\natexlab{}.
\newblock \showarticletitle{Sketched reality: Sketching bi-directional interactions between virtual and physical worlds with ar and actuated tangible ui}. In \bibinfo{booktitle}{\emph{Proceedings of the 35th Annual ACM Symposium on User Interface Software and Technology}}. \bibinfo{pages}{1--12}.
\newblock


\bibitem[\protect\citeauthoryear{Kanade, Rander, and Narayanan}{Kanade et~al\mbox{.}}{1997}]%
        {kanade1997virtualized}
\bibfield{author}{\bibinfo{person}{Takeo Kanade}, \bibinfo{person}{Peter Rander}, {and} \bibinfo{person}{PJ Narayanan}.} \bibinfo{year}{1997}\natexlab{}.
\newblock \showarticletitle{Virtualized reality: Constructing virtual worlds from real scenes}.
\newblock \bibinfo{journal}{\emph{IEEE multimedia}} \bibinfo{volume}{4}, \bibinfo{number}{1} (\bibinfo{year}{1997}), \bibinfo{pages}{34--47}.
\newblock


\bibitem[\protect\citeauthoryear{Kari, Grosse-Puppendahl, Coelho, Fender, Bethge, Sch{\"u}tte, and Holz}{Kari et~al\mbox{.}}{2021}]%
        {kari2021transformr}
\bibfield{author}{\bibinfo{person}{Mohamed Kari}, \bibinfo{person}{Tobias Grosse-Puppendahl}, \bibinfo{person}{Luis~Falconeri Coelho}, \bibinfo{person}{Andreas~Rene Fender}, \bibinfo{person}{David Bethge}, \bibinfo{person}{Reinhard Sch{\"u}tte}, {and} \bibinfo{person}{Christian Holz}.} \bibinfo{year}{2021}\natexlab{}.
\newblock \showarticletitle{TransforMR: Pose-aware object substitution for composing alternate mixed realities}. In \bibinfo{booktitle}{\emph{2021 IEEE International Symposium on Mixed and Augmented Reality (ISMAR)}}. IEEE, \bibinfo{pages}{69--79}.
\newblock


\bibitem[\protect\citeauthoryear{Karrer, Wittenhagen, and Borchers}{Karrer et~al\mbox{.}}{2009}]%
        {karrer2009pocketdragon}
\bibfield{author}{\bibinfo{person}{Thorsten Karrer}, \bibinfo{person}{Moritz Wittenhagen}, {and} \bibinfo{person}{Jan Borchers}.} \bibinfo{year}{2009}\natexlab{}.
\newblock \showarticletitle{Pocketdragon: a direct manipulation video navigation interface for mobile devices}. In \bibinfo{booktitle}{\emph{Proceedings of the 11th International Conference on Human-Computer Interaction with Mobile Devices and Services}}. \bibinfo{pages}{1--3}.
\newblock


\bibitem[\protect\citeauthoryear{Kepplinger, Wallner, Kriglstein, and Lankes}{Kepplinger et~al\mbox{.}}{2020}]%
        {kepplinger2020see}
\bibfield{author}{\bibinfo{person}{Daniel Kepplinger}, \bibinfo{person}{G{\"u}nter Wallner}, \bibinfo{person}{Simone Kriglstein}, {and} \bibinfo{person}{Michael Lankes}.} \bibinfo{year}{2020}\natexlab{}.
\newblock \showarticletitle{See, Feel, Move: player behaviour analysis through combined visualization of gaze, emotions, and movement}. In \bibinfo{booktitle}{\emph{Proceedings of the 2020 CHI Conference on Human Factors in Computing Systems}}. \bibinfo{pages}{1--14}.
\newblock


\bibitem[\protect\citeauthoryear{Kloiber, Settgast, Schinko, Weinzerl, Fritz, Schreck, and Preiner}{Kloiber et~al\mbox{.}}{2020}]%
        {kloiber2020immersive}
\bibfield{author}{\bibinfo{person}{Simon Kloiber}, \bibinfo{person}{Volker Settgast}, \bibinfo{person}{Christoph Schinko}, \bibinfo{person}{Martin Weinzerl}, \bibinfo{person}{Johannes Fritz}, \bibinfo{person}{Tobias Schreck}, {and} \bibinfo{person}{Reinhold Preiner}.} \bibinfo{year}{2020}\natexlab{}.
\newblock \showarticletitle{Immersive analysis of user motion in VR applications}.
\newblock \bibinfo{journal}{\emph{The Visual Computer}} \bibinfo{volume}{36}, \bibinfo{number}{10} (\bibinfo{year}{2020}), \bibinfo{pages}{1937--1949}.
\newblock


\bibitem[\protect\citeauthoryear{Kolbe}{Kolbe}{2004}]%
        {kolbe2004augmented}
\bibfield{author}{\bibinfo{person}{Thomas~H Kolbe}.} \bibinfo{year}{2004}\natexlab{}.
\newblock \showarticletitle{Augmented videos and panoramas for pedestrian navigation}. In \bibinfo{booktitle}{\emph{Proceedings of the 2nd Symposium on Location Based Services \& TeleCartography 2004, 28-29th of January 2004 in Vienna}}.
\newblock


\bibitem[\protect\citeauthoryear{Komiyama, Miyaki, and Rekimoto}{Komiyama et~al\mbox{.}}{2017}]%
        {komiyama2017jackin}
\bibfield{author}{\bibinfo{person}{Ryohei Komiyama}, \bibinfo{person}{Takashi Miyaki}, {and} \bibinfo{person}{Jun Rekimoto}.} \bibinfo{year}{2017}\natexlab{}.
\newblock \showarticletitle{JackIn space: designing a seamless transition between first and third person view for effective telepresence collaborations}. In \bibinfo{booktitle}{\emph{Proceedings of the 8th Augmented Human International Conference}}. \bibinfo{pages}{1--9}.
\newblock


\bibitem[\protect\citeauthoryear{Kunert, Kulik, Beck, and Froehlich}{Kunert et~al\mbox{.}}{2014}]%
        {kunert2014photoportals}
\bibfield{author}{\bibinfo{person}{Andr{\'e} Kunert}, \bibinfo{person}{Alexander Kulik}, \bibinfo{person}{Stephan Beck}, {and} \bibinfo{person}{Bernd Froehlich}.} \bibinfo{year}{2014}\natexlab{}.
\newblock \showarticletitle{Photoportals: shared references in space and time}. In \bibinfo{booktitle}{\emph{Proceedings of the 17th ACM conference on Computer supported cooperative work \& social computing}}. \bibinfo{pages}{1388--1399}.
\newblock


\bibitem[\protect\citeauthoryear{Lawrence, Goldman, Achar, Blascovich, Desloge, Fortes, Gomez, H{\"a}berling, Hoppe, Huibers, et~al\mbox{.}}{Lawrence et~al\mbox{.}}{2021}]%
        {lawrence2021project}
\bibfield{author}{\bibinfo{person}{Jason Lawrence}, \bibinfo{person}{Dan~B Goldman}, \bibinfo{person}{Supreeth Achar}, \bibinfo{person}{Gregory~Major Blascovich}, \bibinfo{person}{Joseph~G Desloge}, \bibinfo{person}{Tommy Fortes}, \bibinfo{person}{Eric~M Gomez}, \bibinfo{person}{Sascha H{\"a}berling}, \bibinfo{person}{Hugues Hoppe}, \bibinfo{person}{Andy Huibers}, {et~al\mbox{.}}} \bibinfo{year}{2021}\natexlab{}.
\newblock \showarticletitle{Project Starline: A high-fidelity telepresence system}.
\newblock  (\bibinfo{year}{2021}).
\newblock


\bibitem[\protect\citeauthoryear{Ledo, Houben, Vermeulen, Marquardt, Oehlberg, and Greenberg}{Ledo et~al\mbox{.}}{2018}]%
        {ledo2018evaluation}
\bibfield{author}{\bibinfo{person}{David Ledo}, \bibinfo{person}{Steven Houben}, \bibinfo{person}{Jo Vermeulen}, \bibinfo{person}{Nicolai Marquardt}, \bibinfo{person}{Lora Oehlberg}, {and} \bibinfo{person}{Saul Greenberg}.} \bibinfo{year}{2018}\natexlab{}.
\newblock \showarticletitle{Evaluation strategies for HCI toolkit research}. In \bibinfo{booktitle}{\emph{Proceedings of the 2018 CHI Conference on Human Factors in Computing Systems}}. \bibinfo{pages}{1--17}.
\newblock


\bibitem[\protect\citeauthoryear{Lee, Lee, Zhang, Tessier, and Khan}{Lee et~al\mbox{.}}{2019}]%
        {lee2019semantic}
\bibfield{author}{\bibinfo{person}{Bokyung Lee}, \bibinfo{person}{Michael Lee}, \bibinfo{person}{Pan Zhang}, \bibinfo{person}{Alexander Tessier}, {and} \bibinfo{person}{Azam Khan}.} \bibinfo{year}{2019}\natexlab{}.
\newblock \showarticletitle{Semantic human activity annotation tool using skeletonized surveillance videos}. In \bibinfo{booktitle}{\emph{Adjunct Proceedings of the 2019 ACM International Joint Conference on Pervasive and Ubiquitous Computing and Proceedings of the 2019 ACM International Symposium on Wearable Computers}}. \bibinfo{pages}{312--315}.
\newblock


\bibitem[\protect\citeauthoryear{Leiva, Gr{\o}nb{\ae}k, Klokmose, Nguyen, Kazi, and Asente}{Leiva et~al\mbox{.}}{2021}]%
        {leiva2021rapido}
\bibfield{author}{\bibinfo{person}{Germ{\'a}n Leiva}, \bibinfo{person}{Jens~Emil Gr{\o}nb{\ae}k}, \bibinfo{person}{Clemens~Nylandsted Klokmose}, \bibinfo{person}{Cuong Nguyen}, \bibinfo{person}{Rubaiat~Habib Kazi}, {and} \bibinfo{person}{Paul Asente}.} \bibinfo{year}{2021}\natexlab{}.
\newblock \showarticletitle{Rapido: Prototyping Interactive AR Experiences through Programming by Demonstration}. In \bibinfo{booktitle}{\emph{The 34th Annual ACM Symposium on User Interface Software and Technology}}. \bibinfo{pages}{626--637}.
\newblock


\bibitem[\protect\citeauthoryear{Leiva, Nguyen, Kazi, and Asente}{Leiva et~al\mbox{.}}{2020}]%
        {leiva2020pronto}
\bibfield{author}{\bibinfo{person}{Germ{\'a}n Leiva}, \bibinfo{person}{Cuong Nguyen}, \bibinfo{person}{Rubaiat~Habib Kazi}, {and} \bibinfo{person}{Paul Asente}.} \bibinfo{year}{2020}\natexlab{}.
\newblock \showarticletitle{Pronto: Rapid augmented reality video prototyping using sketches and enaction}. In \bibinfo{booktitle}{\emph{Proceedings of the 2020 CHI Conference on Human Factors in Computing Systems}}. \bibinfo{pages}{1--13}.
\newblock


\bibitem[\protect\citeauthoryear{Li, Luo, Zheng, Xu, and Fu}{Li et~al\mbox{.}}{2017}]%
        {li2017sweepcanvas}
\bibfield{author}{\bibinfo{person}{Yuwei Li}, \bibinfo{person}{Xi Luo}, \bibinfo{person}{Youyi Zheng}, \bibinfo{person}{Pengfei Xu}, {and} \bibinfo{person}{Hongbo Fu}.} \bibinfo{year}{2017}\natexlab{}.
\newblock \showarticletitle{SweepCanvas: Sketch-based 3D prototyping on an RGB-D image}. In \bibinfo{booktitle}{\emph{Proceedings of the 30th Annual ACM Symposium on User Interface Software and Technology}}. \bibinfo{pages}{387--399}.
\newblock


\bibitem[\protect\citeauthoryear{Liao, Karim, Jadon, Kazi, and Suzuki}{Liao et~al\mbox{.}}{2022}]%
        {liao2022realitytalk}
\bibfield{author}{\bibinfo{person}{Jian Liao}, \bibinfo{person}{Adnan Karim}, \bibinfo{person}{Shivesh Jadon}, \bibinfo{person}{Rubaiat~Habib Kazi}, {and} \bibinfo{person}{Ryo Suzuki}.} \bibinfo{year}{2022}\natexlab{}.
\newblock \showarticletitle{RealityTalk: Real-Time Speech-Driven Augmented Presentation for AR Live Storytelling}.
\newblock \bibinfo{journal}{\emph{arXiv preprint arXiv:2208.06350}} (\bibinfo{year}{2022}).
\newblock


\bibitem[\protect\citeauthoryear{Lindlbauer and Wilson}{Lindlbauer and Wilson}{2018}]%
        {lindlbauer2018remixed}
\bibfield{author}{\bibinfo{person}{David Lindlbauer} {and} \bibinfo{person}{Andy~D Wilson}.} \bibinfo{year}{2018}\natexlab{}.
\newblock \showarticletitle{Remixed reality: Manipulating space and time in augmented reality}. In \bibinfo{booktitle}{\emph{Proceedings of the 2018 CHI Conference on Human Factors in Computing Systems}}. \bibinfo{pages}{1--13}.
\newblock


\bibitem[\protect\citeauthoryear{Liu, Fu, and Tai}{Liu et~al\mbox{.}}{2020}]%
        {liu2020posetween}
\bibfield{author}{\bibinfo{person}{Jingyuan Liu}, \bibinfo{person}{Hongbo Fu}, {and} \bibinfo{person}{Chiew-Lan Tai}.} \bibinfo{year}{2020}\natexlab{}.
\newblock \showarticletitle{Posetween: Pose-driven tween animation}. In \bibinfo{booktitle}{\emph{Proceedings of the 33rd Annual ACM Symposium on User Interface Software and Technology}}. \bibinfo{pages}{791--804}.
\newblock


\bibitem[\protect\citeauthoryear{Montano-Murillo, Nguyen, Kazi, Subramanian, DiVerdi, and Martinez-Plasencia}{Montano-Murillo et~al\mbox{.}}{2020}]%
        {montano2020slicing}
\bibfield{author}{\bibinfo{person}{Roberto~A Montano-Murillo}, \bibinfo{person}{Cuong Nguyen}, \bibinfo{person}{Rubaiat~Habib Kazi}, \bibinfo{person}{Sriram Subramanian}, \bibinfo{person}{Stephen DiVerdi}, {and} \bibinfo{person}{Diego Martinez-Plasencia}.} \bibinfo{year}{2020}\natexlab{}.
\newblock \showarticletitle{Slicing-volume: Hybrid 3d/2d multi-target selection technique for dense virtual environments}. In \bibinfo{booktitle}{\emph{2020 IEEE Conference on Virtual Reality and 3D User Interfaces (VR)}}. IEEE, \bibinfo{pages}{53--62}.
\newblock


\bibitem[\protect\citeauthoryear{Monteiro, Vatsal, Chulpongsatorn, Parnami, and Suzuki}{Monteiro et~al\mbox{.}}{2023}]%
        {monteiro2023teachable}
\bibfield{author}{\bibinfo{person}{Kyzyl Monteiro}, \bibinfo{person}{Ritik Vatsal}, \bibinfo{person}{Neil Chulpongsatorn}, \bibinfo{person}{Aman Parnami}, {and} \bibinfo{person}{Ryo Suzuki}.} \bibinfo{year}{2023}\natexlab{}.
\newblock \showarticletitle{Teachable reality: Prototyping tangible augmented reality with everyday objects by leveraging interactive machine teaching}. In \bibinfo{booktitle}{\emph{Proceedings of the 2023 CHI Conference on Human Factors in Computing Systems}}. \bibinfo{pages}{1--15}.
\newblock


\bibitem[\protect\citeauthoryear{Mori, Ikeda, and Saito}{Mori et~al\mbox{.}}{2017}]%
        {mori2017survey}
\bibfield{author}{\bibinfo{person}{Shohei Mori}, \bibinfo{person}{Sei Ikeda}, {and} \bibinfo{person}{Hideo Saito}.} \bibinfo{year}{2017}\natexlab{}.
\newblock \showarticletitle{A survey of diminished reality: Techniques for visually concealing, eliminating, and seeing through real objects}.
\newblock \bibinfo{journal}{\emph{IPSJ Transactions on Computer Vision and Applications}} \bibinfo{volume}{9}, \bibinfo{number}{1} (\bibinfo{year}{2017}), \bibinfo{pages}{1--14}.
\newblock


\bibitem[\protect\citeauthoryear{Nguyen, Niu, and Liu}{Nguyen et~al\mbox{.}}{2012}]%
        {nguyen2012video}
\bibfield{author}{\bibinfo{person}{Cuong Nguyen}, \bibinfo{person}{Yuzhen Niu}, {and} \bibinfo{person}{Feng Liu}.} \bibinfo{year}{2012}\natexlab{}.
\newblock \showarticletitle{Video summagator: An interface for video summarization and navigation}. In \bibinfo{booktitle}{\emph{Proceedings of the SIGCHI Conference on Human Factors in Computing Systems}}. \bibinfo{pages}{647--650}.
\newblock


\bibitem[\protect\citeauthoryear{Nguyen, Niu, and Liu}{Nguyen et~al\mbox{.}}{2013}]%
        {nguyen2013direct}
\bibfield{author}{\bibinfo{person}{Cuong Nguyen}, \bibinfo{person}{Yuzhen Niu}, {and} \bibinfo{person}{Feng Liu}.} \bibinfo{year}{2013}\natexlab{}.
\newblock \showarticletitle{Direct manipulation video navigation in 3D}. In \bibinfo{booktitle}{\emph{Proceedings of the SIGCHI Conference on Human Factors in Computing Systems}}. \bibinfo{pages}{1169--1172}.
\newblock


\bibitem[\protect\citeauthoryear{Nguyen, Niu, and Liu}{Nguyen et~al\mbox{.}}{2014}]%
        {nguyen2014direct}
\bibfield{author}{\bibinfo{person}{Cuong Nguyen}, \bibinfo{person}{Yuzhen Niu}, {and} \bibinfo{person}{Feng Liu}.} \bibinfo{year}{2014}\natexlab{}.
\newblock \showarticletitle{Direct manipulation video navigation on touch screens}. In \bibinfo{booktitle}{\emph{Proceedings of the 16th international conference on Human-computer interaction with mobile devices \& services}}. \bibinfo{pages}{273--282}.
\newblock


\bibitem[\protect\citeauthoryear{Nuernberger, Ofek, Benko, and Wilson}{Nuernberger et~al\mbox{.}}{2016}]%
        {nuernberger2016snaptoreality}
\bibfield{author}{\bibinfo{person}{Benjamin Nuernberger}, \bibinfo{person}{Eyal Ofek}, \bibinfo{person}{Hrvoje Benko}, {and} \bibinfo{person}{Andrew~D Wilson}.} \bibinfo{year}{2016}\natexlab{}.
\newblock \showarticletitle{Snaptoreality: Aligning augmented reality to the real world}. In \bibinfo{booktitle}{\emph{Proceedings of the 2016 CHI Conference on Human Factors in Computing Systems}}. \bibinfo{pages}{1233--1244}.
\newblock


\bibitem[\protect\citeauthoryear{Orts-Escolano, Rhemann, Fanello, Chang, Kowdle, Degtyarev, Kim, Davidson, Khamis, Dou, et~al\mbox{.}}{Orts-Escolano et~al\mbox{.}}{2016}]%
        {orts2016holoportation}
\bibfield{author}{\bibinfo{person}{Sergio Orts-Escolano}, \bibinfo{person}{Christoph Rhemann}, \bibinfo{person}{Sean Fanello}, \bibinfo{person}{Wayne Chang}, \bibinfo{person}{Adarsh Kowdle}, \bibinfo{person}{Yury Degtyarev}, \bibinfo{person}{David Kim}, \bibinfo{person}{Philip~L Davidson}, \bibinfo{person}{Sameh Khamis}, \bibinfo{person}{Mingsong Dou}, {et~al\mbox{.}}} \bibinfo{year}{2016}\natexlab{}.
\newblock \showarticletitle{Holoportation: Virtual 3d teleportation in real-time}. In \bibinfo{booktitle}{\emph{Proceedings of the 29th annual symposium on user interface software and technology}}. \bibinfo{pages}{741--754}.
\newblock


\bibitem[\protect\citeauthoryear{Pejsa, Kantor, Benko, Ofek, and Wilson}{Pejsa et~al\mbox{.}}{2016}]%
        {pejsa2016room2room}
\bibfield{author}{\bibinfo{person}{Tomislav Pejsa}, \bibinfo{person}{Julian Kantor}, \bibinfo{person}{Hrvoje Benko}, \bibinfo{person}{Eyal Ofek}, {and} \bibinfo{person}{Andrew Wilson}.} \bibinfo{year}{2016}\natexlab{}.
\newblock \showarticletitle{Room2room: Enabling life-size telepresence in a projected augmented reality environment}. In \bibinfo{booktitle}{\emph{Proceedings of the 19th ACM conference on computer-supported cooperative work \& social computing}}. \bibinfo{pages}{1716--1725}.
\newblock


\bibitem[\protect\citeauthoryear{Piumsomboon, Lee, Hart, Ens, Lindeman, Thomas, and Billinghurst}{Piumsomboon et~al\mbox{.}}{2018}]%
        {piumsomboon2018mini}
\bibfield{author}{\bibinfo{person}{Thammathip Piumsomboon}, \bibinfo{person}{Gun~A Lee}, \bibinfo{person}{Jonathon~D Hart}, \bibinfo{person}{Barrett Ens}, \bibinfo{person}{Robert~W Lindeman}, \bibinfo{person}{Bruce~H Thomas}, {and} \bibinfo{person}{Mark Billinghurst}.} \bibinfo{year}{2018}\natexlab{}.
\newblock \showarticletitle{Mini-me: An adaptive avatar for mixed reality remote collaboration}. In \bibinfo{booktitle}{\emph{Proceedings of the 2018 CHI conference on human factors in computing systems}}. \bibinfo{pages}{1--13}.
\newblock


\bibitem[\protect\citeauthoryear{Piumsomboon, Lee, Irlitti, Ens, Thomas, and Billinghurst}{Piumsomboon et~al\mbox{.}}{2019}]%
        {piumsomboon2019shoulder}
\bibfield{author}{\bibinfo{person}{Thammathip Piumsomboon}, \bibinfo{person}{Gun~A Lee}, \bibinfo{person}{Andrew Irlitti}, \bibinfo{person}{Barrett Ens}, \bibinfo{person}{Bruce~H Thomas}, {and} \bibinfo{person}{Mark Billinghurst}.} \bibinfo{year}{2019}\natexlab{}.
\newblock \showarticletitle{On the shoulder of the giant: A multi-scale mixed reality collaboration with 360 video sharing and tangible interaction}. In \bibinfo{booktitle}{\emph{Proceedings of the 2019 CHI conference on human factors in computing systems}}. \bibinfo{pages}{1--17}.
\newblock


\bibitem[\protect\citeauthoryear{PolyCam}{PolyCam}{[n.d.]}]%
        {polycam}
\bibfield{author}{\bibinfo{person}{PolyCam}.} \bibinfo{year}{[n.d.]}\natexlab{}.
\newblock \bibinfo{title}{Polycam}.
\newblock \bibinfo{howpublished}{\url{https://poly.cam/}}.
\newblock


\bibitem[\protect\citeauthoryear{Radu, Joy, and Schneider}{Radu et~al\mbox{.}}{2021}]%
        {radu2021virtual}
\bibfield{author}{\bibinfo{person}{Iulian Radu}, \bibinfo{person}{Tugce Joy}, {and} \bibinfo{person}{Bertrand Schneider}.} \bibinfo{year}{2021}\natexlab{}.
\newblock \showarticletitle{Virtual makerspaces: merging AR/VR/MR to enable remote collaborations in physical maker activities}. In \bibinfo{booktitle}{\emph{Extended Abstracts of the 2021 CHI Conference on Human Factors in Computing Systems}}. \bibinfo{pages}{1--5}.
\newblock


\bibitem[\protect\citeauthoryear{Regenbrecht, Meng, Reepen, Beck, and Langlotz}{Regenbrecht et~al\mbox{.}}{2017}]%
        {regenbrecht2017mixed}
\bibfield{author}{\bibinfo{person}{Holger Regenbrecht}, \bibinfo{person}{Katrin Meng}, \bibinfo{person}{Arne Reepen}, \bibinfo{person}{Stephan Beck}, {and} \bibinfo{person}{Tobias Langlotz}.} \bibinfo{year}{2017}\natexlab{}.
\newblock \showarticletitle{Mixed voxel reality: Presence and embodiment in low fidelity, visually coherent, mixed reality environments}. In \bibinfo{booktitle}{\emph{2017 IEEE International Symposium on Mixed and Augmented Reality (ISMAR)}}. IEEE, \bibinfo{pages}{90--99}.
\newblock


\bibitem[\protect\citeauthoryear{Reipschl{\"a}ger, Brudy, Dachselt, Matejka, Fitzmaurice, and Anderson}{Reipschl{\"a}ger et~al\mbox{.}}{2022}]%
        {reipschlager2022avatar}
\bibfield{author}{\bibinfo{person}{Patrick Reipschl{\"a}ger}, \bibinfo{person}{Frederik Brudy}, \bibinfo{person}{Raimund Dachselt}, \bibinfo{person}{Justin Matejka}, \bibinfo{person}{George Fitzmaurice}, {and} \bibinfo{person}{Fraser Anderson}.} \bibinfo{year}{2022}\natexlab{}.
\newblock \showarticletitle{AvatAR: An Immersive Analysis Environment for Human Motion Data Combining Interactive 3D Avatars and Trajectories}. In \bibinfo{booktitle}{\emph{CHI Conference on Human Factors in Computing Systems}}. \bibinfo{pages}{1--15}.
\newblock


\bibitem[\protect\citeauthoryear{Ribeiro, Kuffner, and Fernandes}{Ribeiro et~al\mbox{.}}{2018}]%
        {ribeiro2018virtual}
\bibfield{author}{\bibinfo{person}{Claudia Ribeiro}, \bibinfo{person}{Rafael Kuffner}, {and} \bibinfo{person}{Carla Fernandes}.} \bibinfo{year}{2018}\natexlab{}.
\newblock \showarticletitle{Virtual reality annotator: A tool to annotate dancers in a virtual environment}. In \bibinfo{booktitle}{\emph{Digital Cultural Heritage: Final Conference of the Marie Sk{\l}odowska-Curie Initial Training Network for Digital Cultural Heritage, ITN-DCH 2017, Olimje, Slovenia, May 23--25, 2017, Revised Selected Papers}}. Springer, \bibinfo{pages}{257--266}.
\newblock


\bibitem[\protect\citeauthoryear{Santosa, Chevalier, Balakrishnan, and Singh}{Santosa et~al\mbox{.}}{2013}]%
        {santosa2013direct}
\bibfield{author}{\bibinfo{person}{Stephanie Santosa}, \bibinfo{person}{Fanny Chevalier}, \bibinfo{person}{Ravin Balakrishnan}, {and} \bibinfo{person}{Karan Singh}.} \bibinfo{year}{2013}\natexlab{}.
\newblock \showarticletitle{Direct space-time trajectory control for visual media editing}. In \bibinfo{booktitle}{\emph{Proceedings of the SIGCHI Conference on Human Factors in Computing Systems}}. \bibinfo{pages}{1149--1158}.
\newblock


\bibitem[\protect\citeauthoryear{Saquib, Huq, and Haque}{Saquib et~al\mbox{.}}{2022}]%
        {saquib2022graphiti}
\bibfield{author}{\bibinfo{person}{Nazmus Saquib}, \bibinfo{person}{Faria Huq}, {and} \bibinfo{person}{Syed~Arefinul Haque}.} \bibinfo{year}{2022}\natexlab{}.
\newblock \showarticletitle{graphiti: Sketch-based Graph Analytics for Images and Videos}. In \bibinfo{booktitle}{\emph{CHI Conference on Human Factors in Computing Systems}}. \bibinfo{pages}{1--15}.
\newblock


\bibitem[\protect\citeauthoryear{Saquib, Kazi, Wei, and Li}{Saquib et~al\mbox{.}}{2019}]%
        {saquib2019interactive}
\bibfield{author}{\bibinfo{person}{Nazmus Saquib}, \bibinfo{person}{Rubaiat~Habib Kazi}, \bibinfo{person}{Li-Yi Wei}, {and} \bibinfo{person}{Wilmot Li}.} \bibinfo{year}{2019}\natexlab{}.
\newblock \showarticletitle{Interactive body-driven graphics for augmented video performance}. In \bibinfo{booktitle}{\emph{Proceedings of the 2019 CHI Conference on Human Factors in Computing Systems}}. \bibinfo{pages}{1--12}.
\newblock


\bibitem[\protect\citeauthoryear{Silva, Cabral, Fernandes, and Correia}{Silva et~al\mbox{.}}{2012}]%
        {silva2012real}
\bibfield{author}{\bibinfo{person}{Jo{\~a}o Silva}, \bibinfo{person}{Diogo Cabral}, \bibinfo{person}{Carla Fernandes}, {and} \bibinfo{person}{Nuno Correia}.} \bibinfo{year}{2012}\natexlab{}.
\newblock \showarticletitle{Real-time annotation of video objects on tablet computers}. In \bibinfo{booktitle}{\emph{Proceedings of the 11th International Conference on Mobile and Ubiquitous Multimedia}}. \bibinfo{pages}{1--9}.
\newblock


\bibitem[\protect\citeauthoryear{Sodhi, Jones, Forsyth, Bailey, and Maciocci}{Sodhi et~al\mbox{.}}{2013}]%
        {sodhi2013bethere}
\bibfield{author}{\bibinfo{person}{Rajinder~S Sodhi}, \bibinfo{person}{Brett~R Jones}, \bibinfo{person}{David Forsyth}, \bibinfo{person}{Brian~P Bailey}, {and} \bibinfo{person}{Giuliano Maciocci}.} \bibinfo{year}{2013}\natexlab{}.
\newblock \showarticletitle{BeThere: 3D mobile collaboration with spatial input}. In \bibinfo{booktitle}{\emph{Proceedings of the SIGCHI Conference on Human Factors in Computing Systems}}. \bibinfo{pages}{179--188}.
\newblock


\bibitem[\protect\citeauthoryear{Sra, Garrido-Jurado, and Maes}{Sra et~al\mbox{.}}{2017}]%
        {sra2017oasis}
\bibfield{author}{\bibinfo{person}{Misha Sra}, \bibinfo{person}{Sergio Garrido-Jurado}, {and} \bibinfo{person}{Pattie Maes}.} \bibinfo{year}{2017}\natexlab{}.
\newblock \showarticletitle{Oasis: Procedurally generated social virtual spaces from 3d scanned real spaces}.
\newblock \bibinfo{journal}{\emph{IEEE transactions on visualization and computer graphics}} \bibinfo{volume}{24}, \bibinfo{number}{12} (\bibinfo{year}{2017}), \bibinfo{pages}{3174--3187}.
\newblock


\bibitem[\protect\citeauthoryear{Sra, Garrido-Jurado, Schmandt, and Maes}{Sra et~al\mbox{.}}{2016}]%
        {sra2016procedurally}
\bibfield{author}{\bibinfo{person}{Misha Sra}, \bibinfo{person}{Sergio Garrido-Jurado}, \bibinfo{person}{Chris Schmandt}, {and} \bibinfo{person}{Pattie Maes}.} \bibinfo{year}{2016}\natexlab{}.
\newblock \showarticletitle{Procedurally generated virtual reality from 3D reconstructed physical space}. In \bibinfo{booktitle}{\emph{Proceedings of the 22nd ACM Conference on Virtual Reality Software and Technology}}. \bibinfo{pages}{191--200}.
\newblock


\bibitem[\protect\citeauthoryear{Studios}{Studios}{[n.d.]}]%
        {holoedit}
\bibfield{author}{\bibinfo{person}{Arcturus Studios}.} \bibinfo{year}{[n.d.]}\natexlab{}.
\newblock \bibinfo{title}{HoloEdit}.
\newblock \bibinfo{howpublished}{\url{https://arcturus.studio/holoedit/}}.
\newblock


\bibitem[\protect\citeauthoryear{Sugita, Higuchi, Yonetani, Kamikubo, and Sato}{Sugita et~al\mbox{.}}{2018}]%
        {sugita2018browsing}
\bibfield{author}{\bibinfo{person}{Yuki Sugita}, \bibinfo{person}{Keita Higuchi}, \bibinfo{person}{Ryo Yonetani}, \bibinfo{person}{Rie Kamikubo}, {and} \bibinfo{person}{Yoichi Sato}.} \bibinfo{year}{2018}\natexlab{}.
\newblock \showarticletitle{Browsing group first-person videos with 3d visualization}. In \bibinfo{booktitle}{\emph{Proceedings of the 2018 ACM International Conference on Interactive Surfaces and Spaces}}. \bibinfo{pages}{55--60}.
\newblock


\bibitem[\protect\citeauthoryear{Suzuki, Karim, Xia, Hedayati, and Marquardt}{Suzuki et~al\mbox{.}}{2022}]%
        {suzuki2022augmented}
\bibfield{author}{\bibinfo{person}{Ryo Suzuki}, \bibinfo{person}{Adnan Karim}, \bibinfo{person}{Tian Xia}, \bibinfo{person}{Hooman Hedayati}, {and} \bibinfo{person}{Nicolai Marquardt}.} \bibinfo{year}{2022}\natexlab{}.
\newblock \showarticletitle{Augmented Reality and Robotics: A Survey and Taxonomy for AR-enhanced Human-Robot Interaction and Robotic Interfaces}. In \bibinfo{booktitle}{\emph{CHI Conference on Human Factors in Computing Systems}}. \bibinfo{pages}{1--33}.
\newblock


\bibitem[\protect\citeauthoryear{Suzuki, Kazi, Wei, DiVerdi, Li, and Leithinger}{Suzuki et~al\mbox{.}}{2020}]%
        {suzuki2020realitysketch}
\bibfield{author}{\bibinfo{person}{Ryo Suzuki}, \bibinfo{person}{Rubaiat~Habib Kazi}, \bibinfo{person}{Li-Yi Wei}, \bibinfo{person}{Stephen DiVerdi}, \bibinfo{person}{Wilmot Li}, {and} \bibinfo{person}{Daniel Leithinger}.} \bibinfo{year}{2020}\natexlab{}.
\newblock \showarticletitle{Realitysketch: Embedding responsive graphics and visualizations in AR through dynamic sketching}. In \bibinfo{booktitle}{\emph{Proceedings of the 33rd Annual ACM Symposium on User Interface Software and Technology}}. \bibinfo{pages}{166--181}.
\newblock


\bibitem[\protect\citeauthoryear{Tait and Billinghurst}{Tait and Billinghurst}{2015}]%
        {tait2015effect}
\bibfield{author}{\bibinfo{person}{Matthew Tait} {and} \bibinfo{person}{Mark Billinghurst}.} \bibinfo{year}{2015}\natexlab{}.
\newblock \showarticletitle{The effect of view independence in a collaborative AR system}.
\newblock \bibinfo{journal}{\emph{Computer Supported Cooperative Work (CSCW)}} \bibinfo{volume}{24}, \bibinfo{number}{6} (\bibinfo{year}{2015}), \bibinfo{pages}{563--589}.
\newblock


\bibitem[\protect\citeauthoryear{Tecchia, Alem, and Huang}{Tecchia et~al\mbox{.}}{2012}]%
        {tecchia20123d}
\bibfield{author}{\bibinfo{person}{Franco Tecchia}, \bibinfo{person}{Leila Alem}, {and} \bibinfo{person}{Weidong Huang}.} \bibinfo{year}{2012}\natexlab{}.
\newblock \showarticletitle{3D helping hands: a gesture based MR system for remote collaboration}. In \bibinfo{booktitle}{\emph{Proceedings of the 11th ACM SIGGRAPH international conference on virtual-reality continuum and its applications in industry}}. \bibinfo{pages}{323--328}.
\newblock


\bibitem[\protect\citeauthoryear{Teo, Lawrence, Lee, Billinghurst, and Adcock}{Teo et~al\mbox{.}}{2019}]%
        {teo2019mixed}
\bibfield{author}{\bibinfo{person}{Theophilus Teo}, \bibinfo{person}{Louise Lawrence}, \bibinfo{person}{Gun~A Lee}, \bibinfo{person}{Mark Billinghurst}, {and} \bibinfo{person}{Matt Adcock}.} \bibinfo{year}{2019}\natexlab{}.
\newblock \showarticletitle{Mixed reality remote collaboration combining 360 video and 3d reconstruction}. In \bibinfo{booktitle}{\emph{Proceedings of the 2019 CHI conference on human factors in computing systems}}. \bibinfo{pages}{1--14}.
\newblock


\bibitem[\protect\citeauthoryear{Thoravi~Kumaravel, Anderson, Fitzmaurice, Hartmann, and Grossman}{Thoravi~Kumaravel et~al\mbox{.}}{2019}]%
        {thoravi2019loki}
\bibfield{author}{\bibinfo{person}{Balasaravanan Thoravi~Kumaravel}, \bibinfo{person}{Fraser Anderson}, \bibinfo{person}{George Fitzmaurice}, \bibinfo{person}{Bjoern Hartmann}, {and} \bibinfo{person}{Tovi Grossman}.} \bibinfo{year}{2019}\natexlab{}.
\newblock \showarticletitle{Loki: Facilitating remote instruction of physical tasks using bi-directional mixed-reality telepresence}. In \bibinfo{booktitle}{\emph{Proceedings of the 32nd Annual ACM Symposium on User Interface Software and Technology}}. \bibinfo{pages}{161--174}.
\newblock


\bibitem[\protect\citeauthoryear{Valentin, Vineet, Cheng, Kim, Shotton, Kohli, Nie{\ss}ner, Criminisi, Izadi, and Torr}{Valentin et~al\mbox{.}}{2015}]%
        {valentin2015semanticpaint}
\bibfield{author}{\bibinfo{person}{Julien Valentin}, \bibinfo{person}{Vibhav Vineet}, \bibinfo{person}{Ming-Ming Cheng}, \bibinfo{person}{David Kim}, \bibinfo{person}{Jamie Shotton}, \bibinfo{person}{Pushmeet Kohli}, \bibinfo{person}{Matthias Nie{\ss}ner}, \bibinfo{person}{Antonio Criminisi}, \bibinfo{person}{Shahram Izadi}, {and} \bibinfo{person}{Philip Torr}.} \bibinfo{year}{2015}\natexlab{}.
\newblock \showarticletitle{Semanticpaint: Interactive 3d labeling and learning at your fingertips}.
\newblock \bibinfo{journal}{\emph{ACM Transactions on Graphics (TOG)}} \bibinfo{volume}{34}, \bibinfo{number}{5} (\bibinfo{year}{2015}), \bibinfo{pages}{1--17}.
\newblock


\bibitem[\protect\citeauthoryear{Walther-Franks, Herrlich, Karrer, Wittenhagen, Schr{\"o}der-Kroll, Malaka, and Borchers}{Walther-Franks et~al\mbox{.}}{2012}]%
        {walther2012dragimation}
\bibfield{author}{\bibinfo{person}{Benjamin Walther-Franks}, \bibinfo{person}{Marc Herrlich}, \bibinfo{person}{Thorsten Karrer}, \bibinfo{person}{Moritz Wittenhagen}, \bibinfo{person}{Roland Schr{\"o}der-Kroll}, \bibinfo{person}{Rainer Malaka}, {and} \bibinfo{person}{Jan Borchers}.} \bibinfo{year}{2012}\natexlab{}.
\newblock \showarticletitle{Dragimation: direct manipulation keyframe timing for performance-based animation}.
\newblock In \bibinfo{booktitle}{\emph{Proceedings of Graphics Interface 2012}}. \bibinfo{pages}{101--108}.
\newblock


\bibitem[\protect\citeauthoryear{Wang, Tsai, Yong, and Chan}{Wang et~al\mbox{.}}{2020}]%
        {wang2020slice}
\bibfield{author}{\bibinfo{person}{Chiu-Hsuan Wang}, \bibinfo{person}{Chia-En Tsai}, \bibinfo{person}{Seraphina Yong}, {and} \bibinfo{person}{Liwei Chan}.} \bibinfo{year}{2020}\natexlab{}.
\newblock \showarticletitle{Slice of light: Transparent and integrative transition among realities in a multi-HMD-user environment}. In \bibinfo{booktitle}{\emph{Proceedings of the 33rd Annual ACM Symposium on User Interface Software and Technology}}. \bibinfo{pages}{805--817}.
\newblock


\bibitem[\protect\citeauthoryear{Wang, Nguyen, Asente, and Dorsey}{Wang et~al\mbox{.}}{2021}]%
        {wang2021distanciar}
\bibfield{author}{\bibinfo{person}{Zeyu Wang}, \bibinfo{person}{Cuong Nguyen}, \bibinfo{person}{Paul Asente}, {and} \bibinfo{person}{Julie Dorsey}.} \bibinfo{year}{2021}\natexlab{}.
\newblock \showarticletitle{Distanciar: Authoring site-specific augmented reality experiences for remote environments}. In \bibinfo{booktitle}{\emph{Proceedings of the 2021 CHI Conference on Human Factors in Computing Systems}}. \bibinfo{pages}{1--12}.
\newblock


\bibitem[\protect\citeauthoryear{Willett, Jansen, and Dragicevic}{Willett et~al\mbox{.}}{2016}]%
        {willett2016embedded}
\bibfield{author}{\bibinfo{person}{Wesley Willett}, \bibinfo{person}{Yvonne Jansen}, {and} \bibinfo{person}{Pierre Dragicevic}.} \bibinfo{year}{2016}\natexlab{}.
\newblock \showarticletitle{Embedded data representations}.
\newblock \bibinfo{journal}{\emph{IEEE transactions on visualization and computer graphics}} \bibinfo{volume}{23}, \bibinfo{number}{1} (\bibinfo{year}{2016}), \bibinfo{pages}{461--470}.
\newblock


\bibitem[\protect\citeauthoryear{Xia, Monteiro, Van, and Suzuki}{Xia et~al\mbox{.}}{2023}]%
        {xia2023realitycanvas}
\bibfield{author}{\bibinfo{person}{Zhijie Xia}, \bibinfo{person}{Kyzyl Monteiro}, \bibinfo{person}{Kevin Van}, {and} \bibinfo{person}{Ryo Suzuki}.} \bibinfo{year}{2023}\natexlab{}.
\newblock \showarticletitle{RealityCanvas: Augmented Reality Sketching for Embedded and Responsive Scribble Animation Effects}. In \bibinfo{booktitle}{\emph{Proceedings of the 36th Annual ACM Symposium on User Interface Software and Technology}}. \bibinfo{pages}{1--14}.
\newblock


\bibitem[\protect\citeauthoryear{Yang, Gao, Li, Gao, Wang, and Zheng}{Yang et~al\mbox{.}}{2023}]%
        {yang2023track}
\bibfield{author}{\bibinfo{person}{Jinyu Yang}, \bibinfo{person}{Mingqi Gao}, \bibinfo{person}{Zhe Li}, \bibinfo{person}{Shang Gao}, \bibinfo{person}{Fangjing Wang}, {and} \bibinfo{person}{Feng Zheng}.} \bibinfo{year}{2023}\natexlab{}.
\newblock \showarticletitle{Track anything: Segment anything meets videos}.
\newblock \bibinfo{journal}{\emph{arXiv preprint arXiv:2304.11968}} (\bibinfo{year}{2023}).
\newblock


\bibitem[\protect\citeauthoryear{Yu, Blackburn-Matzen, Nguyen, Wang, Habib~Kazi, and Bousseau}{Yu et~al\mbox{.}}{2023}]%
        {yu2023videodoodles}
\bibfield{author}{\bibinfo{person}{Emilie Yu}, \bibinfo{person}{Kevin Blackburn-Matzen}, \bibinfo{person}{Cuong Nguyen}, \bibinfo{person}{Oliver Wang}, \bibinfo{person}{Rubaiat Habib~Kazi}, {and} \bibinfo{person}{Adrien Bousseau}.} \bibinfo{year}{2023}\natexlab{}.
\newblock \showarticletitle{Videodoodles: Hand-drawn animations on videos with scene-aware canvases}.
\newblock \bibinfo{journal}{\emph{ACM Transactions on Graphics (TOG)}} \bibinfo{volume}{42}, \bibinfo{number}{4} (\bibinfo{year}{2023}), \bibinfo{pages}{1--12}.
\newblock


\bibitem[\protect\citeauthoryear{Yu, Angerbauer, Mohr, Kalkofen, and Sedlmair}{Yu et~al\mbox{.}}{2020}]%
        {yu2020perspective}
\bibfield{author}{\bibinfo{person}{Xingyao Yu}, \bibinfo{person}{Katrin Angerbauer}, \bibinfo{person}{Peter Mohr}, \bibinfo{person}{Denis Kalkofen}, {and} \bibinfo{person}{Michael Sedlmair}.} \bibinfo{year}{2020}\natexlab{}.
\newblock \showarticletitle{Perspective matters: Design implications for motion guidance in mixed reality}. In \bibinfo{booktitle}{\emph{2020 IEEE International Symposium on Mixed and Augmented Reality (ISMAR)}}. IEEE, \bibinfo{pages}{577--587}.
\newblock


\bibitem[\protect\citeauthoryear{Yue, Yang, Ren, and Wang}{Yue et~al\mbox{.}}{2017}]%
        {yue2017scenectrl}
\bibfield{author}{\bibinfo{person}{Ya-Ting Yue}, \bibinfo{person}{Yong-Liang Yang}, \bibinfo{person}{Gang Ren}, {and} \bibinfo{person}{Wenping Wang}.} \bibinfo{year}{2017}\natexlab{}.
\newblock \showarticletitle{SceneCtrl: Mixed reality enhancement via efficient scene editing}. In \bibinfo{booktitle}{\emph{Proceedings of the 30th annual ACM symposium on user interface software and technology}}. \bibinfo{pages}{427--436}.
\newblock


\end{thebibliography}
